\documentclass{agujournal}

\journalname{JGR-Space Physics}

\begin{document}

\title{Universality of lower hybrid waves at Earth's magnetopause}

%% ------------------------------------------------------------------------ %%
%
%  AUTHORS AND AFFILIATIONS
%
%% ------------------------------------------------------------------------ %%

\authors{D. B. Graham \affil{1}, 
Yu. V. Khotyaintsev \affil{1}, 
C. Norgren \affil{2}, 
A. Vaivads \affil{3}, 
M. Andr\'{e} \affil{1},
J. F. Drake \affil{4},
J. Egedal \affil{5},
M. Zhou \affil{6},
O. Le Contel \affil{7},
J. M. Webster\affil{8}, 
B. Lavraud \affil{9},
I. Kacem \affil{9},
V. G\'{e}not \affil{9},
C. Jacquey \affil{9},
A. C. Rager \affil{10},
D. J. Gershman \affil{10,11},
J. L. Burch \affil{12}, and
R. E. Ergun \affil{13} }

\affiliation{1}{Swedish Institute of Space Physics, Uppsala, Sweden.}
\affiliation{2}{Birkeland Centre for Space Science, Department of Physics and Technology, University of Bergen, Bergen, Norway.}
\affiliation{3}{Space and Plasma Physics, School of Electrical Engineering and Computer Science, KTH Royal Institute of Technology, Stockholm, Sweden.}
\affiliation{4}{IREAP, University of Maryland, College Park, Maryland, USA.}
\affiliation{5}{Department of Physics, University of Wisconsin--Madison, Madison, Wisconsin 53706, USA.}
\affiliation{6}{Institute of Space Science and Technology, Nanchang University, Nanchang 330031, People's Republic of China.}
\affiliation{7}{Laboratoire de Physique des Plasmas, CNRS/Ecole Polytechnique/Sorbonne Universit\'{e}/Univ. Paris Sud/Observatoire de Paris, Paris, France.}
\affiliation{8}{Department of Physics and Astronomy, Rice University, Houston, TX, USA.}
\affiliation{9}{IRAP, Universit\'{e} de Toulouse, CNRS, CNES, UPS, Toulouse, France.}
\affiliation{10}{NASA Goddard Space Flight Center, Greenbelt, MD, USA.}
\affiliation{11}{Department of Astronomy, University of Maryland, College Park, MD, USA.}
\affiliation{12}{Southwest Research Institute, San Antonio, TX, USA.}
\affiliation{13}{Laboratory of Atmospheric and Space Physics, University of Colorado, Boulder, CO, USA.}

\correspondingauthor{D. B. Graham}{dgraham@irfu.se}

\begin{keypoints}
%\item 1: Lower hybrid waves at Earth's magnetopause are consistent with generation by the lower
%hybrid drift instability and/or modified two-stream instability.
\item 1: The velocity and density fluctuations of lower hybrid waves are resolved, showing that electrons 
remain approximately frozen in. 
\item 2: Lower hybrid wave dispersion relation and wave-normal angle
are computed from fields and particle measurements.
\item 3: Single- and multi-spacecraft methods yield consistent lower hybrid wave properties, confirming the accuracy of single-spacecraft methods.
\end{keypoints}

\begin{abstract}
Waves around the lower hybrid frequency are frequently observed at Earth's magnetopause, and readily reach 
very large amplitudes. Determining the properties of lower hybrid waves is crucial because they are thought to 
contribute to electron and ion heating, cross-field particle diffusion, anomalous resistivity, and energy transfer
between electrons and ions. All these processes could play an important role in magnetic reconnection
at the magnetopause and the evolution of the boundary layer.
In this paper,
the properties of lower hybrid waves at Earth's magnetopause are investigated using the
Magnetospheric Multiscale (MMS) mission.
For the first time, the properties of the waves are investigated using fields and
direct particle measurements. The highest-resolution electron moments resolve the velocity and density
fluctuations of lower hybrid waves,
confirming that electrons remain approximately frozen in at lower hybrid wave frequencies.
Using fields and particle moments the dispersion relation is constructed and the wave-normal angle is estimated to be close to $90^{\circ}$ to the background magnetic field. The waves are shown to have a finite parallel wave vector, suggesting that they can interact
with parallel propagating electrons.
The observed wave properties are shown to agree with theoretical predictions, the
previously used
single-spacecraft method, and four-spacecraft timing analyses.
%The lower hybrid wave properties are determined using
%four-spacecraft timing and compared with single-spacecraft methods. 
These results show that single-spacecraft methods can accurately determine lower hybrid wave properties. 

\end{abstract}

\section{Introduction}
Lower hybrid drift waves are waves that develop at frequencies between the ion and electron gyrofrequencies, with
wavelengths between the electron and ion thermal gyroradii \cite[]{krall1,davidson1}.
Under these conditions the electrons remain approximately 
magnetized, while the ions are unmagnetized. In general, lower hybrid waves are treated in the electrostatic approximation, typically
assuming a plasma beta less than unity \cite[]{krall1,davidson2}.
Both observations and simulations show that these waves have properties
consistent with predictions of the electrostatic lower hybrid drift instability, namely wave numbers of
$k \rho_e \sim 0.5$ and frequency $\omega \lesssim \omega_{LH}$, where $\rho_e$ is the electron thermal
gyroradius and $\omega_{LH}$ is the angular lower hybrid frequency \cite[]{graham7,khotyaintsev4,le2,le4}.
Although the lower hybrid wave properties are consistent with
electrostatic predictions, the waves are generally not electrostatic in the sense that the fluctuating magnetic fields
$\delta {\bf B}$ are not zero. Magnetic field fluctuations develop due to the currents associated with waves \cite[]{norgren1}.
Both observations and simulations show that these magnetic field fluctuations
are often primarily in the direction parallel to the background magnetic field, and are frequently observed at Earth's magnetopause \cite[]{bale1,graham4,graham7}.

%The lower hybrid drift instability (LHDI) is driven unstable by cross-field currents in the presence of density,
%temperature, and/or magnetic field inhomogeneities \cite[]{krall1,davidson2,davidson1,huba2,yoon1}.
%Therefore, they are frequently observed at boundary layers
%such as at Earth's magnetopause, where such inhomogeneities are significant \cite[]{bale1}. Lower hybrid waves account
%for some of the most intense electric fields at Earth's magnetopause \cite[]{graham7}.
%In particular,
Lower hybrid waves are thought to play an important role in magnetic reconnection. Lower hybrid waves can be of particular
importance because they can contribute to anomalous resistivity
\cite[]{davidson2,huba1,silin1}, heat electrons and ions \cite[]{mcbride1,cairns1}, transfer energy between electrons and
ions, and produce cross-field
particle diffusion \cite[]{treumann1,vaivads2,graham7}.
In magnetopause reconnection lower hybrid waves are found at the density
gradient on the magnetospheric side of the X line \cite[]{graham4,graham7,khotyaintsev4}, where
the stagnation point is expected to occur \cite[]{cassak1}. Therefore,
lower hybrid waves could potentially play a significant role in reconnection at Earth's magnetopause.
This can modify the predictions of two-dimensional simulations of magnetic reconnection, which suppress lower
hybrid waves. More generally, plasma boundaries, regardless of whether or not magnetic reconnection is occurring,
can be unstable to lower hybrid waves, so it is important to characterize the
observed lower hybrid waves and determine what effects they have on electrons and ions, and how they can modify the boundaries.

During the first magnetopause phase of the Magnetospheric Multiscale (MMS) mission the four spacecraft reached separations
as small as $\sim 15 \, \mathrm{km}$. These separations were either comparable to or larger
than the wavelengths of lower hybrid waves $\sim 10 \, \mathrm{km}$ at Earth's magnetopause
\cite[]{graham4,graham7}.
Therefore, because of the typically broadband (and possibly turbulent) nature
of the waves, timing analysis could not be used to accurately determine the wave properties, such as
phase speed, propagation direction, wavelength, and wave potential. These properties were determined
using a single-spacecraft method \cite[]{norgren1,graham4,khotyaintsev4,graham7}.
However, during the MMS's second
magnetopause phase beginning in September 2016, the spacecraft separations were as small as
$\sim 5 \, \mathrm{km}$. These separations are below the typical wavelength of the quasi-electrostatic lower
hybrid wave and thus enable the lower hybrid wave
properties to be determined using four-spacecraft timing analyses for the first time.
In addition, it is possible with MMS to measure the electron distributions and moments at $7.5$~ms
resolution (corresponding to a Nyquist frequency of $67$~ Hz) \cite{rager1}, which is often sufficient to resolve the lower hybrid frequency at Earth's magnetopause.

In this paper we investigate the properties and generation mechanisms of lower hybrid waves at Earth's magnetopause. For the first time we investigate the lower hybrid waves using direct particle measurements and show that
their properties are consistent with theoretical predictions.
We compare the
single-spacecraft method developed in \cite{norgren1} and single spacecraft methods developed in this paper, based on the measured electron moments with four-spacecraft timing to determine the properties of the lower hybrid waves.
When the spacecraft separations are sufficiently small to enable multi-spacecraft timing to be applied,
the results show good agreement with the single-spacecraft methods, confirming their accuracy.
Lower hybrid waves produced by magnetosheath ions entering the magnetosphere via
the finite gyroradius effect are shown to be consistent
with generation by the modified two-stream instability \cite[]{mcbride1,wu3}.
We show that lower hybrid waves are generated in the ion diffusion region
of magnetopause reconnection and are driven
by a large ${\bf E} \times {\bf B}$ electron drift and a smaller electron diamagnetic drift.
%The free energy source of the waves is the ion pressure divergence.

This paper is organized as follows:
In section \ref{theory} we review the properties of lower hybrid waves based on cold plasma theory.
In section \ref{observ} we introduce the data used.
In sections \ref{28nov2016} and \ref{14Dec2015} we investigate in detail the lower hybrid waves observed at
two magnetopause crossings observed on 28 November 2016 and 14 December 2015.
Section \ref{discussion} contains the discussion and the conclusions are stated in section \ref{conclusions}.

\section{Lower hybrid wave properties} \label{theory}
In this section we review the fields and particle properties of lower hybrid waves 
predicted from cold plasma theory. The derivation of the
cold plasma dispersion equation and the wave properties are well known and derived in several plasma physics
textbooks \cite[e.g.,][]{stix1,swanson1}, so are not repeated here. Electric fields are calculated from the dielectric 
tensor, magnetic fields are computed from Faraday's law, electron and ion velocities are calculated from the 
momentum equation, and density perturbations are calculated from the continuity equation. 
Lower hybrid waves are found for $k_{\perp} \gg k_{\parallel}$ on
the whistler dispersion surface \cite[]{andre4}, where $k_{\parallel}$ and $k_{\perp}$ are the wave numbers parallel
and perpendicular to the background magnetic field ${\bf B}$.
At the magnetopause $f_{pe}/f_{ce} > 1$, where $f_{pe}$ is
the electron plasma frequency and $f_{ce}$ is the electron cyclotron frequency, so the whistler/lower
hybrid dispersion surface does not cross any other dispersion surfaces in cold plasma theory.
In cold plasma theory the lower hybrid wave, for $k_{\parallel} = 0$ has a resonance at $f_{LH} \approx \sqrt{f_{ci} f_{ce}}$, where $f_{ci}$ is the ion cyclotron frequency, while whistler waves with $k_{\perp} = 0$ have a resonance at $f_{ce}$.

\begin{figure*}[htbp!]
\begin{center}
\includegraphics[width=160mm, height=142mm]{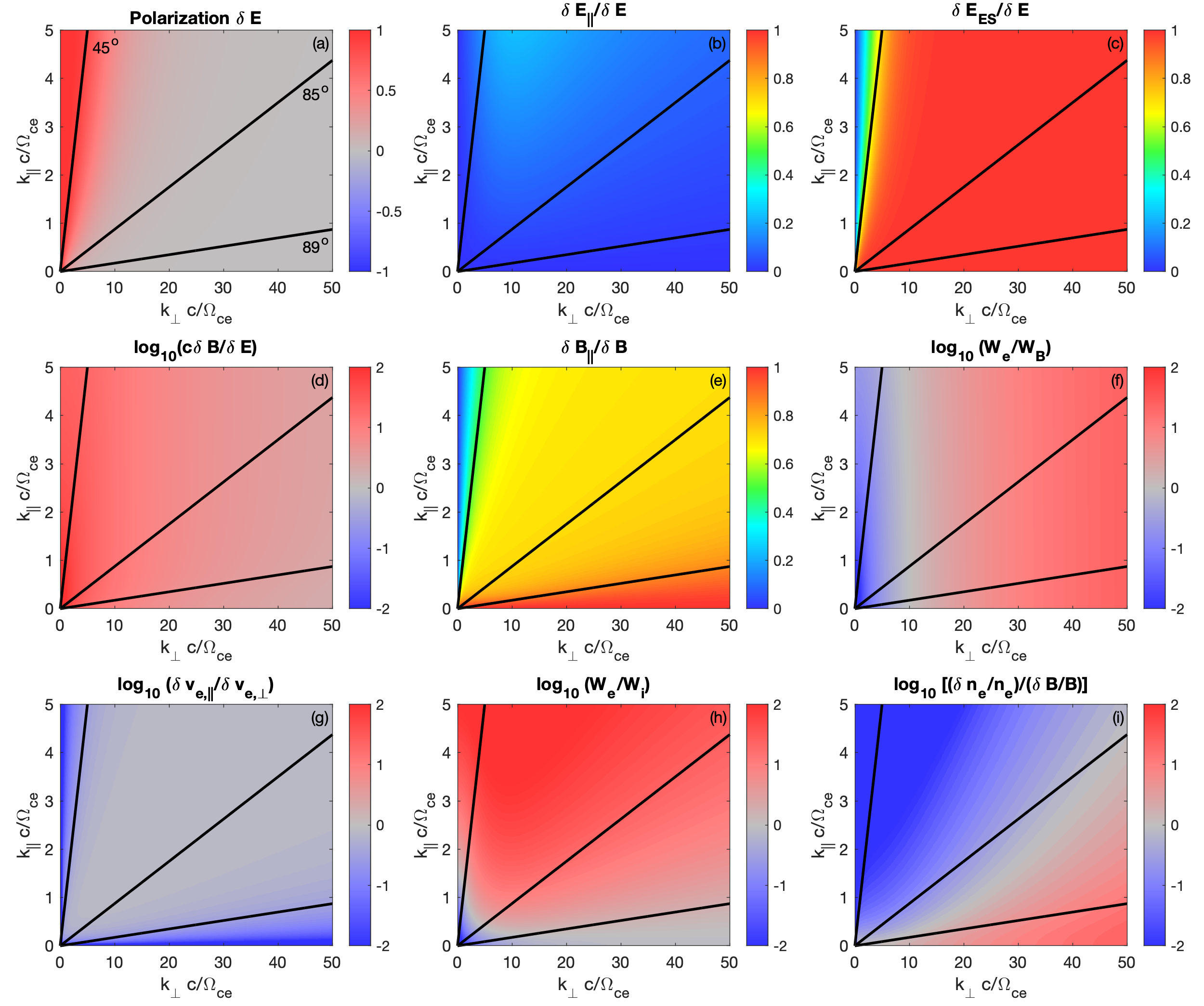}
\caption{Properties of the whistler/lower hybrid dispersion surface as functions of $k_{\parallel} d_e$ and $k_{\perp} d_e$.
The properties are calculated for $f_{pe}/f_{ce} = 10$. (a) Ellipticity of the perpendicular electric field
$\delta {\bf E}$.
(b) Ratio of the parallel to total electric field $\delta {\bf E}_{\parallel}/\delta {\bf E}$.
(c) Ratio of the electrostatic to total electric field $\delta {\bf E}_{ES}/\delta {\bf E}$.
(d) $c \delta {\bf B}/\delta {\bf E}$.
(e) Ratio of parallel to total magnetic field fluctuations $\delta {\bf B}_{\parallel}/\delta {\bf B}$.
(f) Ratio of the electron energy density to magnetic field energy density $W_e/W_B$.
(g) Ratio of parallel to perpendicular electron velocities $\delta {\bf V}_{e,\parallel}/\delta {\bf V}_{e,\perp}$.
(h) Ratio of electron to ion energy densities $W_e/W_i$.
(i) Ratio of density and magnetic field perturbations normalized to their background values
$(\delta n_e/n_e)/(\delta B/B)$.}
\label{Figure1}
\end{center}
\end{figure*}

In Figure \ref{Figure1} we plot the wave properties of the waves on the whistler/lower hybrid wave dispersion surface for $f_{pe}/f_{ce} = 10$, which is representative of the values of $f_{pe}/f_{ce}$ on the low-density side of the
magnetopause, where lower hybrid waves are expected to develop. 
At Earth's magnetopause we estimate the perpendicular wavelengths
$\lambda$ of lower hybrid waves to be $\sim 10 \, \mathrm{km}$ \cite[e.g.,][]{graham7}, which corresponds to
$k_{\perp} d_e \sim 1-3$, where $d_e = c/\omega_{pe}$ is the electron inertial length, $c$ is the speed of light, and $\omega_{pe}$ is the angular electron plasma frequency. We also expect $k_{\perp} \gg k_{\parallel}$, otherwise the lower hybrid
waves should be stabilized by electron Landau damping. 
The plots show the wave properties as functions of $k_{\perp} d_e$ and $k_{\parallel} d_e$. We focus on the range of wave vectors ${\bf k}$
where lower hybrid waves are observed. 
In each panel of Figure \ref{Figure1} the black lines indicate
wave-normal angles $\theta_{kB}$ of $45^{\circ}$, $85^{\circ}$, and $89^{\circ}$.

We now summarize the properties of lower hybrid waves shown in Figure \ref{Figure1} and their relevance to MMS observations are Earth's magnetopause. 

(1) In Figure \ref{Figure1}a we plot the ellipticity of the wave electric field $\delta {\bf E}$ with respect to the background magnetic field ${\bf B}$, where $+1$ indicates right-hand circular polarization, $-1$ indicates left-hand
circular polarization, and $0$ indicates linear polarization. For $\theta_{kB} < 45^{\circ}$, where the waves
are whistler-like we observe clear right-hand polarization. Whereas for $\theta_{kB} > 45^{\circ}$, we
observe linear polarization. Therefore, in the homogeneous approximation considered here, linear polarization is expected for lower hybrid waves. The ellipticity of the wave magnetic field $\delta {\bf B}$ (not shown) is similar to $\delta {\bf E}$.

(2) In Figure \ref{Figure1}b we plot the ratio of the parallel to total electric field
$\delta {\bf E}_{\parallel}/\delta {\bf E}$. For the range of ${\bf k}$ expected for lower hybrid waves $\delta {\bf E}_{\parallel}$ is negligible. Such a small parallel component is extremely difficult to measure accurately at lower hybrid wave frequencies with MMS (often below the uncertainty level for MMS).

(3) In Figure \ref{Figure1}c we plot the ratio of the electrostatic to total electric field
$\delta {\bf E}_{ES}/\delta {\bf E}$, where $\delta {\bf E}_{ES}$ is the electric field aligned with ${\bf k}$.
For $\theta_{kB} > 45^{\circ}$, $\delta {\bf E}_{ES}/\delta {\bf E} \approx 1$ meaning the waves are approximately
electrostatic and the electromagnetic $\delta {\bf E}$ is negligible. When the wave is whistler-like $\delta {\bf E}$
is primarily electromagnetic.

(4) In Figure \ref{Figure1}d we plot the ratio $c \delta {\bf B}/\delta {\bf E}$, which indicates
how large the magnetic field energy density $W_B  = |\delta {\bf B}|^2/(2 \mu_0)$ is compared with the electric field
energy density $W_E = \epsilon_0 |\delta {\bf E}|^2/2$. For $f_{pe}/f_{ce} = 10$, $c \delta {\bf B}/\delta {\bf E} > 1$
for the range of ${\bf k}$ shown in Figure \ref{Figure1}. We find that $c \delta {\bf B}/\delta {\bf E}$ decreases as $k_{\perp}$ increases. The fact that $c \delta {\bf B}/\delta {\bf E}$ scales with $k_{\perp}$ provides a way to estimate
$k_{\perp}$ from $\delta {\bf B}$ and $\delta {\bf E}$ observations (see Appendix \ref{app1}).
For $f_{pe}/f_{ce} = 10$ there is more energy density in the magnetic field
than in the electric field of the lower hybrid waves, despite $\delta {\bf E} \approx \delta {\bf E}_{ES}$. We thus
refer to these waves as quasi-electrostatic. For constant $k_{\perp} d_e$,
$c \delta {\bf B}/\delta {\bf E}$ increases as $f_{pe}/f_{ce}$ increases. For typical magnetopause conditions and lower hybrid wavelengths the ratio $W_B/W_E$ is often greater than one.

(5) In Figure \ref{Figure1}e we plot the ratio $\delta {\bf B}_{\parallel}/\delta {\bf B}$, where
$\delta {\bf B}_{\parallel}$ is the fluctuating magnetic field parallel to the background ${\bf B}$. For $k_{\perp} \gg k_{\parallel}$, $\delta {\bf B}_{\parallel}$ is the largest component of the fluctuating magnetic field,
and for $k_{\parallel} \approx 0$, $\delta {\bf B}_{\parallel} \approx \delta {\bf B}$. The perpendicular
$\delta {\bf B}_{\perp}$ becomes dominant for $k_{\parallel} > k_{\perp}$, when the wave is whistler-like.
For lower hybrid waves observed at the subsolar magnetopause,
which propagate in dawn-dusk direction, a finite $k_{\parallel}$ is expected to produce
$\delta {\bf B}_{\perp}$ in the direction normal to the magnetopause because
$\delta {\bf B} \cdot {\bf k} = 0$ and ${\bf k}$ is tangential to the magnetopause. 

(6) In Figure \ref{Figure1}f we plot the ratio of electron energy density to magnetic field energy density
$W_e/W_B$, where $W_e = n_e m_e |\delta {\bf V}_e|^2/2$ is the electron energy density. For the wave number
range shown in Figure \ref{Figure1}, $W_e/W_B$ depends strongly on $k_{\perp}$, with $W_B \gg W_e$
for low $k_{\perp}$ and $W_e \gg W_B$ for large $k_{\perp}$. Thus, $W_e/W_B$ provides a clear indicator
of $k_{\perp}$. We find that $W_e = W_B$ for $k_{\perp} d_e = 1$ when $k_{\perp} \gg k_{\parallel}$.

(7) In Figure \ref{Figure1}g we plot the ratio of parallel to perpendicular electron fluctuations
$\delta {\bf V}_{e,\parallel}/\delta {\bf V}_{e,\perp}$. For parallel and perpendicular ${\bf k}$ the electron fluctuations
are perpendicular to ${\bf B}$. For oblique $\theta_{kB}$ between $45^{\circ}$ and $89^{\circ}$ the parallel
and perpendicular fluctuations have comparable magnitudes.
For $k_{\perp} \gg k_{\parallel}$, $\delta {\bf V}_{e,\parallel}/\delta {\bf V}_{e,\perp}$ depends strongly on $\theta_{kB}$, which provides a way to estimate $\theta_{kB}$ when $\delta {\bf V}$ is resolved. 

(8) In Figure \ref{Figure1}h we plot the ratio of $W_e$ to $W_i$, where $W_i = n_i m_i |\delta {\bf V}_i|^2/2$ is
the ion energy density. For $\theta_{kB} \gtrsim 89^{\circ}$, $W_e$ and $W_i$ are approximately equal,
meaning that $|\delta {\bf V}_{e}| \gg |\delta {\bf V}_{i}|$ due to the much lower mass of electrons. For
$\theta_{kB} \lesssim 89^{\circ}$, we find that $W_e > W_i$, except at very small $k$.
In general, $\delta {\bf V}_i$ at lower hybrid timescales is often under-resolved by MMS, so it is difficult to compare $W_i$ with $W_e$. However, since $m_i \gg m_e$, $\delta {\bf V}_i$ is expected to be small for lower hybrid waves, except for very low $k$. 

(9) In Figure \ref{Figure1}i we plot the ratio of normalized density perturbations to
normalized magnetic field perturbations,
$(\delta n_e/n_e)/(|\delta {\bf B}|/|{\bf B}|)$. For lower hybrid-like waves $\delta n_e/n_e > |\delta {\bf B}|/|{\bf B}|$,
with $(\delta n_e/n_e)/(|\delta {\bf B}|/|{\bf B}|)$ increasing with $k$. The ratio
$(\delta n_e/n_e)/(|\delta {\bf B}|/|{\bf B}|)$ also increases as $f_{pe}/f_{ce}$ decreases.
For whistler-like waves $\delta n_e/n_e < |\delta {\bf B}|/|{\bf B}|$. In other words,
$(\delta n_e/n_e)/(|\delta {\bf B}|/|{\bf B}|)$ increases as $\theta_{kB}$ increases.
For $k_{\perp} \gg k_{\parallel}$, $(\delta n_e/n_e)/(|\delta {\bf B}|/|{\bf B}|)$ depends strongly on
$\theta_{kB}$, enabling $\theta_{kB}$ to be estimated from observations when $\delta n_e$ is resolved. 
We note that $\delta n_e/n_e$ potentially depends strongly on gradients in $B$ and $n_e$, so $(\delta n_e/n_e)/(|\delta {\bf B}|/|{\bf B}|)$ may differ significantly from the homogeneous case when the waves occur at strong gradients (see 
Appendix \ref{app1}). 

From the properties shown in Figure \ref{Figure1} we can compute important parameters of lower hybrid
waves, including the wave number, dispersion relation, and wave-normal angle from single-spacecraft observations.
In particular, we show that $W_e/W_B$ can be used to determine $k_{\perp}$.
For lower hybrid
waves the electrons are approximately frozen in, i.e., $\delta {\bf E} = - \delta {\bf  V}_{e} \times {\bf B}$ (shown below).
By assuming electrons are frozen in we can calculate $W_e$ and $W_B$ as a function of the electrostatic
potential $\delta \phi$ (see Appendix \ref{app1} for details):
\begin{equation}
W_e = \frac{1}{2} n_e m_e \delta V_e^2 = \frac{1}{2} \frac{n_e m_e}{B_0^2} \left( 1 + \frac{\omega_{ce}^2}{\omega^2}
\frac{k_{\parallel}^2}{k_{\perp}^2} \right) k_{\perp}^2 \delta \phi^2,
\label{We1}
\end{equation}
\begin{equation}
W_B = \frac{1}{2} \frac{\delta B^2}{\mu_0} = \frac{1}{2} \left( 1 + \frac{\omega_{ce}^2}{\omega^2} \frac{k_{\parallel}^2}{k_{\perp}^2} \right) \frac{\delta \phi^2 \mu_0 e^2 n_e^2}{B_0^2}.
\label{WB1}
\end{equation}
By taking the ratio of $W_e$ and $W_B$ we can estimate the dispersion relation in the spacecraft reference frame
using
\begin{equation}
\frac{W_e(\omega)}{W_B(\omega)} = d_e^2 k_{\perp}^2(\omega) \rightarrow k_{\perp}(\omega) = \frac{1}{d_e} \sqrt{\frac{W_e(\omega)}{W_B(\omega)}},
\label{WBWe1}
\end{equation}
where $W_e(\omega)$ and $W_B(\omega)$ are computed in the frequency domain using Fourier or
wavelet methods.
Thus, $k_{\perp}$ can be computed as a
function of frequency (i.e., the dispersion relation) if the electron fluctuations are resolved. 
Similarly, we can estimate $k_{\parallel}$ and $\theta_{kB}$ when $k_{\perp}$ is known using
$|\delta {\bf B}_{\parallel}|/|\delta {\bf B}|$, $(\delta n_e/n_e)/(|\delta {\bf B}|/|{\bf B}|)$, and/or
$\delta |{\bf V}_{e,\parallel}|/|\delta {\bf V}_{e,\perp}|$ as proxies.
Using these parameters we can provide a reasonable
estimate of $\theta_{kB}$ for lower hybrid waves, and potentially investigate whether they can interact with
electrons to produce parallel electron heating. 

\section{MMS Data} \label{observ}
We use data from the MMS spacecraft; we use electric field ${\bf E}$ data from electric field double probes (EDP)
\cite[]{lindqvist1,ergun3}, magnetic field ${\bf B}$ data from fluxgate magnetometer (FGM) \cite[]{russell2}
and search-coil magnetometer (SCM) \cite[]{lecontel1}, and particle
data from fast plasma investigation (FPI) \cite[]{pollock1}.
All data presented in this paper
are high-resolution burst mode data. To study lower hybrid waves we use the highest resolution
electron moments, which are sampled at $133$~Hz \cite[]{rager1}, which is typically sufficient
to resolve fluctuations associated with lower hybrid waves at Earth's magnetopause.
The ion distributions and moments are sampled at $27$~Hz, which is typically not sufficient to fully
resolve lower hybrid waves.
These high time resolution electron distributions and moments are computed with reduced azimuthal coverage 
in the spacecraft spin plane, with the azimuthal coverage being reduced from 11.25$^{\circ}$ to
45$^{\circ}$ \cite[]{pollock1,rager1}.
However, since we are interested in the changes in the bulk distribution, rather than fine structures
in the particle distribution functions, this reduced angular resolution does not present a major problem
to the data analysis here.

%Throughout this paper FGM data is used for the background ${\bf B}$, and all ${\bf B}$ fluctuations are
%measured with SCM. We use the combined FGM/SCM data product for wavelet spectrograms of ${\bf B}$.
To investigate the properties of lower hybrid waves, and the instabilities
generating them, we study two events in detail: A broad magnetopause crossing observed on 28 November 2016 far from any reconnection diffusion region and 
a magnetopause crossing near the electron diffusion region observed on 14 December 2015.
In both events the spacecraft
were in a tetrahedral configuration.

\section{28 November 2016} \label{28nov2016}
\subsection{Event overview}
\begin{figure*}[htbp!]
\begin{center}
\includegraphics[width=140mm, height=160mm]{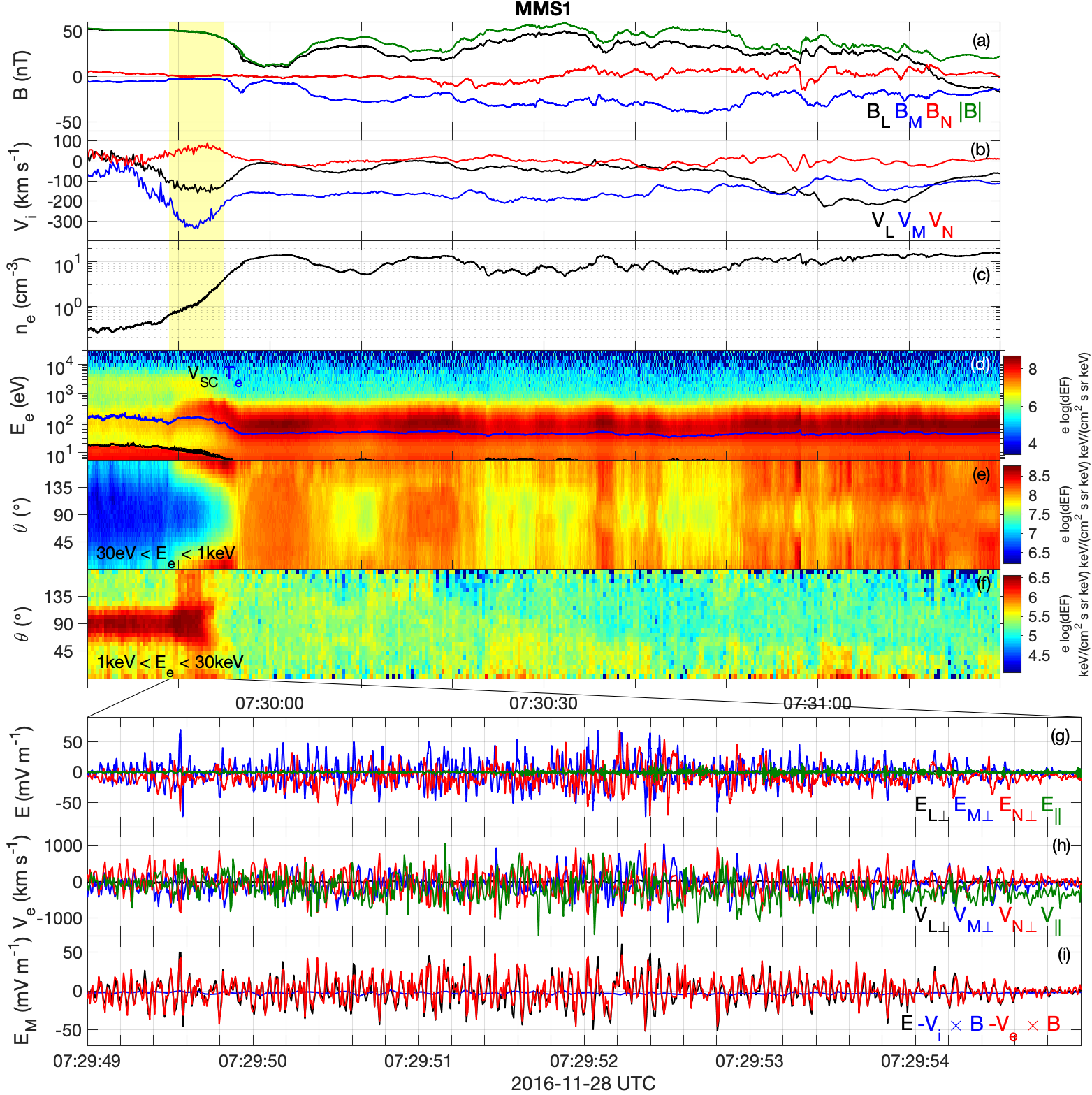}
\caption{Overview of the magnetopause crossing observed on 28 November 2016 observed by MMS1.
(a) ${\bf B}$. (b) ${\bf V}_{i}$. (c) $n_e$.
(d) Electron omni-directional
differential energy flux. (e) and (f) Electron pitch-angle distribution for electron energies
$30 \, \mathrm{eV} < E < 1 \, \mathrm{keV}$ and $1 \, \mathrm{keV} < E < 30 \, \mathrm{keV}$,
respectively. The yellow-shaded region indicates the region of intense lower hybrid wave activity.
Panels (g)--(i) Overview of the lower hybrid waves observed in the yellow shaded region.
(g) Perpendicular and parallel components of ${\bf E}$. (h) Perpendicular and parallel components of
${\bf V}_{e}$.
(i) ${\bf M}$ components of ${\bf E}$ (black), the ion convection term $- {\bf V}_i \times {\bf B}$ (blue),
and the electron convection term $- {\bf V}_e \times {\bf B}$ (red). In panels (b)--(f) we use the standard
burst particle data and in panels (h) and (i) we use the highest resolution FPI data. }
\label{28Novoverview}
\end{center}
\end{figure*}

We first investigate a magnetopause crossing on 28 November 2016 between
07:29:30 UT
and 07:32:00 UT. The spacecraft were located at [10.0, 3.0, -0.3] in Geocentric Solar Ecliptic coordinates 
(GSE), close to the subsolar point.
We transform the vector quantities into LMN coordinates based on minimum variance analysis of the
magnetic field ${\bf B}$, where ${\bf L} = [0.26, 0.09, 0.96]$, ${\bf M} = [0.33, -0.94, -0.01]$,
and ${\bf N} = [0.91, 0.32, -0.28]$ in GSE coordinates.
Based on timing analysis of $B_L$ we estimate that the magnetopause boundary moves at
$\sim 40 \, \mathrm{km} \, \mathrm{s}^{-1}$ in the $-{\bf N}$ direction (Earthward).
The mean spacecraft separation was $\sim 6 \, \mathrm{km}$.
Figures \ref{28Novoverview}a--\ref{28Novoverview}f provide an overview of the magnetopause crossing from the magnetosphere
to the magnetosheath, identified by the increase in electron density $n_e$ (Figure \ref{28Novoverview}c)
and decrease in magnetic field strength (Figure \ref{28Novoverview}a).
Figure \ref{28Novoverview}a shows that the magnetic field ${\bf B}$ remains northward ($B_L>0$) across the boundary
until 07:31:15 UT when $B_L < 0$ is observed.
Across the density gradient we observe an enhancement in the ion bulk velocity ${\bf V}_{i}$
in the $-{\bf M}$ direction (Figure \ref{28Novoverview}b).
This is due to the finite gyroradius effect of magnetosheath ions
 entering the magnetosphere.
Although this is a feature of magnetopause crossings close to the ion diffusion region, we see no clear evidence
of a nearby diffusion region, such as the Hall electric field and electron jets. We observe a southward ion flow $V_L < 0$ where $B_{L}$ changes sign,
suggestive of an ion outflow. The yellow-shaded region in
Figures \ref{28Novoverview}a--\ref{28Novoverview}c indicates when the
lower hybrid waves are observed. This region coincides with the density gradient and enhanced $V_M < 0$
ion flow. In this case the density gradient is relatively weak and the waves are observed over an extended period
of time.

Figure \ref{28Novoverview}d shows the electron omnidirectional energy flux. In the magnetosphere and
near the magnetopause we observed both hot and colder electron populations. When the lower hybrid waves
are observed there is an increase in energy of the colder electrons above the background level in the
magnetosphere and in the magnetosheath.
This corresponds to parallel electron heating, which
can be seen as the large enhancement of electron
fluxes parallel and antiparallel to ${\bf B}$ for electrons with energies $E < 1 \, \mathrm{keV}$
(Figure \ref{28Novoverview}e).
We find that $T_{\parallel}/T_{\perp}$ has a maximum of $4$ at 07:29:54.5 UT,
which is comparable to some of the largest values found in the magnetospheric inflow regions of magnetopause
reconnection \cite[]{graham4,graham7,khotyaintsev4,wang7}.
At high energies $E > 1 \, \mathrm{keV}$ the electrons have a strong perpendicular temperature anisotropy
$T_{\parallel}/T_{\perp} < 1$ in the magnetosphere and as the magnetopause boundary is approached
(Figure \ref{28Novoverview}f). At the beginning of the yellow-shaded region between
07:29:49 UT and 07:29:53 UT there is an enhancement in the flux of high-energy electrons. These high-energy electrons
tend to broaden in pitch angle, although the perpendicular temperature anisotropy remains.

%Despite observing finite-gyroradius magnetosheath ions on the magnetospheric side of the boundary and
%intense parallel electron heating, there are no apparent signs of nearby magnetic reconnection.
%There are no ion outflows or electron jets observed at the boundary,
%the magnetic shear across the boundary is small, and
%there are no clear Hall electric or magnetic fields at the magnetopause boundary.
%Therefore, while parallel electron heating and finite-gyroradius magnetosheath ions are both features
%of asymmetric reconnection close to the X line, these features can occur at Earth's magnetopause without
%nearby ongoing magnetic reconnection.

\subsection{Lower hybrid wave observations}
Figures \ref{28Novoverview}g--\ref{28Novoverview}i provide an overview of the lower hybrid waves
in the yellow-shaded region of Figures \ref{28Novoverview}a--\ref{28Novoverview}c.
Figure \ref{28Novoverview}g shows the perpendicular and parallel components of ${\bf E}$.
The lower hybrid waves
are characterized by large-amplitude fluctuations in $E_{M\perp}$ and $E_{N\perp}$, reaching a peak
amplitude of about $70$~mV~m$^{-1}$. The fact that both $E_{M\perp}$ and $E_{N\perp}$ are
observed and have different traces suggests that the waves are non-planar, and that complex
structures, such as vortices, may be developing ($E_{L\perp}$ is close to field-aligned and therefore very small). 

For this event the electron velocity fluctuations $\delta {\bf V}_e$ are resolved by FPI using the
highest cadence moments.
Figure \ref{28Novoverview}h shows the perpendicular and parallel components of the electron
velocity ${\bf V}_e$. Large-amplitude fluctuations in $V_{N\perp}$, $V_{M\perp}$, and $V_{\parallel}$
are observed, which each reach amplitudes of $\approx 1000$~km~s$^{-1}$. The fact that large
$\delta V_{\parallel}$ are observed indicates that the waves have a finite $k_{\parallel}$
(cf., Figure \ref{Figure1}g). In Figure \ref{28Novoverview}i we plot $E_M$ versus the ${\bf M}$
components of ion and electron convection terms, $-{\bf V}_i \times {\bf B}$ and
$-{\bf V}_e \times {\bf B}$, respectively. For direct comparison we have downsampled the electric field to the same cadence as the electron moments. Throughout the interval
${\bf E}_{\perp} \approx -{\bf V}_e \times {\bf B}$, as expected for lower hybrid waves. This result
also confirms that the high-resolution ${\bf V}_e$ is reliable. Overall, $-{\bf V}_i \times {\bf B}$ remains
small, as expected for lower hybrid waves. However, we note that the resolution of the ion moments
is not sufficient to fully resolve the lower hybrid waves here. We also observe large density perturbations 
associated with the waves (not shown), which reach a peak amplitude of $\delta n_e/n_e \approx 0.2$. 

\begin{figure*}[htbp!]
\begin{center}
\includegraphics[width=140mm, height=160mm]{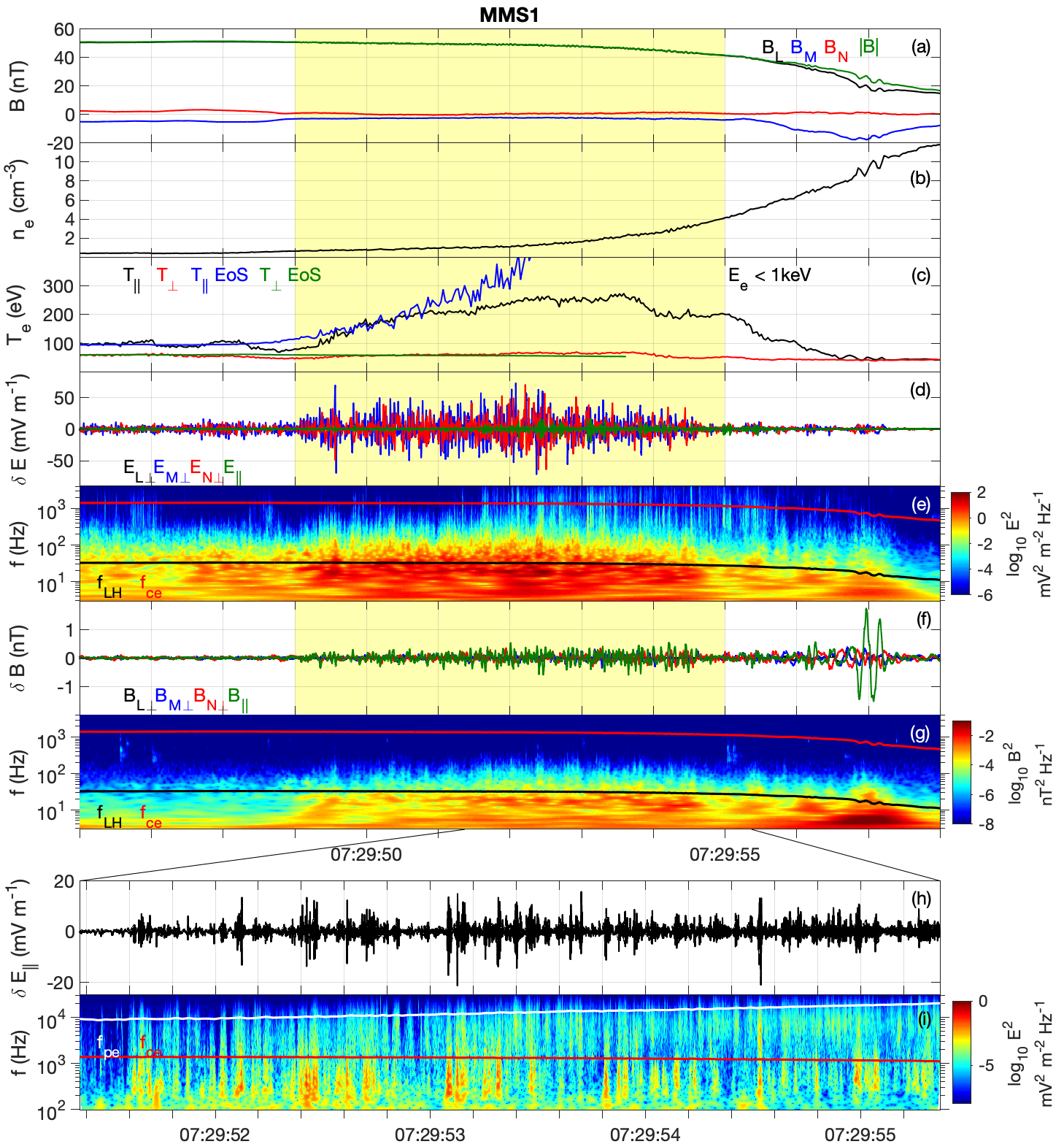}
\caption{Overview of the electric and magnetic fields at the 28 November 2016 magnetopause crossing.
(a) ${\bf B}$. (b) $n_e$. (c) Electron $T_{\parallel}$ and $T_{\perp}$ (black and red lines). The blue and green
lines are $T_{\parallel}$ and $T_{\perp}$ predicted from EoS.
(d) Perpendicular and parallel components of the fluctuating 
($f > 3 \, \mathrm{Hz}$) electric field $\delta {\bf E}$. (e) Spectrogram of ${\bf E}$.
(f) Perpendicular and parallel components of the fluctuating ($f > 3 \, \mathrm{Hz}$) magnetic field
$\delta {\bf B}$. (g) Spectrogram of ${\bf B}$.
In (e) and (g) the black and red curves are $f_{LH}$ and $f_{ce}$,
respectively.
(h) High-frequency $\delta E_{\parallel}$ and (i) the associated spectrogram (the white line is $f_{pe}$).}
\label{28Novfields}
\end{center}
\end{figure*}

The fluctuating $\delta {\bf E}$ and $\delta {\bf B}$ of the lower hybrid waves and the associated
wavelet spectrograms
are shown in Figure \ref{28Novfields}.
The fluctuations are broadband with power peaking just below the local lower hybrid
frequency $f_{LH}$ (Figure \ref{28Novfields}e).
The associated magnetic field fluctuations (Figures \ref{28Novfields}f and \ref{28Novfields}g) are
primarily parallel
to ${\bf B}$. These parallel magnetic field fluctuations $\delta B_{\parallel}$ peak at the same frequency as the
perpendicular electric field fluctuations $\delta {\bf E}_{\perp}$. The combined $\delta {\bf E}_{\perp}$ and
$\delta B_{\parallel}$ are consistent with previous observations of lower hybrid waves
\cite[]{norgren1,khotyaintsev4,graham4,graham7}, and suggest propagation approximately perpendicular to ${\bf B}$. The lower hybrid waves are observed for $\approx 6 \, \mathrm{s}$ on each spacecraft, so the waves occur over a width of $\sim 240 \, \mathrm{km}$ in the direction normal to the magnetopause, based on the estimated
magnetopause velocity of $40$~km~s$^{-1}$, suggesting that the local gradients
are weak.
%At 07:29:57 UT we see large amplitude $\delta B_{\parallel}$, which are anticorrelated with %the fluctuations in
%$n_e$ (described in detail below).

To investigate the electron heating associated with the thermal electron population we calculate
$T_{\parallel}$ and $T_{\perp}$ for thermal electrons with energies $E < 1 \, \mathrm{keV}$ (Figure \ref{28Novfields}c).
The thermal electrons in the magnetosphere have a slight parallel temperature anisotropy.
By comparing Figure \ref{28Novfields}c with Figure \ref{28Novfields}d we see that the lower hybrid waves and
parallel electron heating both start to develop at 07:29:49.0 UT, but lower hybrid activity is reduced when $T_{\parallel}/T_{\perp}$ peaks at 07:29:54.5 UT,
similar to previous observations of asymmetric reconnection \cite[]{graham4,graham7}.
Figure \ref{28Novfields}c also shows the predicted $T_{\parallel}$ and $T_{\perp}$ from the equations of
state (EoS) of the electron trapping model in \cite{le3} and \cite{egedal4}, based on the
upstream magnetospheric plasma conditions. We find good agreement between the predicted and observed
$T_{\parallel}$ and $T_{\perp}$ between 07:29:50.0 UT and  $\sim$07:29:52.0 UT, consistent
with trapping of magnetospheric electrons. After this the EoS prediction,
as well as the Chew-Goldberger-Low (CGL) scalings \cite[]{chew1} (not shown), overestimate $T_{\parallel}$. This is
likely due to the mixing of magnetospheric and magnetosheath electrons. We also see that $T_{\perp}$
is slightly larger than the predicted value after 07:29:52.0 UT,
which could be due to perpendicular electron heating
by the lower hybrid waves \cite[]{daughton2}. Overall, the deviation in the observed $T_{\parallel}$ and $T_{\perp}$
from the predicted values suggests that the lower hybrid waves scatter electrons and enable magnetosheath
electrons to enter the magnetosphere, possibly by cross-field diffusion \cite[]{graham7}.

We also observe smaller-amplitude higher-frequency parallel electric fields $\delta E_{\parallel}$
in the same region as the lower hybrid waves and large $T_{\parallel}/T_{\perp}$.
Figures \ref{28Novfields}h and \ref{28Novfields}i show $\delta E_{\parallel}$ and the associated spectrogram. 
The spectrogram shows that the waves have frequencies
ranging from a few hundred Hz to the local electron plasma frequency $f_{pe}$.
These $\delta E_{\parallel}$ are associated with bipolar electrostatic
solitary waves (ESWs), and more periodic electrostatic waves.
We observe ESWs with distinct time-scales suggesting that both fast and slow ESWs occur in this region \cite[]{graham2}.
The electrostatic waves develop between 07:29:51.5 UT and 07:29:56 UT
as seen in Figures \ref{28Novfields}d and
\ref{28Novfields}e, meaning these waves occur in the region with largest $T_{\parallel}/T_{\perp}$ rather
than span the entirety of the region of lower hybrid waves.
Before 07:29:51.5 UT large-amplitude lower hybrid waves are observed but there are negligible high-frequency
$E_{\parallel}$ fluctuations, thus the $\delta E_{\parallel}$ waves
are more closely correlated to large $T_{\parallel}/T_{\perp}$
than with the lower hybrid waves.
The region where the higher-frequency waves occur roughly coincides with when the observed $T_{\parallel}/T_{\perp}$ deviates
significantly from the EoS prediction, which suggests that the electrostatic waves are associated
with mixing of magnetospheric and magnetosheath electrons.
The electrostatic waves roughly occur over the region where $T_{\parallel}/T_{\perp} > 2$, and may be generated
by parallel electron streaming instabilities, rather than by the lower hybrid waves \cite[]{che2}.
%Although a strong perpendicular temperature anisotropy $T_{\parallel}/T_{\perp} < 1$ for %electrons with
%energies $E_e > 1 \, \mathrm{keV}$ is present in the magnetosphere and near the %magnetopause boundary, we do not see
%any significant whistler emissions close to the magnetopause. 
%Finally, we also observe intense upper hybrid waves between 07:29:44 UT and 07:29:46 UT (discussed briefly in
%the Appendix). The generation mechanism of these waves is unclear; this region approximately corresponds
%to the time when the ion bulk velocity starts to increase but no clear changes in the electron distributions
%are observed in Figure \ref{28Novoverview}.

\subsection{Lower Hybrid Wave Properties} \label{LHWPNov28}
\begin{figure*}[htbp!]
\begin{center}
\includegraphics[width=140mm, height=120mm]{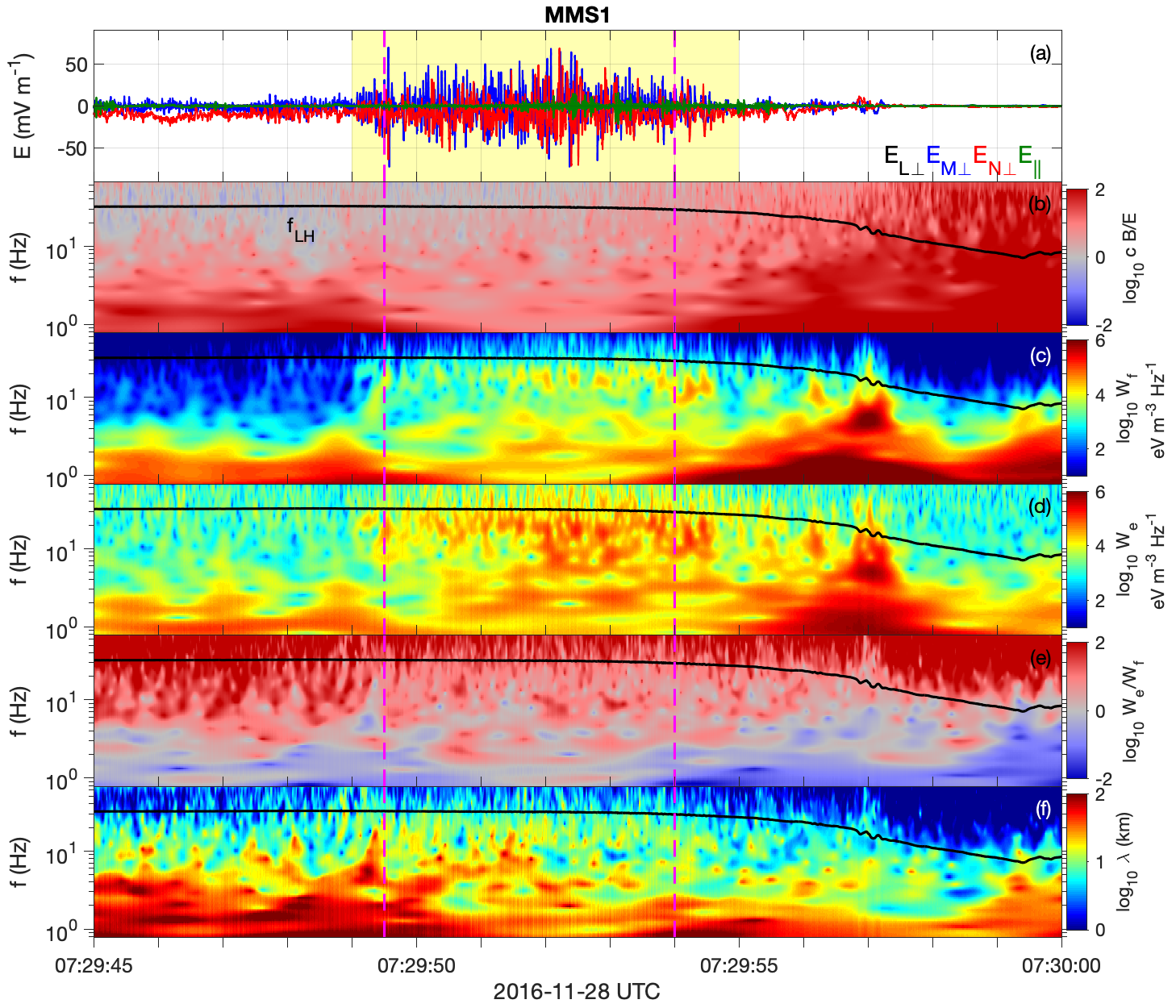}
\caption{Properties of the lower hybrid waves observed on 28 November 2016 by MMS1.
(a) ${\bf E}$ in field-aligned coordinates. (b) Spectrogram of $cB/E$. (c) Spectrogram of $W_f$.
(d ) Spectrogram of electron energy density $W_e$.
(e) Spectrogram of $W_e/W_f$. (f) Spectrogram of $\lambda$. The black lines in panels (b)--(f) indicate the local $f_{LH}$.
The magenta vertical dashed lines mark the bound the interval over which we compute the average
dispersion relation and $\theta_{kB}$. }
\label{LHprops28Nov}
\end{center}
\end{figure*}

In this subsection we investigate the field and particle properties of the lower hybrid waves and compare them with
the predictions in Figure \ref{Figure1}. In Figure \ref{LHprops28Nov} we compute the wavelet spectrograms of the
energy densities of the fields and electrons observed by MMS1.
To directly compare ${\bf B}$ and ${\bf E}$ with the
electrons we have down-sampled ${\bf B}$ and ${\bf E}$ to same cadence as the high-resolution
electron data. Figure \ref{LHprops28Nov}a shows the perpendicular and parallel components of ${\bf E}$ (without down-sampling), associated with the lower hybrid waves.
In Figure \ref{LHprops28Nov}b we plot the spectrogram of $cB/E$. Throughout the interval $cB/E \gtrsim 1$ for
the lower hybrid waves. We find that $cB/E$ tends to decrease as the frequency increases, consistent with
$k_{\perp}$ increasing with frequency (cf., Figure \ref{Figure1}d). We also find that $cB/E$
increases as $n_e$ increases and $B$ decreases, as expected when the plasma becomes more weakly magnetized ($f_{pe}/f_{ce}$ increases). 

In Figures \ref{LHprops28Nov}c and \ref{LHprops28Nov}d we plot spectrograms of the total field 
energy density $W_f = W_E + W_B$ and $W_e$.
Large enhancements in $W_f$ and $W_e$ are observed at frequencies $f \sim 10 - 30$~Hz, just
below the local $f_{LH}$, associated with the waves. We also observe a large enhancement in $W_f$
(due to ${\bf B}$ fluctuations) and $W_e$ at 07:29:57.0 UT. In Figure \ref{LHprops28Nov}e we plot
the spectrogram $W_e/W_f$, which shows that most of the energy density is in the electrons rather than
the fields for these lower hybrid waves. In addition, $W_e/W_f \approx W_e/W_B$ tends to increase with $f$, consistent
with increasing $k_{\perp}$ (cf., Figure \ref{Figure1}f).

A spectrogram of the wavelength $\lambda$ can be calculated from $W_e$ and $W_B$. 
The spectrogram of $\lambda$
is computed using 
\begin{equation}
\lambda(\omega) = 2 \pi d_e \sqrt{\frac{W_B(\omega)}{W_e(\omega)}}
\label{lambdaeq}
\end{equation}
from rearranging equation (\ref{WBWe1}). 
In Figure \ref{LHprops28Nov}f we show the spectrogram of wavelengths $\lambda$ (essentially the dispersion
relation associated with the waves). 
We find that $\lambda$ tends to decrease with increasing
frequency. For the lower hybrid waves we estimate $\lambda \sim 10 - 20$~km in the 10-30~Hz
frequency range, where $W_e$ peaks.

\begin{figure*}[htbp!]
\begin{center}
\includegraphics[width=140mm, height=120mm]{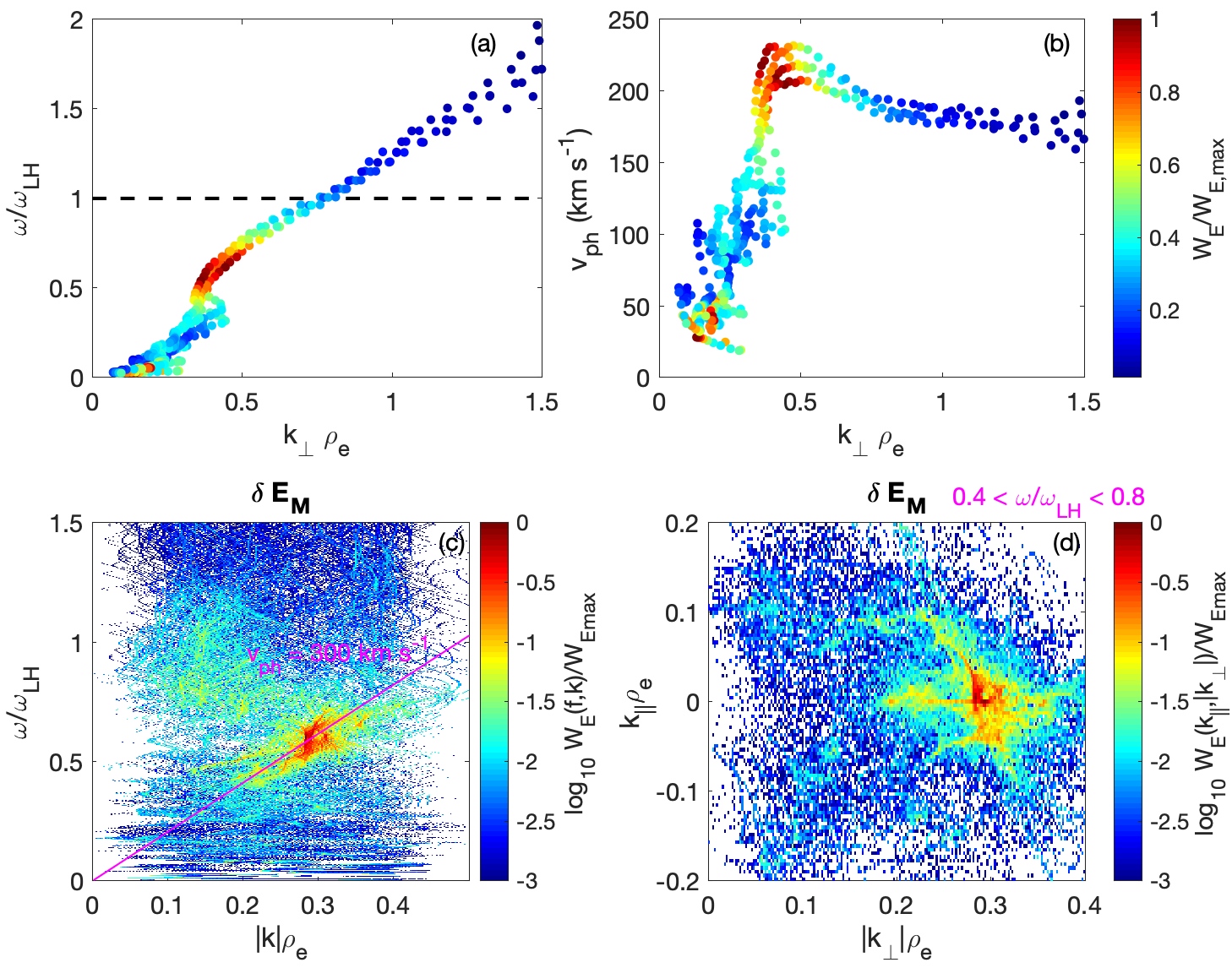}
\caption{Dispersion relation of lower hybrid waves calculated using equation (\ref{WBWe1}) 
[panels (a) and (b)] and dispersion relation computed from the phase differences of $\delta E_M$ between the spacecraft [(c) and (d)].
(a) Dispersion relations of the lower hybrid waves observed between the magenta dashed lines in
Figure \ref{LHprops28Nov}.  The black dashed lined indicates $\omega/\omega_{LH} = 1$.
(b) Phase speed $v_{ph}$ versus $k_{\perp} \rho_e$. In both panels the color of the points
indicates the value of $W_E/W_{E,max}$ of each frequency. 
(c) $W_E/W_{E,max}$ versus $\omega/\omega_{LH}$ and $k \rho_e$. 
(d) $W_E/W_{E,max}$ versus $k_{\parallel} \rho_e$ and $|k_{\perp}| \rho_e$ in the frequency range $0.3 < \omega/\omega_{LH} < 0.8$.}
\label{Disprel28Nov}
\end{center}
\end{figure*}

We now use these spectrograms of $W_e$ and $W_B$ to construct the dispersion relation of the waves for
each spacecraft. To obtain a single dispersion relation we take the median over time of
$W_e/W_B$ for each frequency to compute $k_{\perp}$.
We take this median over the time interval bounded by the magenta lines
in Figure \ref{LHprops28Nov}. The dispersion relations from each spacecraft are shown in
Figure \ref{Disprel28Nov}a. The color of the points indicates $W_E/W_{E,max}$, where $W_{E,max}$
is the maximum median value of $W_E$. As expected $\omega/\omega_{LH}$ increases with $k_{\perp} \rho_e$,
where $\rho_e$ is the median electron thermal gyroradius.
The characteristic frequencies and wave numbers of the lower hybrid waves are indicated by the largest $W_E/W_{E,max}$. We find that the observed waves have
$0.3 \lesssim k_{\perp} \rho_e \lesssim 0.5$ and frequencies $0.5 \lesssim \omega/\omega_{LH} \lesssim 0.8$. This corresponds to $9 \, \mathrm{km} \lesssim \lambda \lesssim 15 \, \mathrm{km}$. 
All spacecraft observe very similar dispersion relations, which is not surprising since the spacecraft are
separated by $\sim 6$~km, smaller than the estimated $\lambda$ of the waves. 
In Figure \ref{Disprel28Nov}b we plot the phase speed $v_{ph} = \omega/k_{\perp}$ versus
$k_{\perp} \rho_e$. In the range where the electric field power is concentrated, $W_E \gtrsim 0.6 W_{E,max}$, we find that $200$~km~s$^{-1} \lesssim v_{ph} \lesssim 240$~km~s$^{-1}$.
Overall, the computed wave properties all agree with expectations for quasi-electrostatic lower hybrid waves.

These calculations suggest that $\lambda$ is larger than the spacecraft separations, so we can compute the frequency/wave number power spectrum using the phase differences between the spacecraft to determine the wave vector ${\bf k}$. Figures \ref{Disprel28Nov}c and \ref{Disprel28Nov}d show the power spectra of $\delta E_{M}$ over the same time interval as Figures \ref{Disprel28Nov}a and \ref{Disprel28Nov}b, using 
the phase differences between the different spacecraft pairs to determine ${\bf k}$. We use the same method 
as \cite{graham5}, but generalized to four points. Figure \ref{Disprel28Nov}c shows $W_E$ versus $\omega/\omega_{LH}$ and $k_{\perp} \rho_e$. We find that $W_E$ peaks at $k_{\perp} \rho_e = 0.29$, which is slightly smaller than the values predicted in Figure \ref{Disprel28Nov}a, and corresponds to $\lambda = 16 \, \mathrm{km}$. For the peak $W_E$ we calculate $v_{ph} = 300$~km~s$^{-1}$, which is slightly larger than the values predicted in Figure \ref{Disprel28Nov}b. Figure \ref{Disprel28Nov}d shows $W_E$ versus $k_{\parallel}$ and $k_{\perp}$. We find the largest $W_E$ for $k_{\perp} \gg k_{\parallel}$, although 
power at finite $k_{\parallel}$ is observed, which is consistent with the observed $\delta V_{e,\parallel}$ in Figure \ref{28Novoverview}h. 

\begin{figure*}[htbp!]
\begin{center}
\includegraphics[width=140mm, height=130mm]{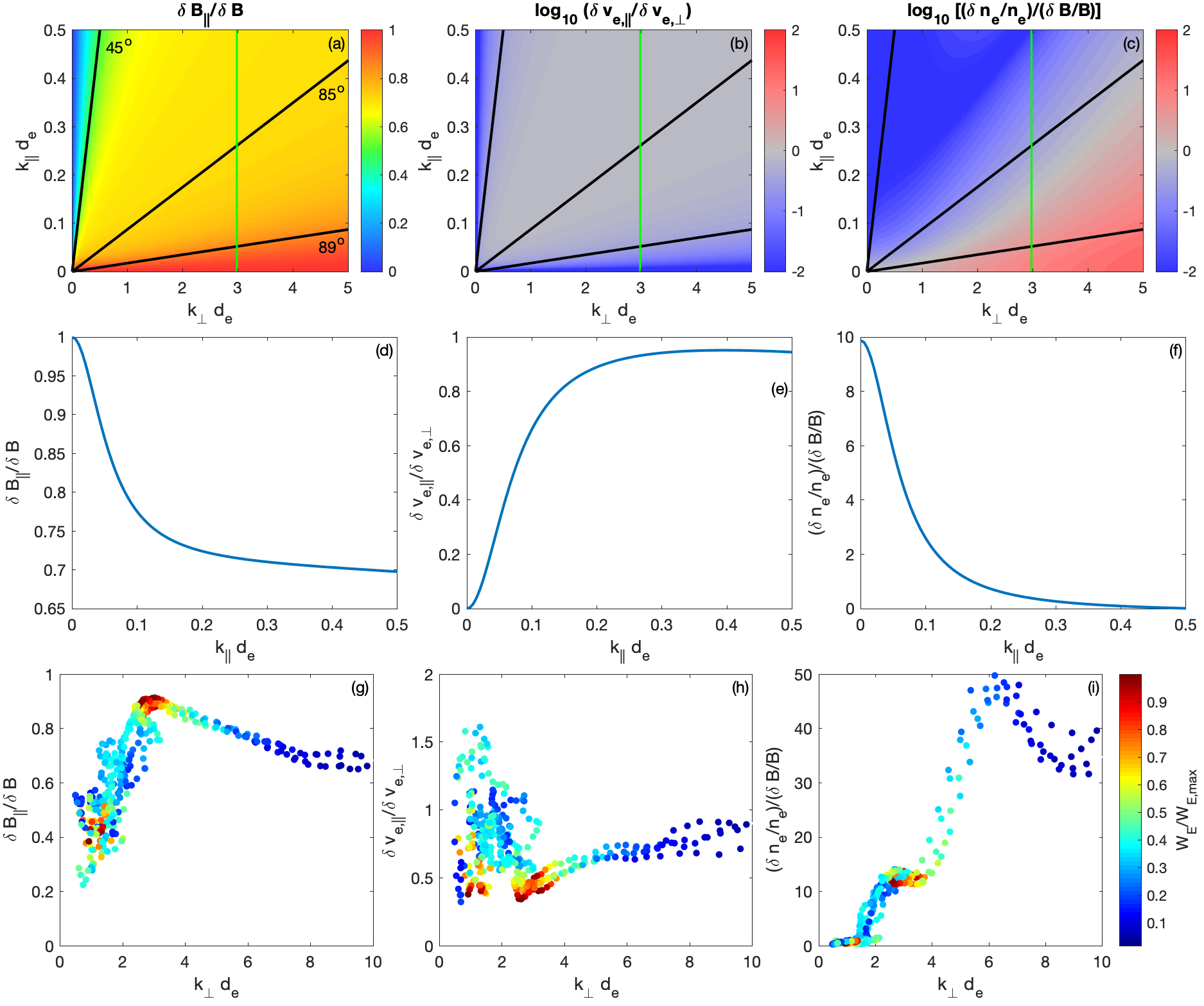}
\caption{Estimates of the wave-normal angle and $k_{\parallel}$ from fields and particle observations of
the lower hybrid waves observed on 28 November 2016.
(a)--(c) $\delta B_{\parallel}/\delta B$, $\delta V_{e,\parallel}/V_{e,\perp}$, and $(\delta n_e/n_e)/(\delta B/B)$
for whistler/lower hybrid waves versus $k_{\parallel}$ and $k_{\perp}$. We use $f_{pe}/f_{ce} = 6.7$.
The green line is $k_{\perp} d_e = 3$, the estimate $k_{\perp}$ of the observed waves.
(d)--(f) $\delta B_{\parallel}/\delta B$, $\delta V_{e,\parallel}/V_{e,\perp}$, and $(\delta n_e/n_e)/(\delta B/B)$
versus $k_{\parallel} d_e$ for $k_{\perp} d_e = 3$.
(g)--(i) Observed $\delta B_{\parallel}/\delta B$, $\delta V_{e,\parallel}/V_{e,\perp}$, and
$(\delta n_e/n_e)/(\delta B/B)$ of the lower hybrid waves versus $k_{\perp} d_e$. The colors of the points 
indicate $W_E/W_{E,max}$.}
\label{kparprops28Nov}
\end{center}
\end{figure*}

We now estimate $k_{\parallel}$ and the wave-normal angle $\theta_{kB}$ over the same interval used
to compute the dispersion relation using the parameters
$\delta B_{\parallel}/\delta B$, $\delta V_{e,\parallel}/V_{e,\perp}$, and $(\delta n_e/n_e)/(\delta B/B)$.
Figures \ref{kparprops28Nov}a--\ref{kparprops28Nov}c show $\delta B_{\parallel}/\delta B$,
$\delta V_{e,\parallel}/V_{e,\perp}$, and $(\delta n_e/n_e)/(\delta B/B)$ versus
$k_{\parallel}$ and $k_{\perp}$. For these figures we use $f_{pe}/f_{ce} = 6.7$, corresponding to the median
observed $f_{pe}/f_{ce}$ for this interval. From the observed dispersion relation we obtain
$k_{\perp} d_e \approx 3$,
indicated by the green lines in Figures \ref{kparprops28Nov}a--\ref{kparprops28Nov}c.
In Figures \ref{kparprops28Nov}d--\ref{kparprops28Nov}f we plot
$\delta B_{\parallel}/\delta B$, $\delta V_{e,\parallel}/V_{e,\perp}$, and $(\delta n_e/n_e)/(\delta B/B)$ versus
$k_{\parallel} d_e$ for $k_{\perp} d_e = 3$. All parameters vary rapidly with $k_{\parallel}$ in the limit
$k_{\parallel} \ll k_{\perp}$.

In Figures \ref{kparprops28Nov}g--\ref{kparprops28Nov}i we plot the observed
$\delta B_{\parallel}/\delta B$, $\delta V_{e,\parallel}/V_{e,\perp}$, and
$(\delta n_e/n_e)/(\delta B/B)$ versus $k_{\perp} d_e$ associated with the lower hybrid waves observed by
each spacecraft. In Figure \ref{kparprops28Nov}g we find
that $\delta B_{\parallel}/\delta B \approx 0.9$, which corresponds to $k_{\parallel} d_e \approx 0.04$ in Figure
\ref{kparprops28Nov}d. Similarly, for $\delta V_{e,\parallel}/\delta V_{e,\perp}$ we obtain
$\approx 0.4$, corresponding to $k_{\parallel} d_e \approx 0.06$. Thus, the two quantities yield consistent
estimates of $k_{\parallel}$. For $(\delta n_e/n_e)/(\delta B/B)$ we obtain $\sim 12$ from observations,
which is slightly larger than the maximum prediction for $k_{\perp} d_e = 3$. This is likely due to the low
plasma density $n_e \approx 1$~cm$^{-3}$. For lower densities the signal to noise level can be large, due
to lower counting statistics, causing $\delta n_e/n_e$ to be overestimated. Thus, higher $n_e$ should be more
favorable for computing $\delta n_e/n_e$.

Based on the observations in Figures \ref{kparprops28Nov}g and \ref{kparprops28Nov}h we estimate
$k_{\parallel} d_e \approx 0.05$, corresponding to a wave-normal angle of 
$\theta_{kB} = \tan^{-1}(k_{\perp}/k_{\parallel}) \approx 89^{\circ}$. This value is consistent with the four spacecraft observation in Figure \ref{Disprel28Nov}d. Since $k_{\parallel}$ is known
we can estimate the parallel resonance speed/energy $v_{\parallel} = \omega/k_{\parallel}$.
From the estimates in Figures \ref{Disprel28Nov} and \ref{kparprops28Nov} we obtain
$v_{\parallel} \sim 500$~eV. This energy is above the peak parallel electron thermal energy
$T_{e,\parallel} \sim 250$~eV, which suggests that the waves can interact with suprathermal electrons.
As $k_{\parallel}$ increases $v_{\parallel}$ decreases, which will result in stabilization by Landau damping. 
Overall, the estimated $\theta_{kB}$ is in excellent agreement
with values predicted for quasi-electrostatic lower hybrid waves.

\subsection{Single-Spacecraft and Multi-Spacecraft Observations of Lower Hybrid Waves}
We now use the single-spacecraft method developed in \cite{norgren1} to calculate
${\bf v}_{ph}$ and compare with the results in section \ref{LHWPNov28} and four-spacecraft
timing analysis, as well as investigate how the wave properties change as the magnetopause is approached. 
The wave potential is related to the magnetic field fluctuations parallel to ${\bf B}$ by \cite[]{norgren1}
\begin{equation}
\delta \phi_{B} = \frac{|{\bf B}| \delta B_{\parallel}}{e n_e \mu_0}.
\label{phiB}
\end{equation}
The wave potential is also determined from the fluctuating electric field $\delta {\bf E}$, using
\begin{equation}
\delta \phi_{E} = \int \delta {\bf E} \cdot {\bf v}_{ph} dt.
\label{phiE}
\end{equation}
The phase speed and direction are found by determining the best fit of $\delta \phi_{E}$ to $\delta \phi_B$. The wavelength and $k$ are found using $\lambda = v_{ph}/f$, where $f$ is the wave frequency.
Using
this method we have assumed the waves propagate perpendicular to ${\bf B}$,
which is justified because the estimated $k_{\parallel}$ is small compared with $k_{\perp}$.
Equation (\ref{phiB}) assumes electrons are frozen-in,
which is justified based on Figure \ref{28Novoverview}i. We bandpass the fields above $10$~Hz. We also estimate ${\bf v}_{ph}$ using the time offsets between the four spacecraft. 

\begin{figure*}[htbp!]
\begin{center}
\includegraphics[width=140mm, height=70mm]{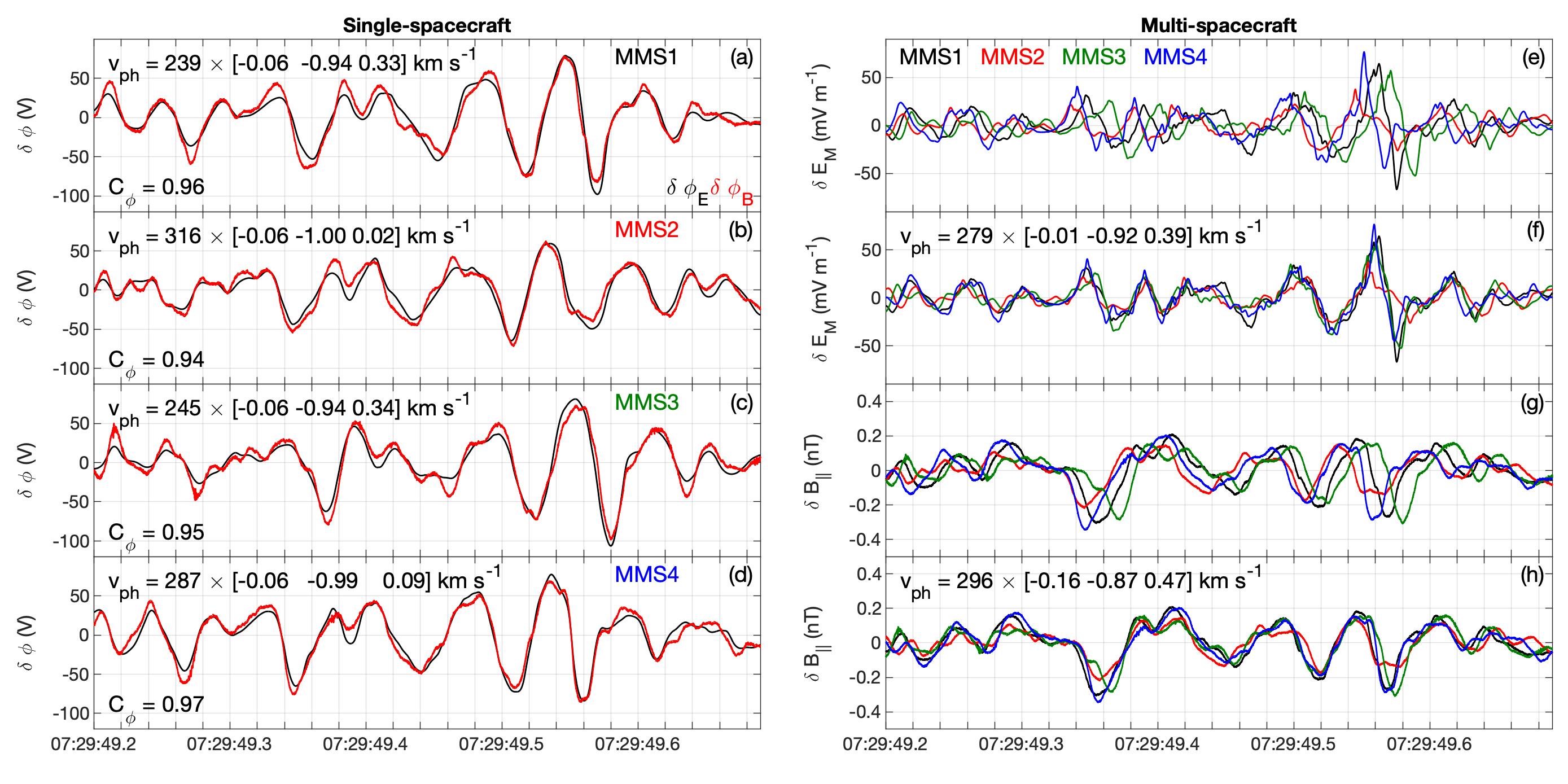}
\caption{Part of the lower hybrid wave interval observed by the four spacecraft. (a)--(d) $\delta \phi_E$ and $\delta \phi_B$
observed by each spacecraft.
The computed phase velocities (in LMN coordinates) and $C_{\phi}$ are stated in each panel.
(e) $\delta E_{M}$ observed by the four spacecraft. (f) $\delta E_{M}$ with time offsets applied
to estimate ${\bf v}_{ph}$. (g) $\delta B_{\parallel}$ observed by the four spacecraft. (h) $\delta B_{\parallel}$
with time offsets applied to estimate ${\bf v}_{ph}$. We bandpass filter above $10 \, \mathrm{Hz}$ to obtain
$\delta E_M$ and $\delta B_{\parallel}$.}
\label{timingeg}
\end{center}
\end{figure*}

As an example we compare the single-spacecraft method [equations (\ref{phiB}) and (\ref{phiE})]
with four-spacecraft timing for a short interval of
lower hybrid wave activity, shown in Figure \ref{timingeg}.
Figures \ref{timingeg}a--\ref{timingeg}d show $\delta \phi_{B}$ and $\delta \phi_{E}$ as well as the calculated ${\bf v}_{ph}$
for MMS1--4, respectively. For all spacecraft $\delta \phi_{B}$ and $\delta \phi_{E}$ show excellent agreement with
correlation coefficients $C_{\phi}$ between $\delta \phi_{B}$ and $\delta \phi_{E}$ close to $1$. All spacecraft yield
propagation directions close to the $-{\bf M}$ direction; the same direction as the cross-field
ion flow.
The phase speeds range from $v_{ph} = 240 \, \mathrm{km} \, \mathrm{s}^{-1}$ to
$320 \, \mathrm{km} \, \mathrm{s}^{-1}$, with a mean of $270 \, \mathrm{km} \, \mathrm{s}^{-1}$.
The approximate wave frequency is $f \approx 18 \, \mathrm{Hz}$, whence we calculate
$\lambda \approx 15 \, \mathrm{km}$, in agreement with the estimates in section \ref{LHWPNov28}.

Figures \ref{timingeg}e and \ref{timingeg}f show $\delta E_{M}$ from the four spacecraft
without time offsets and $\delta E_{M}$ with time
offsets applied to find the best overlap of the waveforms over the interval.
The velocity of the waves past the spacecraft is then determined from the
time offsets. We calculate ${\bf v}_{ph} \approx 280 \, \mathrm{km} \, \mathrm{s}^{-1}$ in the
$- {\bf M}$ direction, in
excellent agreement with the mean ${\bf v}_{ph}$ from the single-spacecraft method.
The angle between ${\bf k}$ and ${\bf B}$ is $\theta_{kB} = 87^{\circ}$, consistent with near perpendicular
propagation. 
We apply the same timing analysis to $\delta B_{\parallel}$ in Figures \ref{timingeg}g and \ref{timingeg}h
and find very good agreement with ${\bf v}_{ph}$
computed from $\delta E_{M}$ timing and the single-spacecraft estimates.
Based on the $\delta B_{\parallel}$ timing we calculate 
${\bf v}_{ph} \approx 300 \, \mathrm{km} \, \mathrm{s}^{-1}$ and $\theta_{kB} = 84^{\circ}$.
Thus, $\delta E_M$ and $\delta B_{\parallel}$ propagate
together at approximately the same velocity, as expected for lower hybrid waves.
For both $\delta E_{M}$ and $\delta B_{\parallel}$ with time offsets applied the waveforms remain in phase
and overlap well over multiple wave periods, which shows that the timing analyses are reliable.
These results
show that single-spacecraft methods used to calculate the lower hybrid waves can be
reproduced using four-spacecraft methods, which confirms their reliability.

\begin{figure*}[htbp!]
\begin{center}
\includegraphics[width=140mm, height=160mm]{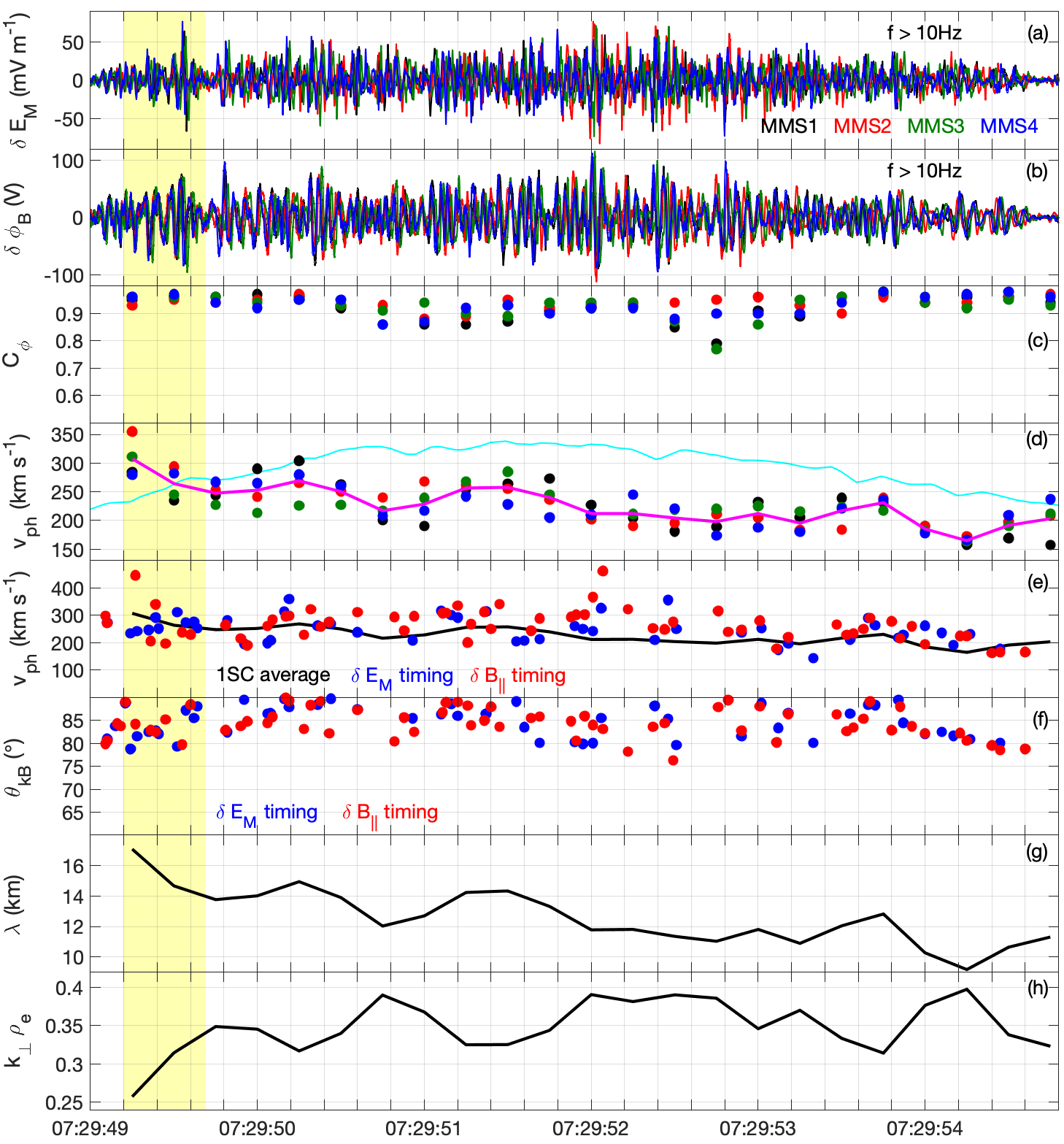}
\caption{Lower hybrid wave properties calculated using the single-spacecraft method and multi-spacecraft timing
over the yellow-shaded regions in Figures \ref{28Novoverview}--\ref{LHprops28Nov}.
(a) $\delta E_{M}$ observed by MMS1 (black), MMS2 (red), MMS3 (green), and MMS4 (blue).
(b) $\delta \phi_B$ computed for MMS1--4.
(c) $C_{\phi}$ for MMS1--MMS4 using the one-spacecraft method. (d) $v_{ph}$ for MMS1--MMS4 using the
one-spacecraft method. The magenta curve is the average from the four spacecraft and the cyan curve is
four-spacecraft averaged bulk ion speed in the ${\bf M}$ direction, $|V_{i,M}|$. (e) $v_{ph}$ from the four-spacecraft average of the single-spacecraft
method (black), and from timing analysis of $\delta E_{M}$ (blue) and $\delta B_{\parallel}$ (red).
(f) $\theta_{kB}$ from timing analysis of $\delta E_{M}$ (blue) and $\delta B_{\parallel}$ (red). (g) and (h)
$\lambda$ and $k_{\perp} \rho_e$ computed from the four-spacecraft average of the single-spacecraft method using
$f = 18 \, \mathrm{Hz}$. The yellow-shaded region is the time interval used in Figure \ref{timingeg}. }
\label{LHwaves28Nov}
\end{center}
\end{figure*}

We can investigate how the properties change across the boundary because the lower hybrid waves are
observed over an extended period of time. Figure \ref{LHwaves28Nov} shows the results based on the
single-spacecraft method and four-spacecraft timing in the yellow-shaded region of
Figures \ref{28Novoverview}--\ref{LHprops28Nov}. For the single-spacecraft method we use 0.5~s intervals
and perform the calculations for each spacecraft every 0.25~s. For the four-spacecraft timing of $\delta E_M$
and $\delta B_{\parallel}$ we calculate ${\bf v}_{ph}$ by estimating the time delays in the peaks in the waveforms.
Figures \ref{LHwaves28Nov}a and \ref{LHwaves28Nov}b show $\delta E_{M}$ and $\delta \phi_B$
from the four spacecraft. The waveforms remain similar to each other across the boundary but tend
to be more similar at earlier times, further from the boundary.
We find that $\delta \phi_B$ remains very large throughout the interval with a peak of $\delta \phi_{\mathrm{max}} \approx 120 \, \mathrm{V}$,
corresponding to $e\delta \phi_{\mathrm{max}}/k_{B} T_{e} \approx 0.7$. Such large values of $\delta \phi_B$ suggest that the waves
have amplitudes close to saturation.

Figure \ref{LHwaves28Nov}c shows
that throughout the interval the correlation coefficient $C_{\phi}$ between $\phi_E$ and $\phi_B$ remains
close to 1, indicating that the single-spacecraft method is very reliable. The phase speeds $v_{ph}$ calculated
from the single-spacecraft method are shown in Figure \ref{LHwaves28Nov}d. Each spacecraft shows similar
results, with $v_{ph}$ tending to decrease toward the boundary (the magenta line shows $v_{ph}$ averaged
over the four spacecraft). The propagation direction is consistently in the $- {\bf M}$ direction in the 
spacecraft frame. However, throughout most of the interval $v_{ph}$ is less than ${\bf V}_i$ in the $- {\bf M}$ direction. Therefore, in the bulk ion frame the waves tend to propagate in the ${\bf M}$ direction. 
Figures \ref{LHwaves28Nov}g and \ref{LHwaves28Nov}h show the wavelength
$\lambda$ and $k_{\perp} \rho_e$ computed from the averaged $v_{ph}$ and 
$f  \approx 18 \, \mathrm{Hz}$.
The predicted $\lambda$ decreases toward the magnetopause as $v_{ph}$ decreases,
while $k_{\perp} \rho_e$
remains relatively constant with $0.3 \lesssim k_{\perp} \rho_e \lesssim 0.4$, which agrees with the observations in
Figure \ref{Disprel28Nov}. These values are slightly smaller
than the typical $k_{\perp} \rho_e \approx 0.5-1$ observed at the magnetopause
\cite[]{graham4,graham7,khotyaintsev4}, 
but consistent with lower hybrid waves. Throughout the region $\lambda$
remains larger than the spacecraft separations, enabling timing analysis to be used although
the uncertainty in the timing analysis increases with decreasing $\lambda$ because the
differences in the waveforms between the spacecraft become more substantial.

Figures \ref{LHwaves28Nov}e and \ref{LHwaves28Nov}f show $v_{ph}$ and $\theta_{kB}$ based on timing
analysis of $\delta E_M$ and $\delta B_{\parallel}$. Throughout the region $v_{ph}$ calculated from timing of
$\delta E_M$ and $\delta B_{\parallel}$ agree well with each other and the single-spacecraft observations.
Statistically, there is negligible difference between $v_{ph}$ and $\theta_{kB}$ calculated from $\delta E_M$ and
$\delta B_{\parallel}$, confirming that both the $\delta E_M$ and $\delta B_{\parallel}$ perturbations propagate at the same ${\bf v}_{ph}$.
The waves propagate approximately
perpendicular to ${\bf B}$ (for all points the propagation direction was close to the $- {\bf M}$ direction).
We find that $75^{\circ} \lesssim \theta_{kB} < 90^{\circ}$, with an average of
$\theta_{kB} \approx 85^{\circ}$. The spread in values of $\theta_{kB}$ likely provide an indicator of the
uncertainty in the four-spacecraft timing, rather than the actual $\theta_{kB}$.

\begin{figure*}[htbp!]
\begin{center}
\includegraphics[width=140mm, height=90mm]{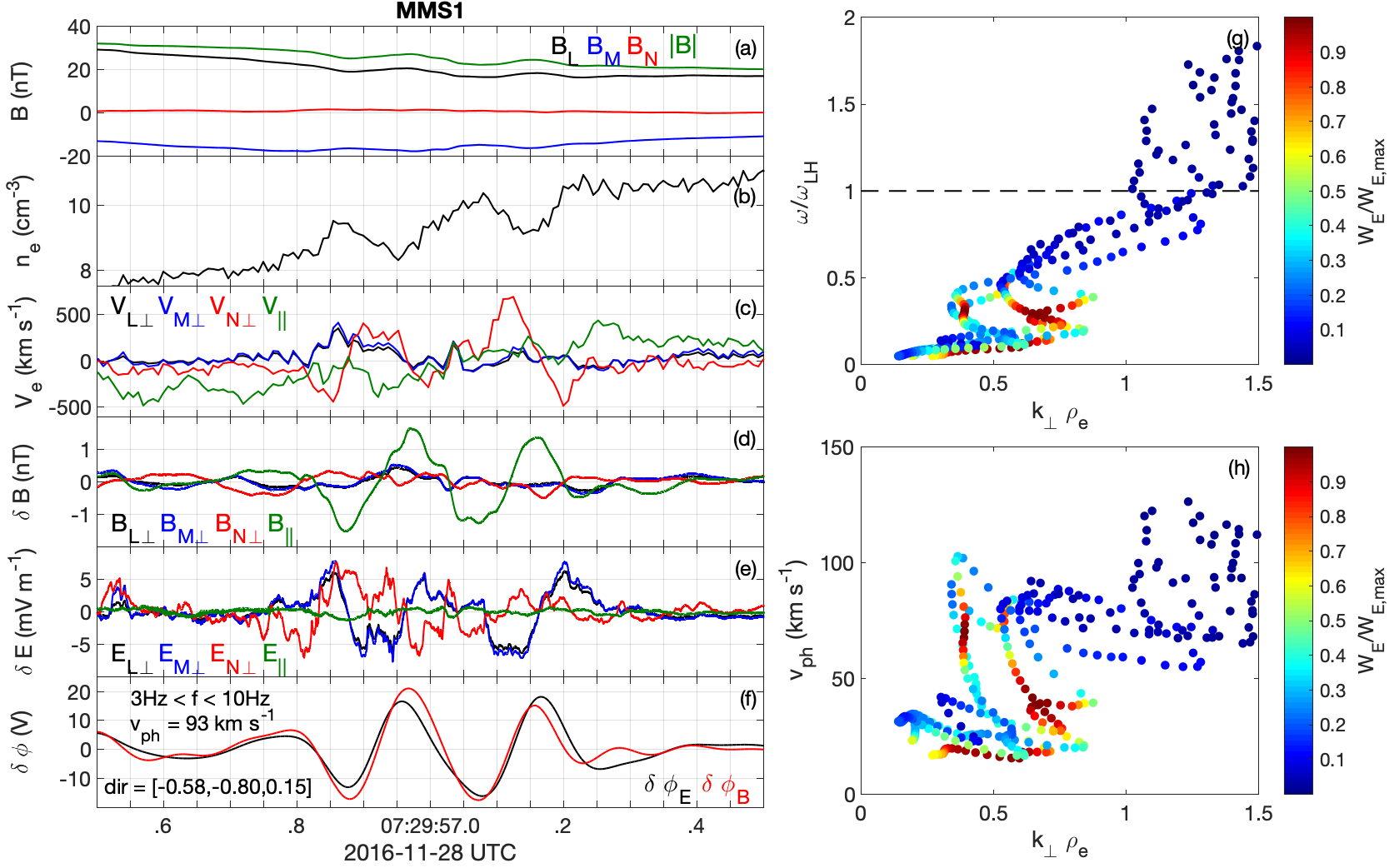}
\caption{Lower hybrid waves observed close to the magnetopause boundary. (a) ${\bf B}$.
(b) $n_e$. (c) Perpendicular and parallel components of ${\bf V}_e$. (d) Perpendicular and
parallel components of $\delta {\bf B}$ ($f > 2 \, \mathrm{Hz}$).
(e) Perpendicular and parallel components of $\delta {\bf E}$ ($f > 2 \, \mathrm{Hz}$).
(f) $\delta \phi_E$ and $\delta \phi_B$. The results from the single-spacecraft analysis are stated in the panel.
(g) Dispersion relation from all four spacecraft. (h) Phase speed $v_{ph}$ versus $k_{\perp}$ from all
four spacecraft.}
\label{LHex28Nov}
\end{center}
\end{figure*}

Figure \ref{LHex28Nov} shows the waves characterized by large $\delta B_{\parallel}$ observed at 07:29:57 UT
in Figure \ref{28Novfields}.
The waves have frequency $f \approx 5 \, \mathrm{Hz}$, so the fluctuations in $n_e$ and electron velocity
${\bf V}_e$ associated
with the wave are well resolved by FPI. The waves are observed at relatively high ion plasma beta,
$\beta_i \sim 4$, in contrast to the waves described above.
Figures \ref{LHex28Nov}a and \ref{LHex28Nov}d show that
$\delta B_{\parallel}$ is sufficiently large to significantly modify the total magnetic field $|{\bf B}|$.
Figure \ref{LHex28Nov}d shows that there is negligible $\delta {\bf B}$ perpendicular to the background ${\bf B}$,
so the amplitude of ${\bf B}$ is changed, rather than the direction. 
In addition, density fluctuations $\delta n_{e}$ are observed, which
are anticorrelated with $\delta B_{\parallel}$. Similar fluctuations in the ion density $\delta n_i$ are observed (not shown), while the ion velocity fluctuations are negligible. 
This behavior is consistent with lower hybrid waves found in simulations \cite[]{pritchett4,le2}.
The fluctuations in ${\bf V}_e$ are primarily in the ${\bf N}$ direction, consistent with
$\delta {\bf E} \times {\bf B}$ drifting electrons, due to the wave electric field (Figure \ref{LHex28Nov}e).
The electric field associated with the waves is significantly smaller
than the lower hybrid waves observed earlier.
We apply the single-spacecraft method to the waves in Figure \ref{LHex28Nov}f, to determine the wave properties.
We find good correlation between $\delta \phi_B$ and $\delta \phi_E$, with $C_{\phi} = 0.92$.
Despite the small amplitude of $\delta {\bf E}$ the waves have a peak potential of
$\delta \phi_{\mathrm{max}} \approx 20 \, \mathrm{V}$,
corresponding to $e \delta \phi_{\mathrm{max}}/k_B T_e \approx 0.4$.
We estimate a phase speed of $v_{ph} \approx 90 \, \mathrm{km} \, \mathrm{s}^{-1}$ close to the $- {\bf M}$ direction, whence we calculate
$\lambda \approx 19 \, \mathrm{km}$ for $f \approx 5 \, \mathrm{Hz}$. Despite this large $\lambda$ we
are not able to perform four-spacecraft timing analysis, which might suggest that the waves are highly localized in the ${\bf N}$ direction. 
This $\lambda$ corresponds to $k_{\perp} \rho_e \approx 0.3$, which is comparable to values for the lower hybrid waves observed
earlier. Therefore, the waves are consistent with lower hybrid waves; the much larger $\delta B_{\parallel}$
develop because the waves are observed in a more weakly magnetized plasma.

In Figures \ref{LHex28Nov}g and \ref{LHex28Nov}h we plot the dispersion relations and $v_{ph}$ versus
$k_{\perp} \rho_e$ for each spacecraft using equation (\ref{WBWe1}). For MMS1 we obtain
$k_{\perp} \rho_e \sim 0.4$ and $v_{ph} \approx 70$~km~s$^{-1}$, consistent with the observations in
Figure \ref{LHex28Nov}f. We find that $k_{\perp} \rho_e$ and $v_{ph}$ differ quite significantly between
the spacecraft.

In conclusion, we have estimated the lower hybrid wave properties using three different methods:
(1) Determining the dispersion relation from fields and particle measurements. (2) Computing $\delta \phi_B$ and
$\delta \phi_E$ from equations (\ref{phiB}) and (\ref{phiE}). (3) Four-spacecraft timing analysis of $\delta E_M$ and
$\delta B_{\parallel}$.
All three methods yield consistent results. Methods (1) and (2) primarily rely on the assumption that
electrons remain frozen-in. Based on Figure \ref{28Novoverview}i this assumption is well satisfied.
Thus, single spacecraft
methods are reliable for determining lower hybrid wave properties.

\subsection{Instability analysis}
\begin{figure*}[htbp!]
\begin{center}
\includegraphics[width=160mm, height=135mm]{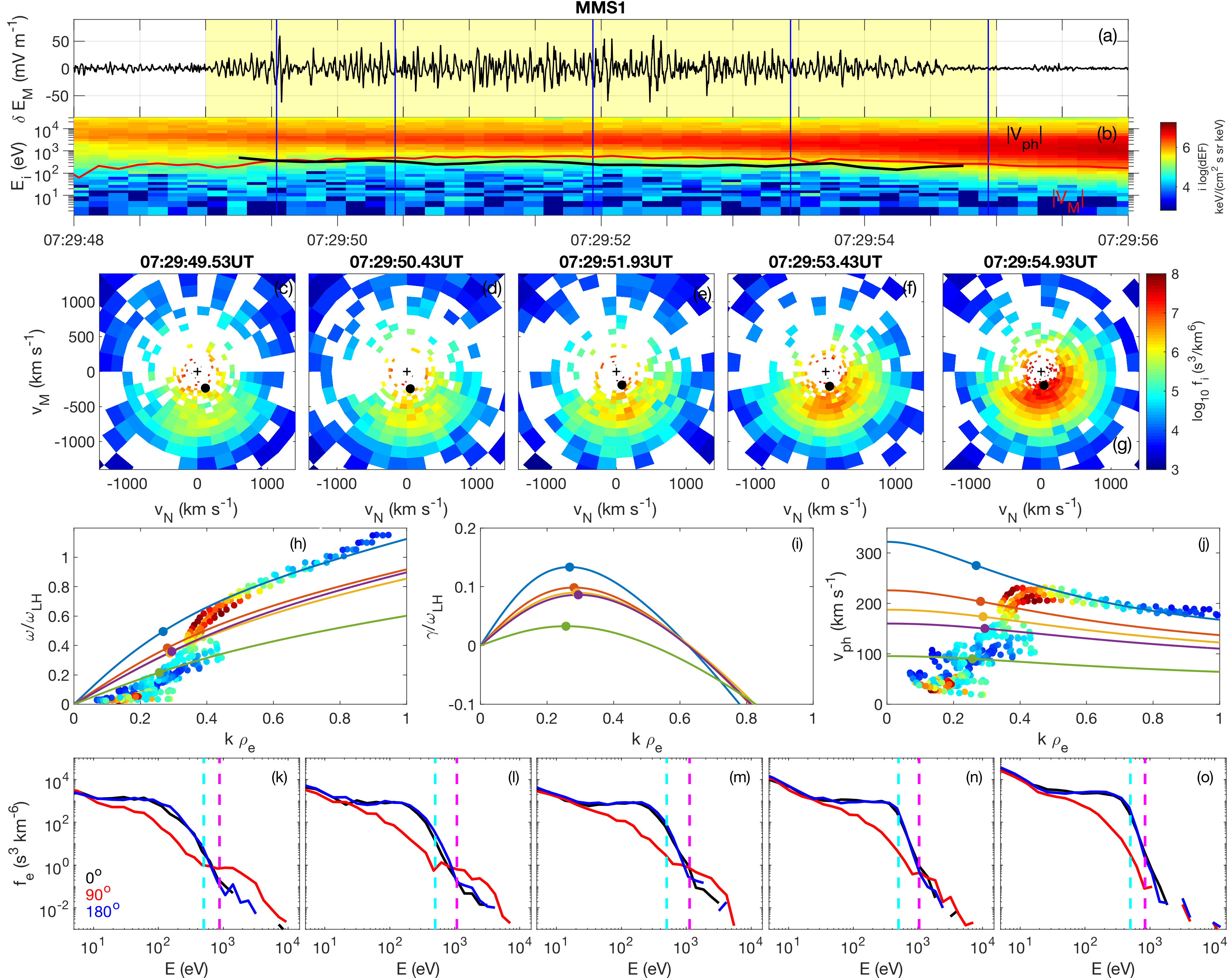}
\caption{Ion and electron distributions and MTSI dispersion relations based on MMS1 data when the lower hybrid waves are
observed. (a) $\delta E_{M}$. (b) Ion differential energy flux. (c)--(g) Ion distributions in the $v_N$--$v_M$ plane
perpendicular to ${\bf B}$ at the times indicated by the blue vertical lines in panels (a) and (b). The distributions
are shown in the spacecraft frame (the plusses indicate zero velocity and the circles indicate ${\bf v}_{ph}$ around
the time the ion distributions are observed).
(h)--(j) Frequencies, growth rates, and phase speeds versus $k$, respectively, based on the ion distributions
in panels (c) blue, (d) red, (e) gold, (f) purple, and (g) green. The dispersion relations are found by solving
equation (\ref{MTSI}) using the parameters in Table \ref{MTSItable}. We also plot the results from
Figure \ref{Disprel28Nov} in panels (h) and (i).
(k)--(o) Electron pitch-angle distributions measured by MMS1 at the same time as the ion distributions in panels
(c)--(g), respectively. The phase-space densities
$f_e$ are plotted as a function of $E$ for pitch angles $\theta = 0^{\circ}$ (black), $90^{\circ}$ (red),
and $180^{\circ}$. The magenta dashed lines indicate the parallel resonant parallel energies calculated from the predicted
dispersion relations in panels (h)--(j). The cyan lines indicate $v_{\parallel} = 500$~eV
estimated from equation (\ref{kparprops28Nov}).}
\label{DisprelsNov28}
\end{center}
\end{figure*}

To investigate the instability of the plasma we select 5 intervals across the lower hybrid
wave region, indicated by the vertical lines in Figures \ref{DisprelsNov28}a and \ref{DisprelsNov28}b.
Two-dimensional cuts of the three-dimensional ion distributions in the ${\bf N}-{\bf M}$ plane are
shown in Figures \ref{DisprelsNov28}c--\ref{DisprelsNov28}g. The distributions are shown in the spacecraft frame. In these panels the finite gyroradius ions
are the beam-like distributions centered close to the $-{\bf M}$ direction. Such distributions are similar to those found in the magnetospheric inflow region of asymmetric reconnection \cite[]{graham7}.
In each panel some hot magnetospheric ions
remain. As the magnetopause is approached the density of magnetosheath ions increases, while the bulk
velocity of magnetosheath ions decreases.
The black circles indicate ${\bf v}_{ph}$ of the lower hybrid waves at the times
of the observed distributions. In each case the lower hybrid waves propagate in approximately the same
direction as the drifting ions, but at a slower speed. Thus, in the frame of the magnetosheath ions the waves 
propagate in the ${\bf M}$ direction, the opposite direction to the spacecraft frame. 

We use these 5 ion distributions and the local plasma conditions as the basis of the following instability analysis.
The large cross-field ion drift and finite $k_{\parallel}$ of the waves,
suggests that the modified two-stream instability (MTSI) is likely active. The region
over which the lower hybrid waves are observed is broad, corresponding to weak gradients over most of the interval. 
Therefore, the electron diamagnetic drift is negligible,
especially at the start of the region where the waves are first observed.
The local electrostatic dispersion equation of the modified two-stream instability is \cite[]{mcbride1,wu3}:
\begin{equation}
0 = 1 - \frac{\omega_{pih}^2}{k^2 v_{ih}^2}Z'\left(\frac{\omega}{k v_{ih}}\right)
- \frac{\omega_{pic}^2}{k^2 v_{ic}^2}Z'\left(\frac{\omega - k_{\perp} V_{ic}}{k v_{ic}}\right)
+ \frac{2 \omega_{pe}^2}{k^2 v_{e}^2} \left[1 + \zeta_e Z(\zeta_e) \exp{(-b)} I_{0}(b) \right],
\label{MTSI}
\end{equation}
where $\omega_{pic,pih,e}$ are the cold ion, hot ion, and electron plasma frequencies,
$v_{ic,ih,e}$ are the cold ion, hot ion,
and electron thermal speeds, $Z$ is the plasma dispersion function,
$\zeta_e = \omega/(k_{\parallel} v_{e\parallel})$,
$b = k_{\perp}^2 v_{e\perp}^2/(2 \Omega_{ce}^2)$, and $I_0$
is the modified Bessel function of first kind of order zero. We model the ions with two populations associated
with the finite gyroradius magnetosheath ions propagating perpendicular to ${\bf B}$ (cold ions)
and stationary hot magnetospheric ions. The
electrons are modeled as a single stationary population. The particle moments and current density estimated
using the Curlometer technique show that there is a cross-field current associated with the ion motion;
the electrons move slower in the cross-field direction in the spacecraft frame. We find that the electrons propagate on average at
about $V_{e \perp} \sim 100 \, \mathrm{km} \, \mathrm{s}^{-1}$ in the $- {\bf M}$ direction,
much smaller than the cross-field ion drift associated with the magnetosheath ions.
Throughout most of the region
with lower hybrid waves the large-scale parallel ion and electron speeds are comparable.
The parameters used in equation (\ref{MTSI})
are summarized in Table \ref{MTSItable}, where cases 1--5 correspond to the ion distributions in
Figures \ref{DisprelsNov28}c--\ref{DisprelsNov28}g, respectively.
Throughout the region of lower hybrid waves the ion plasma beta
$\beta_i<1$, and the electron plasma beta satisfies $\beta_e \ll 1$,
justifying the electrostatic approximation for the instability analysis.

\begin{table}[ht]
\centering
\begin{center}\vspace{0.2cm}
\begin{tabular}{|c | c| c| c| c| c| c| c|}
\hline
Case & Time (UT) & $n_{ic}$ (cm$^{-3}$) & $V_{ic}$ ($\mathrm{km} \, \mathrm{s}^{-1}$) & $T_{ic}$ (eV) & B (nT) & $T_{e\parallel}$ (eV) & $T_{e\perp}$ (eV) \\
\hline
1 & 07:29:49.53  & 0.5 & 600 & 860 & 50 & 130 & 120 \\
2 & 07:29:50.43 & 0.6 & 500 & 850 & 49 & 210 & 120 \\
3 & 07:29:51.93 & 0.8 & 460 & 820 & 49 & 250 & 110 \\
4 & 07:29:53.43 & 1.5 & 400 & 710 & 47 & 260 & 80 \\
5 & 07:29:54.43 & 3.6 & 250 & 650 & 42 & 200 & 60 \\
\hline
\end{tabular}
\end{center}\vspace{0.5cm}
\caption{Parameters used to solve equation (\ref{MTSI}) based on observed values at the stated times.
The hot magnetospheric ion background is assumed
to be the same in each case with $n_{ih} = 0.2 \, \mathrm{cm}^{-3}$ and $T_{ih} = 3500 \, \mathrm{eV}$. The electron
number density is $n_{e} = n_{ic} + n_{ih}$.}
\label{MTSItable}
\end{table}

The solutions to equation (\ref{MTSI}) for the parameters in Table \ref{MTSItable} are shown in
Figures \ref{DisprelsNov28}h--\ref{DisprelsNov28}j, which show the dispersion relations, growth rates $\gamma$
as a function of $k \rho_e$, and $v_{ph}$ as a function of $k \rho_e$, respectively.
The solutions shown correspond to the values of $\theta_{kB}$ that yield the largest $\gamma$.
The results from Figure \ref{Disprel28Nov} (replotted in Figures \ref{DisprelsNov28}h and \ref{DisprelsNov28}j)
are in good agreement with the numerical predictions. 
We find that the $\theta_{kB}$ that yields the largest $\gamma$ increases as the magnetopause is approached from the magnetospheric side,
with $\theta_{kB}$ ranging from $89.1^{\circ}$ (case 1) to $89.7^{\circ}$ (case 5). Thus $\theta_{kB}$ tends to approach
$90^{\circ}$ as the ion flow decreases, although MTSI is stabilized for $\theta_{kB} = 90^{\circ}$ unless
the effects of density gradients are included. Similarly, the range of unstable $\theta_{kB}$ decreases toward
the magnetopause with MTSI being unstable for $88^{\circ} \lesssim \theta_{kB} < 90^{\circ}$ (case 1)
furthest from the magnetopause, and $89.4^{\circ} \lesssim \theta_{kB} < 90^{\circ}$ (case 5) close
to the magnetopause where the instability begins to stabilize. These $\theta_{kB}$ are consistent with
the estimated $\theta_{kB} \approx 89^{\circ}$ from Figure \ref{kparprops28Nov}.

The maximum growth rate $\gamma_{\mathrm{max}}$ decreases as the magnetopause
is approached, due to the decrease in cross-field drift of magnetosheath ions. Figure \ref{DisprelsNov28}i
shows that $0.25 \lesssim k \rho_e \lesssim 0.3$ for $\gamma_{\mathrm{max}}$, and does not change strongly
across the magnetopause. This $ k \rho_e$ is in good agreement with the observations in
Figures \ref{kparprops28Nov}c and \ref{LHwaves28Nov}h.
The predicted range of wavelengths is $13.7 \, \mathrm{km} \lesssim \lambda \lesssim 17.3 \, \mathrm{km}$;
the longest wavelength is predicted for case 1 and the shortest wavelength is predicted for case 4. These
values of $\lambda$ and the tendency of $\lambda$ to decrease toward the magnetopause are in good agreement
with the observations in Figure \ref{LHwaves28Nov}g.
Figure \ref{DisprelsNov28}h predicts
$0.2 \lesssim \omega/\omega_{LH} \lesssim 0.5$ corresponding to $\gamma_{\mathrm{max}}$, with
$\omega/\omega_{LH}$ decreasing toward the magnetopause. This change in frequency is difficult to see in
Figures \ref{28Novfields}e and \ref{28Novfields}g.
Figure \ref{DisprelsNov28}j shows that $v_{ph}$ should decrease toward the magnetopause, as the bulk
speed of magnetosheath ions decreases, and is consistent with the observations in Figure \ref{LHwaves28Nov}e.
We therefore conclude that the observed waves are consistent with generation by the modified two-stream
instability.

%The fact that the observed lower hybrid waves are consistent with MTSI suggests that they have a small but finite
%parallel wavenumber $k_{\parallel} = k \cos{\theta_{kB}}$, and thus the lower hybrid waves can
%accelerate electrons parallel to ${\bf B}$ via Landau resonance \cite[]{cairns1}.
The parallel resonant energies $v_{\parallel} = \omega/k_{\parallel}$
are $\sim 1 \, \mathrm{keV}$ for the 5
cases, based on the predicted wave properties in Figure \ref{DisprelsNov28}. The values of $v_{\parallel}$
only depend weakly on $\theta_{kB}$ over the range of $\theta_{kB}$ where
$\gamma_{\mathrm{max}} > 0$ is found. This value is in good agreement with $v_{\parallel} \sim 500$~eV,
estimated in section \ref{LHWPNov28}.
Therefore, the predicted resonant energies are above the thermal energies of the electrons
($\sim 100 - 250 \, \mathrm{eV}$).
Figures \ref{DisprelsNov28}k--\ref{DisprelsNov28}o show the electron phase-space densities
$f_e$ at pitch angles $\theta = 0^{\circ}$,
$90^{\circ}$, and $180^{\circ}$ at times corresponding to cases 1--5 in Table \ref{MTSItable}.
In each case a clear temperature anisotropy
$T_{\parallel}/T_{\perp} > 1$ occurs for the thermal electron population. In Figures
\ref{DisprelsNov28}m--\ref{DisprelsNov28}o, $f_e$ at $\theta = 0^{\circ}$ and $180^{\circ}$ are characterized by
approximately flat-top distributions
over a wide range of energies, consistent with trapping and acceleration by large-scale parallel electric fields.
The distributions are nearly identical to those found in the magnetospheric inflow regions of
magnetopause reconnection \cite[]{graham1,graham4,wang7}. Figures \ref{DisprelsNov28}k--\ref{DisprelsNov28}o
show that the parallel resonant
energies associated with the lower hybrid waves
are above the energy range of the flat-top $f_e$, suggesting that the observed
waves are not directly responsible
for electron heating in the thermal energy range. In this case the wavelengths are too large to directly interact
with the thermal population. If any shorter wavelength waves develop and contribute to the observed parallel
electron heating, they are likely quickly dissipated.

The distributions in Figures \ref{DisprelsNov28}m--\ref{DisprelsNov28}o are observed
in the interval where high-frequency
electrostatic waves are seen in Figure \ref{28Novfields}. The approximately flat-top distributions for
$\theta = 0^{\circ}$ and $180^{\circ}$ suggests marginal stability. Therefore, any modifications to the distributions
resulting in beam-like features are potentially unstable to parallel
streaming instabilities, resulting in the observed high-frequency electrostatic waves.
Once generated, the effect of the waves is to return the distribution to the
marginally stable flat-top distribution \cite[]{egedal5}. This scenario accounts for the simultaneous observation of the
flat-top distributions and high-frequency electrostatic waves over an extended interval.

In Figures \ref{DisprelsNov28}k--\ref{DisprelsNov28}m we observe a hot electron distribution for
$\theta = 90^{\circ}$ corresponding to the enhancement of hot electron fluxes in Figure \ref{28Novoverview}j.
In Figures \ref{DisprelsNov28}k and \ref{DisprelsNov28}l there is evidence of a positive slope in $f_e$ at
$\theta = 90^{\circ}$, suggesting that ring distributions are developing. At these energies there is negligible
$f_e$ at $\theta = 0^{\circ}$ and $180^{\circ}$, so we do not expect these distributions to develop as a result
of wave-particle interactions, although the distributions only develop when the lower hybrid waves are observed.
This may suggest that the high-energy electron fluxes are enhanced as a result of large-scale electric fields,
possibly set up by the finite-gyroradius effect of the magnetosheath ions.

In summary, we investigated the lower hybrid waves at an extended magnetopause crossing.
The electron velocity and density fluctuations associated with the lower hybrid waves are resolved. 
The spacecraft
separations are sufficiently small that the phase speed and propagation direction of the lower hybrid waves
can be determined using four-spacecraft timing of the electric and magnetic field fluctuations. We find excellent
agreement between the four-spacecraft timing and single-spacecraft methods for determining the
lower hybrid wave properties.
Comparison of observations with linear theory shows that the lower hybrid waves are consistent with generation
by MTSI due to the cross-field ion drift associated with the finite gyroradius magnetosheath ions entering
the magnetosphere.
This suggests that these ion distributions, which are often associated with asymmetric reconnection,
are unstable and generate lower hybrid waves.

\section{14 December 2015} \label{14Dec2015}
In this section we investigate the lower hybrid waves observed near the EDR encounter on 14 December 2015 observed at approximately 01:17:40 UT \cite[]{graham11,ergun4,chen5}. In \cite{ergun4} the waves observed close to the neutral point were interpreted as a long wavelength corrugation of the current sheet. Here, we reinvestigate the wave properties using the highest resolution electron moments and compare the results with the lower hybrid waves observed in section \ref{28nov2016}. 

\subsection{Overview}
\begin{figure*}[htbp!]
\begin{center}
\includegraphics[width=140mm, height=160mm]{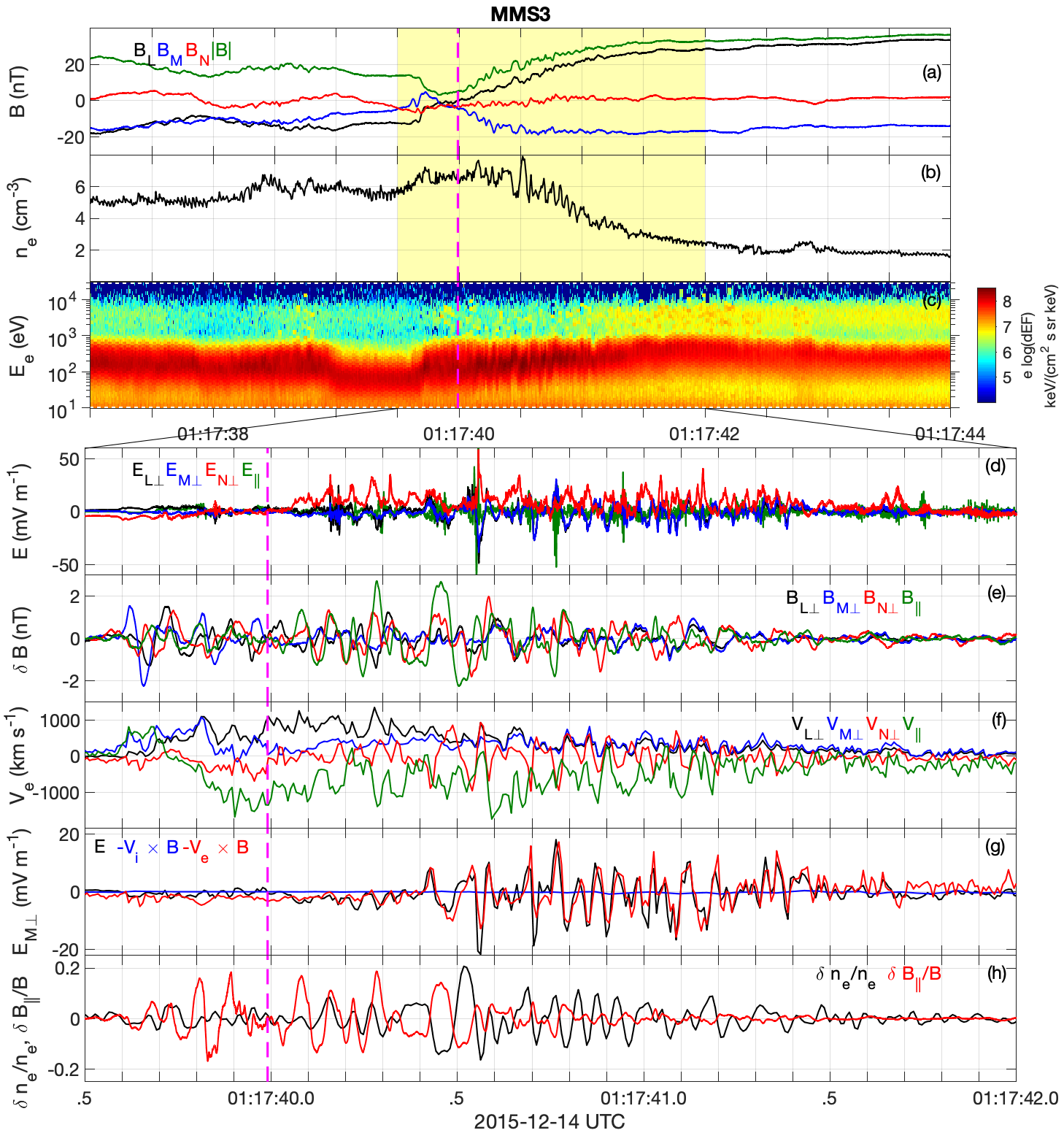}
\caption{Overview of the magnetopause crossing observed on 14 December 2015 observed by MMS3.
(a) ${\bf B}$ (Combined FGM/SCM data). (b) $n_e$ (133~Hz data).
(c) Electron omni-direction differential energy flux. Panels (d)--(h) show properties
of the lower hybrid waves in the yellow-shaded region in panels (a)--(c).
(d) Perpendicular and parallel components of ${\bf E}$. (e) Parallel and perpendicular components of the
fluctuating ($f > 5$~ Hz) magnetic field $\delta {\bf B}$.
(f) Perpendicular and parallel components of ${\bf V}_e$.
(g) $E_{M\perp}$ (black) and the M components
of the ion and electron convection terms, $(-{\bf V}_i \times {\bf B})_M$ (blue) and
$(-{\bf V}_e \times {\bf B})_M$ (red). (h) $\delta n_e/n_e$ (black) and $\delta B_{\parallel}/|{\bf B}|$ (red).}
\label{14Decoverview}
\end{center}
\end{figure*}

For this magnetopause crossing the spacecraft were located at [10.1, -4.3, -0.8] $R_E$ (GSE)
and separated
by $\sim 15 \, \mathrm{km}$.
We rotate the vector quantities into an LMN coordinate system given by ${\bf L} = [0.02, -0.52, 0.86]$,
${\bf M} = [-0.51, -0.74, -0.44]$,
${\bf N} = [0.86, -0.43, -0.27]$ in GSE coordinates.
Based on timing analysis of $B_L$ we estimate the magnetopause boundary
velocity to be $\approx 35 \times [-0.28, -0.10, 0.96] \, \mathrm{km} \, \mathrm{s}^{-1}$ (LMN).
This reconnection event has a relatively small guide-field,
$\sim 30 \%$ of the reconnection magnetic field.
Figures \ref{14Decoverview}a--\ref{14Decoverview}c provide an overview of the reconnection event from MMS3, which
crosses the magnetopause from the magnetosheath to the magnetosphere. At the beginning of the interval
the spacecraft is in
the southward reconnection outflow.
The spacecraft crosses the current sheet neutral point at about 01:17:40.0 UT where $B_L = 0$
(indicated by the magenta vertical 
dashed line) and then enters the magnetospheric inflow region. Around this region agyrotropic electron
distributions are observed, indicating close proximity to the electron diffusion region \cite[]{graham11}.
Like previous observations, the magnetospheric inflow region
is characterized by increased electric field fluctuations near $f_{LH}$ and parallel electron heating
(not shown). On the magnetospheric side of the magnetopause we observed both hot ($E \gtrsim 1$~keV) and colder ($E \lesssim 1$~keV) electron populations in Figure \ref{14Decoverview}c. The colder magnetosheath population tends to increase in temperature and decrease in density toward the magnetosphere 
within the yellow-shaded interval in Figure \ref{14Decoverview}.

In the magnetospheric inflow region we observe large perturbations in ${\bf B}$ (Figures \ref{14Decoverview}a and \ref{14Decoverview}h) and $n_e$ (Figures \ref{14Decoverview}b and \ref{14Decoverview}h). The density
perturbations are seen in the electron omnidirectional energy flux (Figure \ref{14Decoverview}c).
These perturbations are largest at the density gradient, suggestive of lower hybrid drift waves. Below we
investigate the properties of the waves, in particular, their dispersion relation and wave-normal angle. 

\subsection{Lower hybrid wave properties} \label{14DecLHwaves}
Figures \ref{14Decoverview}d--\ref{14Decoverview}h show fields and particle observations in the
yellow-shaded region of Figures \ref{14Decoverview}a and \ref{14Decoverview}b.
Figure \ref{14Decoverview}d shows the components of ${\bf E}$ perpendicular and parallel to ${\bf B}$.
Large amplitude fluctuations are seen in all components of ${\bf E}$. Lower frequency
fluctuations are seen in ${\bf E}_{\perp}$, and higher frequency ${\bf E}_{\parallel}$ are also observed. In addition,
there is a large-scale Hall electric field $E_N > 0$. Here, lower hybrid fluctuations are seen in
$E_{M\perp}$ and $E_{L\perp}$ due to the guide-field.

Figure \ref{14Decoverview}e shows that $\delta {\bf B}$ is primarily aligned with ${\bf B}$ and is largest amplitude when the $E_{M\perp}$ and $E_{L\perp}$ fluctuations are observed. We also observe
significant $\delta B_{N\perp}$, consistent with a finite $k_{\parallel}$. Large-amplitude $\delta {\bf B}$ are
also observed on the magnetosheath side of the neutral point, where ${\bf E}$ is small. 
Close to the neutral point between 01:17:40.0 UT and 01:17:40.4 UT there are fluctuations in $E_{N\perp}$  
and $\delta B_{\parallel}$. These fluctuations are inconsistent with the usual lower hybrid wave 
predictions. 

Figure \ref{14Decoverview}f shows perpendicular and parallel components of ${\bf V}_e$. Large fluctuations
in $V_{N\perp}$ are observed, consistent with lower hybrid waves. We also observe large fluctuations in
$V_{\parallel}$, indicating a finite $k_{\parallel}$, and some fluctuations in $V_{L\perp}$ and $V_{M\perp}$.
In addition, we observe large-scale parallel and perpendicular $V_e$ associated with the current sheet.
In Figure \ref{14Decoverview}g we plot $E_{M\perp}$ and the ${\bf M}$ components of the ion and electron
convection terms, $(-{\bf V}_i \times {\bf B})_{M}$ and $(-{\bf V}_e \times {\bf B})_{M}$, respectively.
Throughout the interval ${\bf E}_{\perp} \approx -{\bf V}_e \times {\bf B}$ meaning electrons remain
approximately frozen in. In contrast, $-{\bf V}_i \times {\bf B}$ remains close to zero (although the sampling
rate for ions only partially resolves the lower hybrid fluctuations). We interpret these fluctuations in $E_M$ 
between 01:17:40.4 UT and 14:17:41.5 UT as lower hybrid waves. 

In Figure \ref{14Decoverview}h we plot $\delta n_e/n_e$ and $\delta B_{\parallel}/B$, where the fluctuating quantities are assumed to have $f > 5$~Hz. Both quantities reach maximum
values of $\approx 0.2$. The largest $\delta n_e/n_e$ are colocated with largest $E_{M\perp}$, suggesting
that the density perturbations are associated with the lower hybrid waves on the lower-density side of the
current sheet. In contrast, $\delta B_{\parallel}/B$ become larger as the plasma becomes more weakly
magnetized and are largest near the center of the current sheet, where ${\bf B}$ is close to the ${\bf M}$
direction. Thus, at low densities $(\delta n_e/n_e)/(\delta B_{\parallel}/B) > 1$, while at higher densities
$(\delta n_e/n_e)/(\delta B_{\parallel}/B) < 1$. While this trend is qualitatively consistent with cold plasma 
predictions, the gradients in $n$ and $B$ will modify the predictions (see Appendix \ref{app1}). 
We find that $\delta n_e$
and $\delta B_{\parallel}$ tend to be anticorrelated, where the lower hybrid waves are observed, while 
close to the neutral point $\delta n_e$ and $\delta B_{\parallel}$ are close to in phase.
We note that since fluctuations in $E_{N\perp}$, $V_{L\perp}$ and
$V_{M\perp}$ are observed, the waves are non-planar, and possibly vortex-like structures 
\cite[]{tanaka2,norgren1,price1}. 
%Such vortex like structures are likely due to the localization of the waves in the ${\bf N}$ direction. 
We 
conclude that the waves observed between 01:17:40.4 UT and 01:17:41.5 UT on MMS3 are lower hybrid waves. 

For this event we can investigate whether the differences between ${\bf E}$ and $- {\bf V}_e \times {\bf B}$ are due to electron pressure fluctuations associated with the observed $\delta n_e/n_e$. The electron momentum equation is given by 
\begin{equation}
{\bf E} + {\bf V}_e \times {\bf B} = - \frac{\nabla \cdot {\bf P}_e}{n e} - \frac{m_e}{e} \left[ \frac{\partial {\bf V}_e}{\partial t} + \left( {\bf V}_e \cdot \nabla \right) {\bf V}_e \right],
\label{emomentum}
\end{equation}
where ${\bf P}_e$ is the electron pressure tensor. We can estimate the pressure divergence term in the ${\bf M}$ direction with a single-spacecraft method using $- \nabla \cdot {\bf P}_e/ne \approx - \nabla P_{e,\perp}/ne \approx (n e v_{ph})^{-1} \partial P_{e,\perp}/\partial t$, where $v_{ph}$ is the speed of the pressure fluctuations past the spacecraft in the ${\bf M}$ direction, and $P_{e,\perp}$ is the perpendicular electron pressure. We use $v_{ph} = 220$~km~s$^{-1}$, which is determined by the best fit of $- \nabla P_{e,\perp}/ne$ to ${\bf E} + {\bf V}_e \times {\bf B}$. This provides an estimate of $v_{ph}$ for the waves. 
This $v_{ph}$ is calculated in the spacecraft frame, which approximately corresponds to the ion stationary frame. 

\begin{figure*}[htbp!]
\begin{center}
\includegraphics[width=140mm, height=100mm]{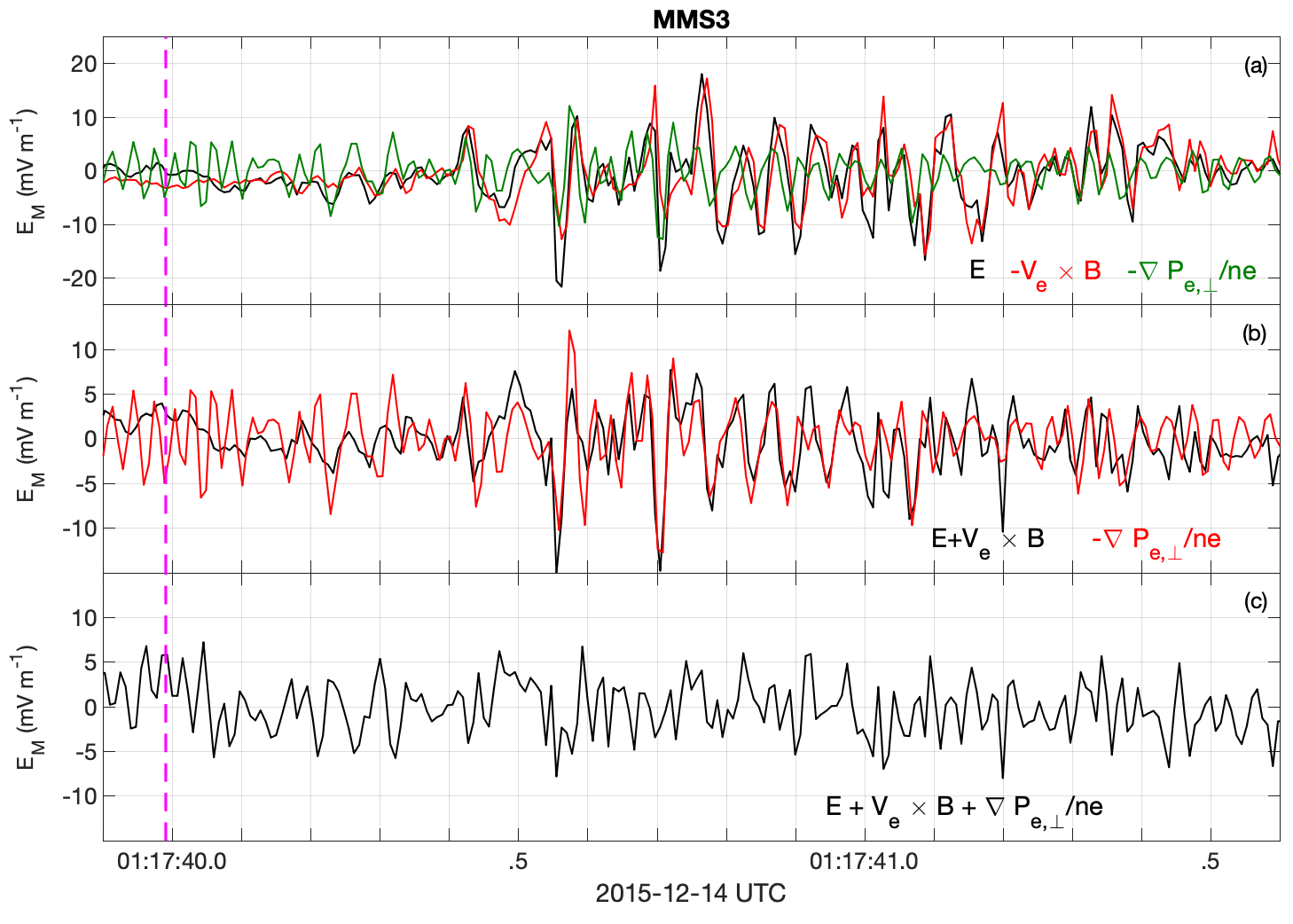}
\caption{Terms in the electron momentum equation for lower hybrid waves observed by MMS3. (a) ${\bf M}$ components of ${\bf E}$ (black), $- {\bf V}_e \times {\bf B}$ (red), and 
$- \nabla P_{e,\perp}/ne$ (green). (b) ${\bf M}$ components of ${\bf E} + {\bf V}_e \times {\bf B}$ (black) and $- \nabla P_{e,\perp}/ne$ (red). (c) ${\bf M}$ component of 
${\bf E} + {\bf V}_e \times {\bf B} + \nabla P_{e,\perp}/ne$. The magenta line indicates when $B_L = 0$. }
\label{OhmsLH}
\end{center}
\end{figure*}

In Figure \ref{OhmsLH}a we plot the ${\bf M}$ components of ${\bf E}$, $- {\bf V}_e \times {\bf B}$, and $- \nabla P_{e,\perp}/ne$. We find that $- \nabla P_{e,\perp}/ne$ reaches large amplitudes ($> 10$~mV~m$^{-1}$) while the waves are observed, and in some places is comparable in magnitude to ${\bf E}$ and $- {\bf V}_e \times {\bf B}$. In general, $- \nabla P_{e,\perp}/ne$ is out of phase with both ${\bf E}$ and $- {\bf V}_e \times {\bf B}$ when the waves are observed. This results in some phase difference between ${\bf E}$ and $- {\bf V}_e \times {\bf B}$, while the relative amplitudes of ${\bf E}$ and $- {\bf V}_e \times {\bf B}$ remain comparable. In Figure \ref{OhmsLH}b we plot the ${\bf M}$ components of ${\bf E} + {\bf V}_e \times {\bf B}$ and $- \nabla P_{e,\perp}/ne$. Overall, we find that ${\bf E} + {\bf V}_e \times {\bf B} \approx - \nabla P_{e,\perp}/ne$, which is most clearly seen between 01:17:40.5 UT and 01:17:41.0 UT. The amplitudes and phases are similar, indicating that the pressure fluctuations associated with the waves can account for the observed differences between ${\bf E}$ and $- {\bf V}_e \times {\bf B}$. 

In Figure \ref{OhmsLH}c we plot the ${\bf M}$ component of ${\bf E} + {\bf V}_e \times {\bf B} + \nabla P_{e,\perp}/ne$. We find that this quantity fluctuates with amplitudes of $\sim 5$~mV~m$^{-1}$, typically smaller than the values of ${\bf E}$, $-{\bf V}_e \times {\bf B}$, and $-\nabla P_{e,\perp}/ne$. This quantity provides an indicator of the overall uncertainties, rather than the values of the remaining terms in equation (\ref{emomentum}). The main sources of uncertainty are: (1) ${\bf E}$ is down-sampled to the cadence of the electron moments, (2) ${\bf V}_e$ and ${\bf P}_e$ are computed from distributions with reduced angular coverage \cite[]{rager1}, and (3) the pressure divergence terms must be approximated using the single-spacecraft method. For comparison, rough estimates of the remaining terms in equation (\ref{emomentum}) [not shown] yield values less than $1$~mV~m$^{-1}$, and are thus  unlikely to account for the fluctuations in Figure \ref{OhmsLH}c.
For this example, we conclude that deviations of ${\bf E}$ from $- {\bf V}_e \times {\bf B}$ result from fluctuations in ${\bf P}_e$ associated with the waves, and to a lesser extend the uncertainties associated with the measurements of ${\bf E}$ and the electron moments. 

\begin{figure*}[htbp!]
\begin{center}
\includegraphics[width=130mm, height=90mm]{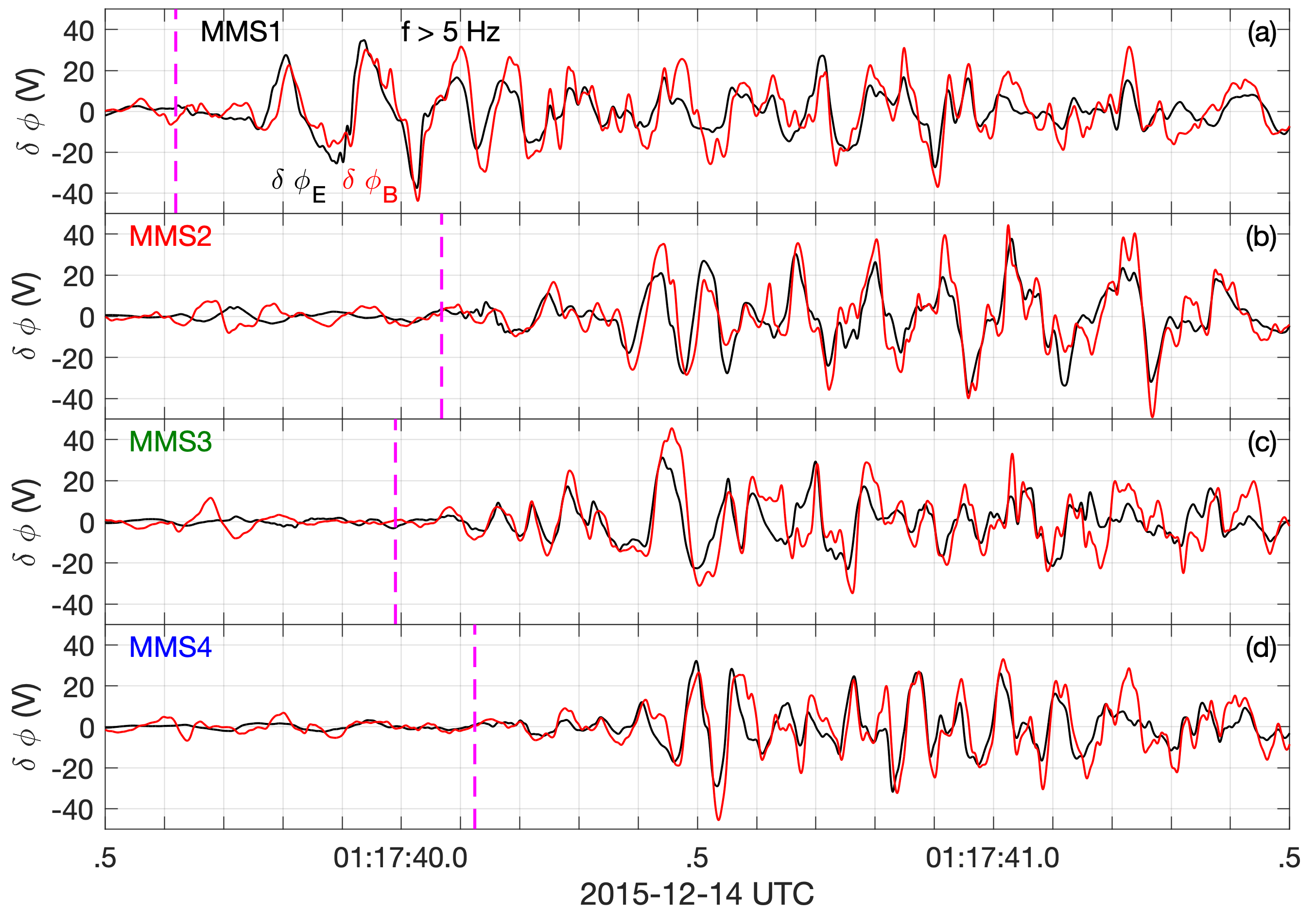}
\caption{Lower hybrid waves observed in the magnetospheric inflow region observed by the four spacecraft. 
(a)--(d) $\phi_B$ (red) and the best fit of $\delta \phi_E$ (black) to $\delta \phi_B$ for
MMS1--4, respectively. Fluctuating ${\bf E}$ and ${\bf B}$ are obtained for $f > 5 \, \mathrm{Hz}$. The lower
hybrid wave properties are summarized in Table \ref{dec14lhprops}. The magenta dashed lines indicate the
neutral point where $B_L = 0$ for each spacecraft. }
\label{Dec14LHDI}
\end{center}
\end{figure*}

We now investigate the wave properties in more detail. 
Figure \ref{Dec14LHDI} shows the calculated $\delta \phi_{B}$ using equation (\ref{phiB}) and
the best fit of $\delta \phi_E$ to $\delta \phi_{B}$ for MMS1--MMS4, respectively. For reference, $B_L = 0$ is indicated
by the vertical magenta lines in each panel. To compute $\delta \phi_{B}$ and $\delta \phi_E$ we bandpass
${\bf B}$ and ${\bf E}$ between $5 \, \mathrm{Hz}$ and $100 \, \mathrm{Hz}$, which corresponds to the frequencies where the wave power is maximal. For each spacecraft
we find good correlations between $\delta \phi_E$ to $\delta \phi_{B}$ throughout the interval.
We find that the maximum wave potentials are $\delta \phi_{\mathrm{max}} \approx 50 \, \mathrm{V}$ on each spacecraft, corresponding to
$e \delta \phi_{\mathrm{max}}/k_B T_{e} \sim 0.6$. In each case the largest $\delta \phi$ are found on the low-density
side of the neutral point and $\delta \phi$ becomes negligible as the neutral point is approached. Thus, quasi-electrostatic 
lower hybrid waves do not penetrate into the electron diffusion region. 

We note that the waveforms of $\delta \phi_{B}$ and $\delta \phi_E$ differ significantly
for each spacecraft, prohibiting multi-spacecraft timing analysis of the waves to determine their properties.
Based on the magnetopause boundary speed the lower hybrid
waves occupy a width of $\sim 50 \, \mathrm{km}$, corresponding to $\sim 0.7 d_i$
(consistent with \cite{pritchett4}), where
$d_i \approx 70 \, \mathrm{km}$ is the magnetosheath ion inertial length. The most intense lower hybird waves 
occur at $\gtrsim 12$~km $ = 0.2$~$d_i$ from the neutral point. 

The lower hybrid wave properties determined from the analysis in Figure \ref{Dec14LHDI} are
summarized in Table \ref{dec14lhprops} for each spacecraft. Even though
the waveforms differ significantly the estimated properties are very similar on each spacecraft. We find
that the waves propagate in the ${\bf M}$ direction (dawnward),
corresponding to the direction of both the large-scale ${\bf E} \times {\bf B}$ drift and
electron diamagnetic drift (shown below).
The lower hybrid waves are predicted to propagate approximately perpendicular to ${\bf B}$, so their propagation
direction is oblique to the out-of-plane direction, due to the guide field in this event.
On average we find that the waves have
$v_{ph} \approx 160 \, \mathrm{km} \, \mathrm{s}^{-1}$, slightly smaller than the estimate from the fluctuations in ${\bf P}_e$. 
We calculate $f \approx 13 \, \mathrm{Hz}$
for the lower hybrid waves based on the power spectra of $E_{M}$ over the interval the waves are observed.
We estimate the wavelength $\lambda \approx 12$~km, which is smaller than the spacecraft separations,
accounting for the lack of correlation between $\delta \phi_E$ (and $\delta \phi_{B}$) observed by the different spacecraft.
From this $\lambda$ we estimate $k \rho_e \sim 0.6$, corresponding to quasi-electrostatic lower hybrid waves,
consistent with the predictions for lower hybrid waves in the electrostatic limit,
and in agreement with previous observations \cite[]{khotyaintsev4,graham7}.
This supports the conclusion that the fluctuations in ${\bf E}$, ${\bf B}$, ${\bf V}_e$ and $n_e$ observed
on the low-density side of the neutral point are primarily due to lower hybrid waves.

\begin{table}[ht]
\centering
\begin{center}\vspace{0.2cm}
\begin{tabular}{|c | c| c| c| c|}
\hline
MMS & v $(\mathrm{km} \, \mathrm{s}^{-1})$ & direction (LMN) & $C_{\phi}$ & $\lambda$ (km) \\
\hline
1 & $144$ & [0.47, 0.81, 0.35] & $0.78$ & $11$ \\
2 & $172$ & [0.66, 0.74, 0.12] &  $0.84$ & $13$ \\
3 & $162$ & [0.56, 0.57, 0.60] & $0.72$ & $12$ \\
4 & $166$ & [0.66, 0.73, 0.21] & $0.78$ & $13$ \\
\hline
\end{tabular}
\end{center}\vspace{0.5cm}
\caption{Properties of the lower hybrid waves observed in the ion diffusion region on 14 December 2015.
The properties are calculated for electric and magnetic field fluctuations above $5 \, \mathrm{Hz}$.}
\label{dec14lhprops}
\end{table}

\begin{figure*}[htbp!]
\begin{center}
\includegraphics[width=140mm, height=120mm]{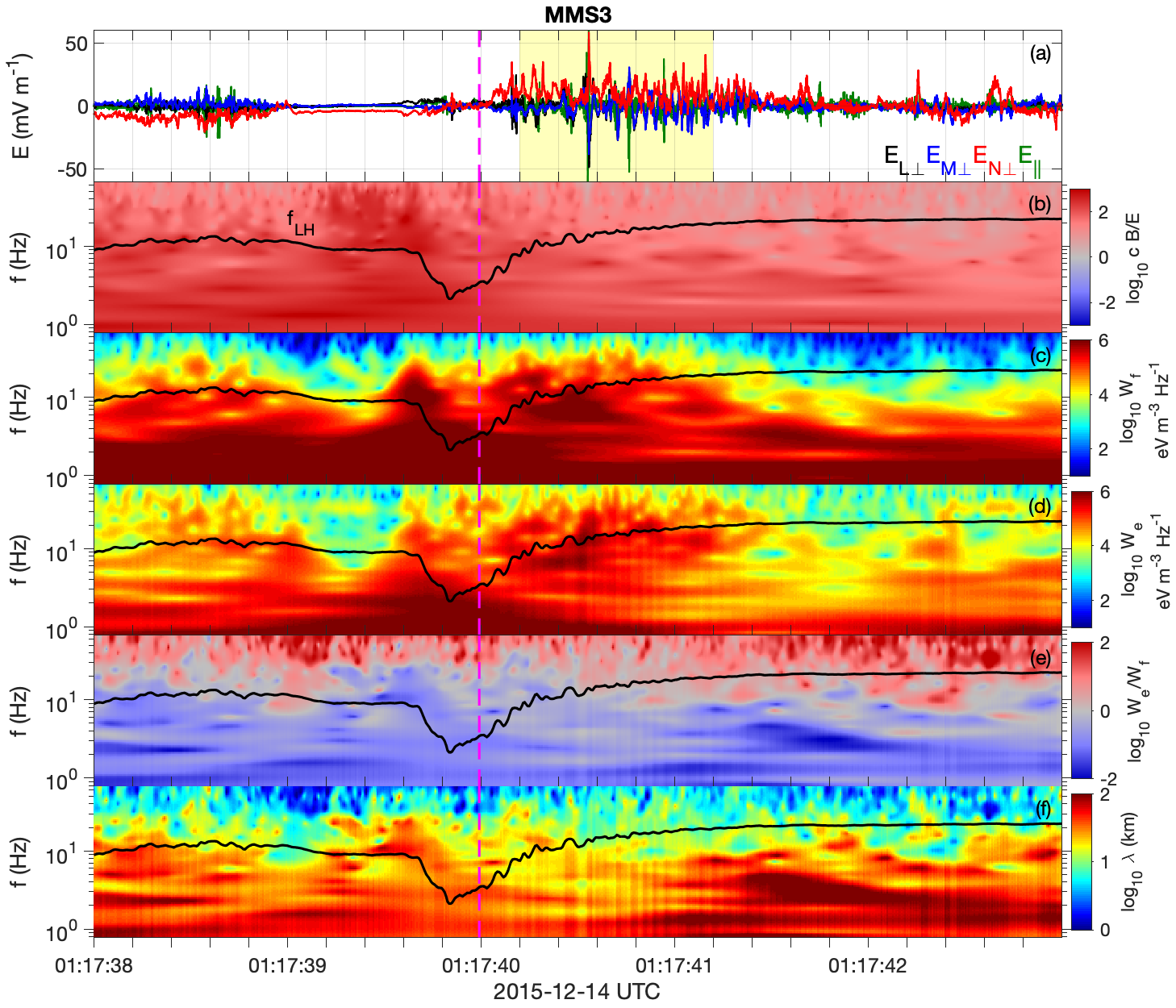}
\caption{Properties of the lower hybrid waves observed on 14 December 2015 by MMS3.
(a) Perpendicular and parallel components of ${\bf E}$. (b) Spectrogram of $cB/E$.
(c) Spectrogram of $W_f$.
(d) Spectrogram of $W_e$.
(e) Spectrogram of $W_e/W_f$.
(f) Spectrogram of $\lambda$.}
\label{LHprops14Dec}
\end{center}
\end{figure*}

We now compare the fields and electron energy densities of the lower hybrid waves
using MMS3 in Figure \ref{LHprops14Dec}.
Figure \ref{LHprops14Dec}b shows that for these waves most of the field energy density is in ${\bf B}$
since $c B/E > 1$, thus $W_f \approx W_B$. Figures \ref{LHprops14Dec}c and \ref{LHprops14Dec}d
show spectrograms of $W_f$ and $W_e$. Both spectrograms are similar, with most of the energy density
being found close to but below $f_{LH}$ on the low-density side of the neutral point.
Figure \ref{LHprops14Dec}e shows that for $f < f_{LH}$ there is more energy density in the fields than
electrons, in contrast to the 28 November 2016 event (Figure \ref{LHprops28Nov}). This occurs because the waves have a smaller $k_{\perp} d_e$ (Figure \ref{Figure1}f). Figure \ref{LHprops14Dec}f shows the spectrogram
of $\lambda$ using equation (\ref{lambdaeq}). For the lower hybrid waves shown in
Figure \ref{LHprops14Dec}a we estimate $\lambda \sim 10 - 20$~km, which agrees well with the
results in Table \ref{dec14lhprops}.

\begin{figure*}[htbp!]
\begin{center}
\includegraphics[width=140mm, height=50mm]{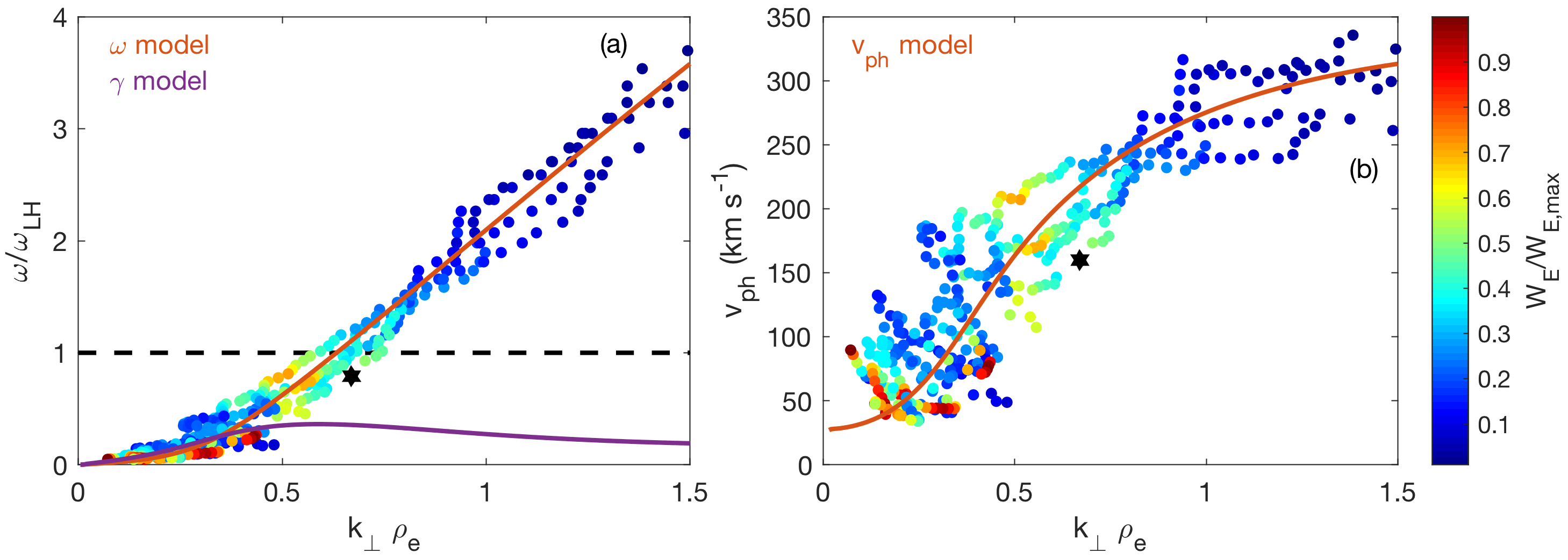}
\caption{Dispersion relation of lower hybrid waves calculated using equation (\ref{WBWe1}) for all four
spacecraft.
(a) Dispersion relations of the lower hybrid waves observed in the yellow-shaded region of
Figure \ref{LHprops14Dec}a.  The black dashed lined indicates $\omega/\omega_{LH} = 1$. The red and
purple lines are the dispersion relation and growth rate calculated using equation (\ref{disprelLHDI}).
(b) Phase speed $v_{ph}$ versus $k_{\perp} \rho_e$. The black stars in panels (a) and (b)
are the averages of the wave properties estimated in Table \ref{dec14lhprops}. }
\label{Disprel14Dec}
\end{center}
\end{figure*}

In Figure \ref{Disprel14Dec}a we plot the dispersion relations from the four spacecraft using equation
(\ref{WBWe1}). For MMS3 we take the median over the time interval indicated by the yellow-shaded region
in Figure \ref{LHprops14Dec}a, where the $E_{M\perp}$ fluctuations are observed. We use similarly long time intervals for the remaining spacecraft, although
the start and end times differ because the spacecraft cross the neutral point and region with
lower hybrid waves at different times.
All four spacecraft yield similar results. The lower hybrid waves are characterized by
$0.5 \lesssim k_{\perp} \rho_e \lesssim 0.7$ (corresponding to $k_{\perp} d_e \approx 1$) and frequencies of
$0.5 \lesssim \omega/\omega_{LH} \lesssim 1$, or equivalently $8$~Hz $\lesssim f \lesssim 16$~Hz. This smaller $k_{\perp} d_e$ accounts for the smaller $W_e/W_f$ observed here compared with the 28 November 2016 event (cf., Figure \ref{Figure1}f). From Figure \ref{Disprel14Dec}b we estimate
100~km~s$^{-1} \lesssim v_{ph} \lesssim 250$~km~s$^{-1}$, which agrees with the results in Table \ref{dec14lhprops} and the value estimated from the fluctuations in ${\bf P}_e$.
Compared with Figure \ref{Disprel28Nov}b we find a much broader range of $v_{ph}$, which is likely
because the waves here are more broadband in frequency. In addition, the spacecraft separation is larger
here compared with $\lambda$, resulting in larger differences in the
dispersion relations between each spacecraft.

\begin{figure*}[htbp!]
\begin{center}
\includegraphics[width=140mm, height=120mm]{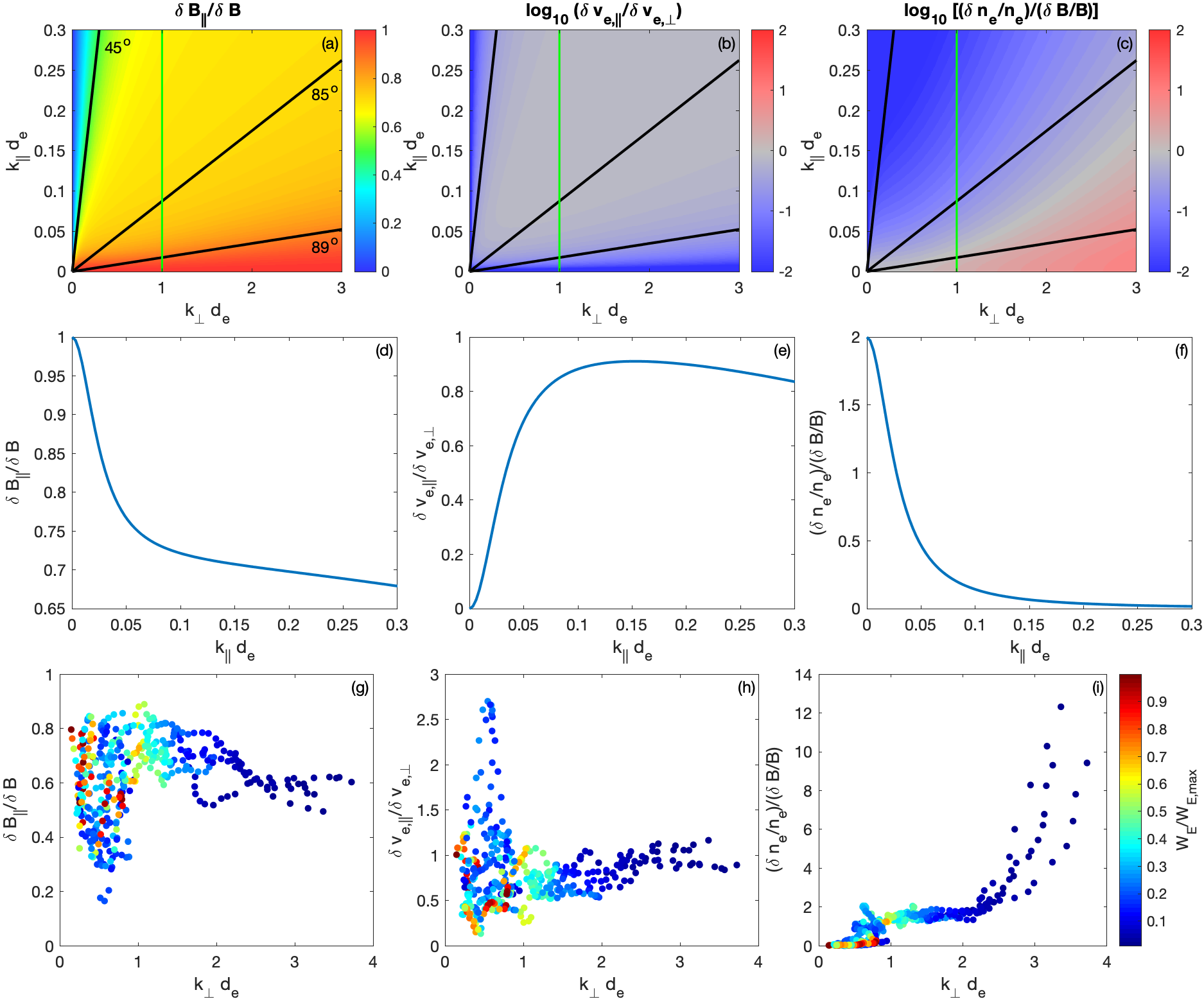}
\caption{Estimates of the wave-normal angle and $k_{\parallel}$ from fields and particle observations of
the lower hybrid waves observed 14 December 2015.
(a)--(c) $\delta B_{\parallel}/\delta B$, $\delta V_{e,\parallel}/V_{e,\perp}$, and $(\delta n_e/n_e)/(\delta B/B)$
for whistler/lower hybrid waves versus $k_{\parallel}$ and $k_{\perp}$. We use $f_{pe}/f_{ce} = 30$.
The green line is $k_{\perp} d_e = 1$, the estimate $k_{\perp}$ of the observed waves.
(d)--(f) $\delta B_{\parallel}/\delta B$, $\delta V_{e,\parallel}/V_{e,\perp}$, and $(\delta n_e/n_e)/(\delta B/B)$
versus $k_{\parallel} d_e$ for $k_{\perp} d_e = 1$.
(g)--(i) Observed $\delta B_{\parallel}/\delta B$, $\delta V_{e,\parallel}/V_{e,\perp}$, and
$(\delta n_e/n_e)/(\delta B/B)$ of the lower hybrid waves versus $k_{\perp} d_e$. }
\label{kparprops14Dec}
\end{center}
\end{figure*}

We now estimate $k_{\parallel}$ and $\theta_{kB}$ for these waves using
$\delta B_{\parallel}/\delta B$, $\delta V_{e,\parallel}/V_{e,\perp}$, and $(\delta n_e/n_e)/(\delta B/B)$.
Figures \ref{kparprops14Dec}a--\ref{kparprops14Dec}c show
$\delta B_{\parallel}/\delta B$, $\delta V_{e,\parallel}/V_{e,\perp}$, and $(\delta n_e/n_e)/(\delta B/B)$
versus $k_{\parallel}$ and $k_{\perp}$ predicted from homogeneous theory. 
We use $f_{pe}/f_{ce} = 30$, corresponding to the median
$f_{pe}/f_{ce}$ over the interval used to calculate the dispersion relations. We note that $f_{pe}/f_{ce}$
varies with position here so the estimates of $k_{\parallel}$ and $\theta_{kB}$ are approximate. We find
that the lower hybrid waves have $k_{\perp} d_e \approx 1$, indicated by the vertical green lines
in Figures \ref{kparprops14Dec}a--\ref{kparprops14Dec}c.

In Figures \ref{kparprops14Dec}d--\ref{kparprops14Dec}f we plot
$\delta B_{\parallel}/\delta B$, $\delta V_{e,\parallel}/\delta V_{e,\perp}$, and
$(\delta n_e/n_e)/(\delta B/B)$ versus
$k_{\parallel} d_e$ for $k_{\perp} d_e = 1$. Qualitatively, the dependence of
$\delta B_{\parallel}/\delta B$ and $\delta V_{e,\parallel}/V_{e,\perp}$ on $k_{\parallel}$
are very similar to the 28 November 2016 case, where $f_{pe}/f_{ce}$ is much smaller. In contrast, a
substantially smaller $(\delta n_e/n_e)/(\delta B/B)$ is predicted here because $\delta {\bf B}$ increases as $f_{pe}/f_{ce}$ increases.
In Figures \ref{kparprops14Dec}g--\ref{kparprops14Dec}i we plot
$\delta B_{\parallel}/\delta B$, $\delta V_{e,\parallel}/V_{e,\perp}$, and $(\delta n_e/n_e)/(\delta B/B)$ obtained
from the four spacecraft versus $k_{\perp} d_e$. For $\delta B_{\parallel}/\delta B$ we obtain values of
$0.6 - 0.9$, with an average of $\approx 0.75$ around $k_{\perp} d_e = 1$,
corresponding to $k_{\parallel} d_e \approx 0.06$  in Figure \ref{kparprops14Dec}d.
For $\delta V_{e,\parallel}/\delta V_{e,\perp}$ we obtain $0.3 - 1.0$, with an average of $0.7$ around
$k_{\perp} d_e = 1$, corresponding to $k_{\parallel} d_e \approx 0.05$ in Figure \ref{kparprops14Dec}e. For
$(\delta n_e/n_e)/(\delta B/B)$ we obtain $1 - 2$, which is consistent with the predictions
in Figure \ref{kparprops14Dec}f. Here, $n_e$ is much larger than in Figure \ref{kparprops28Nov}, so the
spectrum of $\delta n_e/n_e$ should be more reliable.
From the average of $(\delta n_e/n_e)/(\delta B/B)$ around $k_{\perp} d_e = 1$ we obtain $1.3$,
corresponding to $k_{\parallel} d_e \approx 0.02$. This $k_{\parallel} d_e$ is smaller than
the predictions from $\delta B_{\parallel}/\delta B$ and $\delta V_{e,\parallel}/\delta V_{e,\perp}$. 
We note that the values of $(\delta n_e/n_e)/(\delta B/B)$ predicted from homogeneous theory 
are likely not valid here due to the dependence of $\delta n_e$ on the 
gradients in $n_e$ and $B$ (see Appendix \ref{app1}). 
The average $k_{\perp} d_e$ is then $\approx 0.05$, whence we calculate $\theta_{kB} \approx 87^{\circ}$.
Thus, the estimated $\theta_{kB}$ is consistent with lower hybrid waves. The spread of data in Figures
\ref{kparprops14Dec}g--\ref{kparprops14Dec}i suggests that $\theta_{kB}$ may change with frequency
or time/position. From $\theta_{kB} \approx 87^{\circ}$ we obtain a parallel resonant energy of
$v_{\parallel} \sim 40$~eV. This $v_{\parallel}$ is below the local electron thermal energy, although
there is a large uncertainty in the estimated $v_{\parallel}$. The estimated $v_{\parallel}$ is therefore not inconsistent
with the waves interacting with the thermal electrons. In summary, the quantities $\delta B_{\parallel}/\delta B$ and 
$\delta V_{e,\parallel}/\delta V_{e,\perp}$ indicate that the lower hybrid waves have finite $k_{\parallel}$. 

\subsection{Cross-field drifts and instability analysis}
To investigate the instability of the lower hybrid waves we study the force balance of the current sheet using the ion and electron
momentum equations and investigate the nature of the associated cross-field particle drifts.
The ion and electron pressure
divergences are calculated from the four-spacecraft differences using the full ion and electron pressure tensors,
${\bf P}_i$ and ${\bf P}_e$, respectively. The large-scale electric field is found by resampling
${\bf E}$ to the cadence of the electron moments (30~ms) on each spacecraft
and averaging the field over the four spacecraft.
This sampling rate tends to under-resolve lower hybrid fluctuations. In addition, the four-spacecraft averaging
tends to average out the lower hybrid waves because for this event
the spacecraft separations are comparable or larger than the lower hybrid wavelength, Therefore, the computed terms approximate the non-fluctuating component of ${\bf E}$.

\begin{figure*}[htbp!]
\begin{center}
\includegraphics[width=120mm, height=120mm]{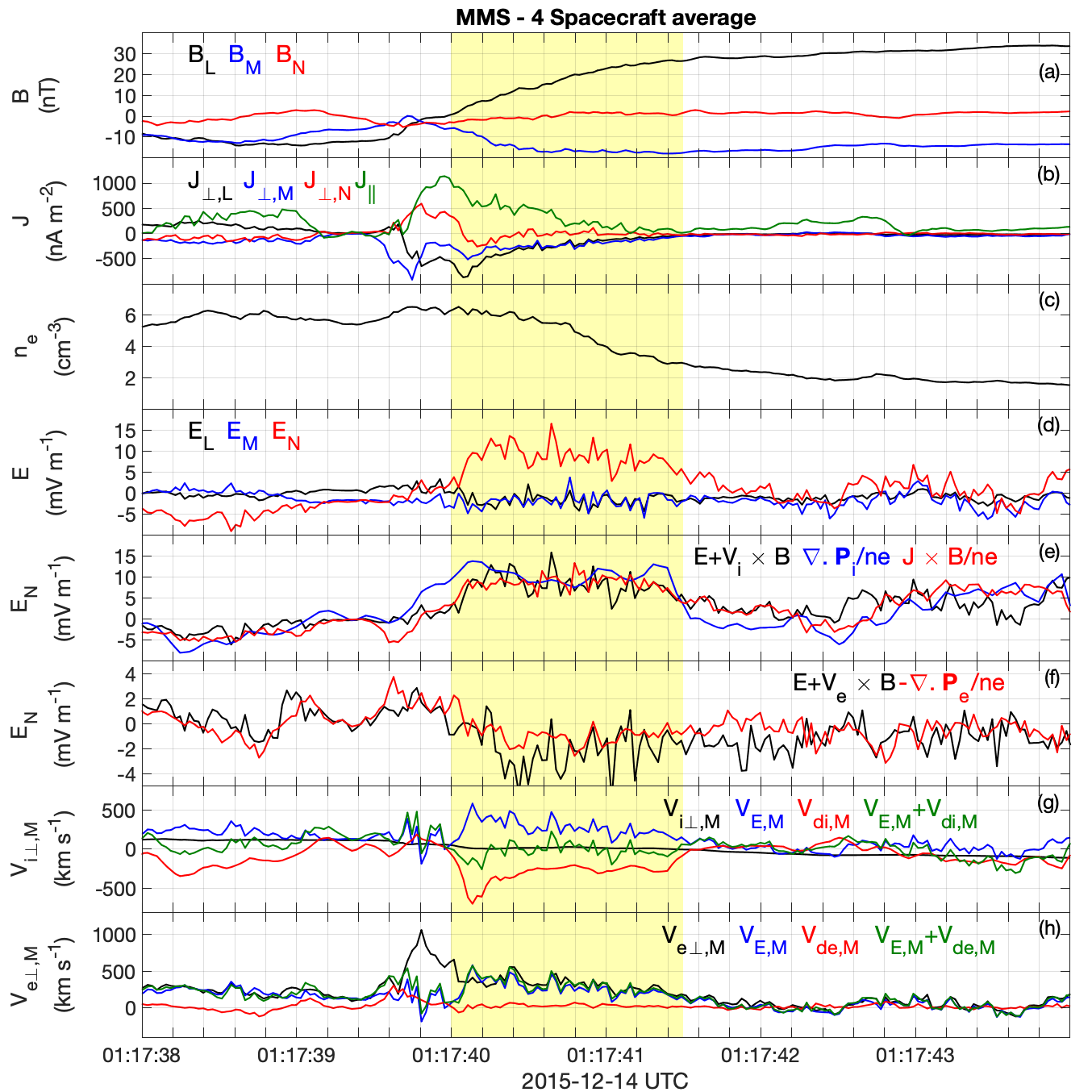}
\caption{Large-scale fields and particle drifts obtained using four-spacecraft methods for the 14 December 2015
magnetopause crossing. (a) ${\bf B}$. (b) ${\bf J}$ computed using the Curlometer technique. (c) $n_{e}$.
(d) ${\bf E}$. (e) ${\bf N}$ components of ${\bf E} + {\bf V}_{i} \times {\bf B}$ (black), $\nabla \cdot {\bf P}_{i}/ne$ (blue),
and ${\bf J} \times {\bf B}/ne$. (f) ${\bf N}$ components of ${\bf E} + {\bf V}_{e} \times {\bf B}$ (black),
$-\nabla \cdot {\bf P}_e/ne$ (blue). (g) ${\bf M}$ components of ion drifts perpendicular to ${\bf B}$; $V_{i\perp,y}$ (black),
$V_{E,M}$ (blue), $V_{di,M}$ (red), and $V_{E,M} + V_{di,M}$ (green).
(h) ${\bf M}$ components of electron drifts perpendicular to ${\bf B}$; $V_{e\perp,M}$ (black),
$V_{E,M}$ (blue), $V_{de,M}$ (red), and $V_{E,M} + V_{de,M}$ (green). }
\label{Dec14fields}
\end{center}
\end{figure*}

Figure \ref{Dec14fields} shows the results of the four-spacecraft analysis. Figure \ref{Dec14fields}a shows the
four-spacecraft averaged ${\bf B}$. Compared with Figure \ref{14Decoverview}a
most of the fluctuations have been removed. The parallel and perpendicular components of ${\bf J}$, shown
in Figure \ref{Dec14fields}b, are calculated using the Curlometer technique. This ${\bf J}$ approximates the
large-scale non-fluctuating ${\bf J}$, because the spacecraft separations are too large to
resolve $\delta {\bf J}$ intrinsic to the lower hybrid waves \cite[]{graham4}.
The current density peaks close to the neutral point, rather than where the lower hybrid waves are observed. 
Comparable parallel and perpendicular 
${\bf J}$ magnitudes 
(primarily in the ${\bf M}$ and ${\bf L}$ directions, except near the center of the current sheet) are
observed in the yellow-shaded region, where the lower hybrid waves occur.
On the magnetospheric side of the current sheet (yellow-shaded region) a large-scale normal electric field $E_N$ develops, typical of the ion diffusion region of magnetopause reconnection. In Figure \ref{14Decoverview}d, 
$E_N$ is due to the large-scale Hall electric field and the fluctuations associated with the waves. 

Neglecting anomalous terms, inertial terms, and temporal
changes, the ion and electron momentum equations are
\begin{linenomath}
\begin{equation}
{\bf E} + {\bf V}_{i} \times {\bf B} \approx \frac{\nabla \cdot {\bf P}_{i}}{ne},
\label{ionm}
\end{equation}
\begin{equation}
{\bf E} + {\bf V}_{e} \times {\bf B} \approx -\frac{\nabla \cdot {\bf P}_{e}}{ne},
\label{elem}
\end{equation}
\end{linenomath}
respectively.
Figures \ref{Dec14fields}e and \ref{Dec14fields}f show
that these equations are approximately satisfied for ions and electrons, respectively. Moreover, we find
that $\nabla \cdot {\bf P}_{i} \approx {\bf J} \times {\bf B}$, thus the cross-field current is produced by the ion
pressure divergence to maintain force balance across the current sheet. In contrast, the electron pressure
divergence has a much smaller contribution, due to the small $T_e/T_i$, yielding a maximum
$- \nabla \cdot {\bf P}_{e}/ne \approx - 2 \, \mathrm{mV} \, \mathrm{m}^{-1}$ in the ${\bf N}$ direction on the magnetospheric side of the current sheet. This value is significantly smaller than the fluctuating component in the out-of-plane direction associated with the lower hybrid waves (Figure \ref{OhmsLH}).

By taking the cross products of equations (\ref{ionm}) and (\ref{elem}) with
${\bf B}/B^2$ we obtain ${\bf V}_{i \perp} \approx {\bf V}_{E} + {\bf V}_{di}$ and
${\bf V}_{e\perp} \approx {\bf V}_{E} + {\bf V}_{de}$. Here ${\bf V}_{di}$ and ${\bf V}_{de}$ are the
ion and electron diamagnetic drifts, given by
\begin{equation}
{\bf V}_{di,e} = \pm \frac{{\bf B} \times \nabla \cdot {\bf P}_{i,e}}{B^2 ne}.
\label{diamagd}
\end{equation}
In the out-of-plane direction we find that ${\bf V}_{i \perp} \approx {\bf V}_{E} + {\bf V}_{di}$ and
${\bf V}_{e\perp} \approx {\bf V}_{E} + {\bf V}_{de}$ are both approximately satisfied throughout
the ion diffusion region, as seen in Figures \ref{Dec14fields}g and \ref{Dec14fields}h.
For this event $V_{i\perp,M} \approx 0$, meaning $V_{E,M} \approx - V_{di,M}$.
The cross-field current is therefore due to the ${\bf E} \times {\bf B}$ drift of electrons in the
${\bf M}$ direction.

At about 01:17:39.8 UT we find that ${\bf V}_{e\perp} \neq  {\bf V}_{E} + {\bf V}_{de}$, likely because
the EDR is observed around this time, which is smaller than the spacecraft separations.
Therefore, the spacecraft separations may be too large
to accurately compute $\nabla \cdot {\bf P}_e$ and spacecraft averaged quantities, thus the computed drifts
may not be reliable here.

The results in Figure \ref{Dec14fields} are simply a consequence of the electron and ion momentum
equations being satisfied in the limit when temporal changes and local acceleration can be neglected at ion
spatial scales (larger than the typical lower hybrid wavelength).
These results should not be particularly surprising, but they show that in the region where
lower hybrid waves are observed,
the cross-field current develops due to $\nabla \cdot {\bf P}_i$.
Thus, the likely energy source of the observed waves is the cross-field current produced by $\nabla \cdot {\bf P}_i$,
which can be unstable to LHDI.

To investigate the instability of the observed waves we consider the local dispersion equation
for LHDI in the ion stationary frame \cite[]{davidson1}
\begin{equation}
0 = 1 - \frac{\omega_{pi}^2}{k^2 v_i^2} Z' \left( \frac{\omega}{k v_i} \right) + \frac{\omega_{pe}^2}{\Omega_{ce}^2} \left( 1 + \frac{\omega_{pe}^2}{c^2 k^2} \right) + \frac{2 \omega_{pe}^2}{k^2 v_e^2} \left( 1 + \frac{\beta_i}{2} \right) \frac{k V_{de}}{\omega - k V_E},
\label{disprelLHDI}
\end{equation}
where ${\bf k} \cdot {\bf B} = 0$.
For the local plasma conditions we use $B = 25 \, \mathrm{nT}$,  $n_e = 5 \, \mathrm{cm}^{-3}$,
$T_e = 90 \, \mathrm{eV}$ and $T_i = 500 \, \mathrm{eV}$, and
$V_E = -V_{di} = 400 \, \mathrm{km} \, \mathrm{s}^{-1}$, and
$V_{de} = 50 \, \mathrm{km} \, \mathrm{s}^{-1}$, based on the median values over the interval the waves
are observed. The effect of the weak pressure gradient is included through $V_{de}$. 

Figure \ref{Disprel14Dec}a shows the dispersion relation and growth rate, overplotted with the observed
dispersion relations. The dispersion relation predicted by equation (\ref{disprelLHDI}) is in excellent
agreement with the observed dispersion relations. Similarly, the predicted $v_{ph}$, shown in Figure
\ref{Disprel14Dec}b is in excellent agreement with observations. 
For LHDI $\gamma_{\mathrm{max}}$ corresponds to $k_{\perp} \rho_e = 0.6$, in agreement with
where $W_E/W_{E,max}$ peaks.
At $\gamma_{\mathrm{max}}$, $v_{ph} = 190 \, \mathrm{km} \, \mathrm{s}^{-1}$, which agrees with the observed dispersion relations and the value of $v_{ph}$ predicted in Figure \ref{OhmsLH}, and is only slightly larger than values in table \ref{dec14lhprops}.
Therefore, the LHDI predictions agree with observations so we conclude that the observed
waves are produced by LHDI. 
%For this event $\delta {\bf E}_{ES} \approx \delta {\bf E}$, despite $W_B$ exceeding $W_E$ (cf., Figures \ref{Figure1}c and \ref{Figure1}d) based on the cold plasma predictions, which justifies the use of the electrostatic dispersion equation. 
However, close to the neutral point we expect 
equation (\ref{disprelLHDI}) to become unreliable due to strong gradients in ${\bf B}$ and the neglect 
of electromagnetic effects. 

In summary, the results show that the fluctuations in the ion diffusion region on the low-density side 
of the neutral point are consistent with lower hybrid
drift waves. All the measured fluctuations, phase speed, dispersion relation and wave-normal angle are
consistent with predictions for lower hybrid waves. We find that some deviation from the cold plasma predictions can occur due to the fluctuations in electron pressure associated with the waves and the density gradient 
where the waves occur. The single-spacecraft methods used to estimate
the wave properties are in good agreement with each other.
The primary
free energy source of the lower hybrid waves is the ion pressure divergence,
which is responsible for the cross-field current that excites the lower hybrid waves by LHDI. 

\section{Discussion} \label{discussion}
We have investigated in detail two examples of lower hybrid waves at the magnetopause. In both 
cases we find that the waves have $k \rho_e$ close to $0.5$ and frequencies $0.5 \lesssim f_{}/f_{LH} \lesssim 1$, 
consistent with quasi-electrostatic lower hybrid waves. Although in both examples the waves are consistent with lower hybrid waves, the waves have distinct electric field, magnetic field, and electron energy densities. These differences can be explained by cold plasma theory. For the 28 November 2016 event we find that the lower hybrid waves have $k_{\perp} d_e \approx 3$ for $f_{pe}/f_{ce} \approx 6.7$, and for the 14 December 2015 event we find $k_{\perp} d_e \approx 1$ and $f_{pe}/f_{ce} \approx 30$. We find that these differences account for the different wave properties between the two events. Specifically: (1) The value of $k_{\perp} d_e$ determines $W_e/W_B$ based on Figure \ref{Figure1}f and equation (\ref{WBWe1}), and accounts for the differences in $W_e/W_B$ between the two events. (2) As $f_{pe}/f_{ce}$ increases and/or $k_{\perp} d_e$ decreases, $c B/E$ is predicted to increase based on cold plasma predictions. This is due to the approximately frozen in motion of electrons, which is observed in both events. As $f_{pe}/f_{ce}$ increases $\delta {\bf J}$ is predicted to increase for a given $\delta {\bf E}$. This results in larger $\delta {\bf B}$ according to Ampere's law, resulting in $c B/E$ increasing, which is consistent with the observed differences between the two events. 
In both events $\delta n_e/n_e$ are comparable, and reach peak values of $\delta n_e/n_e \approx 0.2$. Thus, the 
smaller $(\delta n_e/n_e)/(\delta B/B)$ is due to the larger $\delta B/B$ observed on 14 December 2015. 
%We find that gradients in $n$ and $B$ contribute significantly to $\delta n_e$, so predictions of $\delta n_e$ 
%from homogeneous theory may be underestimates.  
We conclude that the observed differences in lower hybrid waves properties in the two events are due 
to distinct $f_{pe}/f_{ce}$ and $k_{\perp} d_e$. 

Another important difference between the two events is that on 14 December 2015 the waves were much more 
localized and the density gradient is more significant. Therefore, we need to consider the effect of gradients on 
the wave properties. Based on the continuity equation it is straightforward to show that
\begin{equation}
\delta n_e \approx \frac{n k_{\perp}}{B (\omega - k_{\perp} V_{e,M})} \left( \frac{n'}{n} - \frac{B'}{B} \right) \delta \phi,
\label{necontd}
\end{equation}
when electrons are frozen in. Here the primes denote derivatives in the ${\bf N}$ direction. At Earth's 
magnetopause, where the lower hybrid waves are observed, $n' > 0$ and $B' < 0$ and $\omega - k_{\perp} 
V_{e,M} < 0$, so $\delta n_e$ is expected to be anticorrelated with $\delta \phi$. Thus, gradient terms 
may be the dominant contribution to $\delta n_e$. By substituting equation (\ref{phiB}) into 
equation (\ref{necontd}) we obtain:
\begin{equation}
(\delta n_e/n_e)/(\delta B_{\parallel}/B) \approx \frac{f_{ce}}{f_{pe}} \frac{c k_{\perp} d_e}{(\omega - k_{\perp} V_{e,M})} \left( \frac{n'}{n} - \frac{B'}{B} \right).
\label{dnndBBd}
\end{equation}
Equation (\ref{dnndBBd}) predicts that $(\delta n_e/n_e)/(\delta B_{\parallel}/B)$ increases toward the magnetopause 
as $f_{pe}/f_{ce}$ and $n'/n - B'/B$ increase. This is consistent with the observations in Figure \ref{14Decoverview}, so we conclude that the gradients in $n$ and $B$ provide important contributions to $\delta n_e$. Finally, we note that the localization in the ${\bf N}$ direction will result in $\delta E_N$ because 
$\delta \phi' \neq 0$. Since electrons are approximately frozen in $\delta V_{e,M}$ will also occur. These 
fluctuations are observed in Figure \ref{14Decoverview}. 

Our interpretation of these localized lower hybrid waves is that they correspond to the 
ripple structures found in three-dimensional 
simulations of asymmetric reconnection \cite[]{pritchett6,pritchett4,pritchett7}. In \cite{pritchett4} and  \cite{pritchett7} these ripple structures and the associated electric field fluctuations were interpreted as waves generated by LHDI. In addition, rippling waves have been found 
in simulations \cite[]{pritchett9,divin2} and observations \cite[]{pan1} of dipolarization fronts in Earth's magnetotail. 
In general, these ripples have been interpreted as resulting from LHDI or a closely related instability \cite[]{pritchett9,price2}. 

The 14 December 2015 event was investigated in detail by \cite{ergun4}. They investigated the waves at $7.5$~Hz near the neutral point. They found that the polarization properties were consistent 
with a corrugation of the current sheet, which explains the fluctuations in $\delta E_N$, $\delta n_e$, and 
$\delta B_{\parallel}$. They concluded that these waves were an electromagnetic drift wave, with phase speed of $\sim 600$~km~s$^{-1}$ and wavelength $\sim 80$~km. These values of $v_{ph}$ and $\lambda$ are significantly larger than the values we calculate here 
for lower hybrid waves. 
Qualitatively, the main difference is that \cite{ergun4} predicts that the current sheet is corrugated, 
while here we interpret the fluctuations as smaller-scale ripples localized to the low-density side 
of the current sheet. However, both models predict very similar fluctuations in $\delta {\bf E}$, $\delta n_e$, 
and $\delta B_{\parallel}$, and thus both processes could be active at the current sheet; both 
processes may be manifestations of the same underlying instability. 

The two events detailed in this paper show that the observed lower hybrid waves are consistent with
generation by the lower hybrid drift instability (LHDI) and the closely related modified two-stream
instability. Technically both instabilities are approximations to a more general dispersion equation for lower hybrid waves \cite[]{hsia1,silveira1}. 
In addition to the instabilities investigated in sections \ref{28nov2016} and \ref{14Dec2015},
\cite{graham7} found that when cold magnetospheric
ions are present the ion-ion cross-field instability could develop
between cold magnetospheric ions and finite gyroradius
magnetosheath ions. The wave properties and propagation direction developing for this instability are similar to the MTSI predictions
(without cold magnetospheric ions).
The primary difference between the two instabilities is that the ion-ion cross-field instability is
unstable for ${\bf k} \cdot {\bf B} = 0$, due to the second ion population, whereas MTSI is stabilized.
In either case the ion drift associated with the finite gyroradius magnetosheath ions provides the free energy of
the lower hybrid waves.
In both cases the waves propagate duskward in the cross-field ion drift direction, but at a slow speed 
than the bulk ion velocity of the magnetosheath ions. Therefore, 
in the frame of these ions, the waves propagate dawnward. 
The width of the region
over which the instabilities can occur is determined by the gyroradius of magnetosheath ions, which is much
larger than the predicted and observed wavelengths of the lower hybrid waves and thus the gradients
can be weak, thus justifying the MTSI and local approximations.
The results in section
\ref{28nov2016} suggest that finite gyroradius ion effects
are not necessarily associated with the ion diffusion region of ongoing magnetic reconnection.

For magnetopause reconnection near the subsolar point we expect lower hybrid drift waves
to be produced in the ion diffusion region by the ion pressure divergence. 
In such cases the ion diamagnetic drift velocity is approximately balanced by
${\bf E} \times {\bf B}$, resulting in negligible ion motion in the reconnection out-of-plane direction 
in the spacecraft frame.
In contrast, the electrons propagate at approximately the ${\bf E} \times {\bf B}$ velocity,
with a smaller contribution from
electron diamagnetic drift in the same direction. As a result the lower hybrid drift waves propagate in the
${\bf E} \times {\bf B}$ and electron diamagnetic drift directions (dawnward). Thus, in the spacecraft frame 
the waves propagate the opposite direction to the waves associated with finite gyroradius ions. 

In both cases we find that the lower hybrid waves have a finite $k_{\parallel}$, and can thus interact with thermal
electrons via Landau resonance \cite[]{cairns1}. The finite $k_{\parallel}$ is most evident from the fluctuations
in the parallel electron velocity, and are observed in many other magnetopause crossings (not shown).
From the estimates of the observed $k_{\parallel}$ we find that lower hybrid waves can interact
with parallel propagating thermal and suprathermal electrons.
The observed lower hybrid waves reach large amplitudes and can occur over an extended region,
so they can plausibly contribute to the observed electron heating. 
Both events show wave potentials reaching 
$e \delta \phi_{\mathrm{max}}/k_B T_{e} \sim 0.5-1$. Similarly large potentials have been reported in other
magnetopause reconnection events \cite[]{khotyaintsev4,graham7}.
Future work is required to investigate
the importance of lower hybrid waves for parallel electron heating. Parallel electron heating is expected
in the ion diffusion region and magnetospheric inflow regions due to electron trapping
\cite[e.g.,][]{egedal3}. In three-dimensional simulations when lower hybrid waves are excited,
\cite{le2} found that parallel electron heating was further enhanced compared with the two dimensional case.
However, the precise mechanisms and role of the lower hybrid waves in parallel electron heating were
not clear.

%Both events show lower hybrid waves reaching
%very high amplitudes, with wave potentials
%of $e \phi_{\mathrm{max}}/k_B T_{e} \sim 0.5-1$. Similarly large potentials have been reported in other
%magnetopause reconnection events \cite[]{khotyaintsev4,graham7}. In each case, except for the 28 November 2016
%event, the lower hybrid waves were found near the reconnection diffusion region.
%Therefore, these large-amplitude lower hybrid waves seem to be a recurring feature of
%asymmetric reconnection at the magnetopause.
%When the current sheet is reconnecting it has likely reached a quasi-steady stage, which would best correspond
%to the later stages of three-dimensional simulations of asymmetric reconnection.
%These potentials are consistent with maximal potentials found in three-dimensional simulations of asymmetric
%reconnection typically found during an early transient phase of the simulations when the
%electrostatic instability develops rapidly and 
%then saturates. However, this transient behavior is not likely to be representative of the reconnection events
%referenced above. Typically, the lower hybrid waves are much smaller, especially in the ion diffusion region,
%once steady reconnection develops,
%which likely represents the magnetic reconnection events observed by MMS. Thus, the lower hybrid waves
%observed by MMS seem to have larger amplitudes than expected in many simulations.

\section{Conclusions} \label{conclusions}
In this paper we have investigated the properties and generation of lower hybrid waves at Earth's
magnetopause based on two case studies. For the first time we use electron moments, which resolve
fluctuations at lower hybrid wave frequencies, to investigate the wave properties in unprecedented detail.
The key results of this paper are:

(1) Electron number density and electron velocity fluctuations associated with lower hybrid waves are resolved.
The electrons are shown to remain frozen in at frequencies where the amplitude of lower hybrid waves is
maximal. Large parallel electron velocity fluctuations are observed, indicating that the waves have a finite
parallel wave vector.

(2) The spectrogram of electron energy density associated with lower hybrid waves is computed and
compared with energy density of the electric and magnetic field. The ratio of the electron to field energy
density increases with frequency, consistent with theoretical predictions. The ratio of electron to
magnetic field energy density is used to construct the dispersion relations of the waves, which are in excellent
agreement with theoretical predictions.

(3) Comparison of the observed wave properties with theoretical predictions shows that the lower
hybrid waves have a finite parallel wave number and wave-normal angle close to $89^{\circ}$. This allows
lower hybrid waves to interact with thermal and suprathermal electrons, potentially contributing to parallel
electron heating near the magnetopause. The estimated wave properties are in excellent agreement with
the single-spacecraft method developed in \cite{norgren1}.

(4) For spacecraft separations below the wavelength of lower hybrid waves,
four-spacecraft timing analysis can
be used to determine the wave properties. The phase speed and propagation directions agree very well with
single-spacecraft methods,
thus showing that when multi-spacecraft observations are unavailable the lower hybrid wave
properties can be accurately determined from single-spacecraft observations. 

(5) The observed waves are consistent with generation by the lower hybrid drift instability or the 
modified two-stream instability. In both cases the source of instability is the cross-field current at the 
magnetopause. 

(6) The differences between lower hybrid wave properties, such as the ratio of magnetic field energy density to 
electric field energy density and the relative amplitudes of magnetic field and density fluctuations, are 
determined by the ratio of the electron plasma frequency to electron cyclotron frequency and the wave 
number. The ratio of field to particle energy densities is determined by the perpendicular wave number of the 
waves. These predictions are well approximated by cold plasma theory and account for the differences in lower 
hybrid wave properties observed at the magnetopause.

\acknowledgments
We thank the entire MMS team and instrument PIs for data access and support.
This work was supported by
the Swedish National Space Board, grants 175/15 and 128/17.
The work at IRAP as well as the LPP involvement for the SCM instrument on MMS 
are supported by CNRS and CNES. We acknowledge support
from the ISSI team MMS and Cluster Observations of Magnetic Reconnection.
MMS data are available at https://lasp.colorado.edu/mms/sdc/public and the highest-resolution particle moments and distributions are available on request. 

\appendix
\section{Derivation of single spacecraft methods to determine lower hybrid wave properties} \label{app1}
In this section we derive the equations used to determine the lower hybrid wave properties and dispersion relation
using a single spacecraft. We also consider the sources of uncertainty in the methods used. 
To model the wave properties we make the following assumptions: 
(1) The waves are quasi-electrostatic. 
(2) Electrons are frozen-in, while ions are unmagnetized. This is justified because we are interested in the frequency range $f_{ci} \ll f \ll f_{ce}$.

We assume the fluctuating quantities have the form: 
\begin{equation}
\delta {\bf Q} = \delta {\bf Q}(N) \exp{(-i \omega t + i k_{\perp} M + i k_{\parallel} L)}, 
\label{Q}
\end{equation}
where the LMN coordinate system is used, and the wave vector is along primarily along the ${\bf M}$ direction 
and $k_\perp \gg k_{\parallel}$. 
We assume that the waves are quasi-electrostatic and the electric field is modeled by an electrostatic potential of the form:
\begin{equation}
\delta \phi = \delta \phi(N) \exp{(-i \omega t + i k_\perp M + i k_{\parallel} L)}.
\label{LHphi}
\end{equation}
From equation (\ref{LHphi}) we obtain
\begin{equation}
\delta {\bf E} = \left( -i k_{\parallel} \delta \phi, - i k_{\perp} \delta \phi, - \delta \phi' \right), 
\label{dELH}
\end{equation}
where primes denotes the spatial derivative in the ${\bf N}$ direction. 

The perpendicular electron velocity fluctuations are given by $\delta {\bf V}_e = \delta {\bf E} \times {\bf B}/|{\bf B}|^2$, where 
${\bf B} = (B, 0, 0)$ is along the ${\bf L}$ direction. We then obtain 
\begin{equation}
\delta V_{e,L} = - \frac{e}{m_e} \frac{k_{\parallel} \delta \phi}{(\omega - k_{\perp} V_{e,M})},
\label{dVeL}
\end{equation}
\begin{equation}
\delta V_{e,M} = - \frac{\delta \phi'}{B},
\label{dVeM}
\end{equation}
\begin{equation}
\delta V_{e,N} = i \frac{k_{\perp} \delta \phi}{B},
\label{dVeN}
\end{equation}
\begin{equation}
\delta V_{e,N}' = i \frac{k_{\perp} \delta \phi'}{B} - i \frac{B' k \delta \phi}{B^2}.
\label{dVeNp}
\end{equation}
Here $V_{e,M}$ is the background cross-field electron drift in the spacecraft frame. 

From the electron continuity equation we obtain: 
\begin{equation}
\delta n_e = \frac{n}{\omega - k_{\perp} V_{e,M}} \left( -i \delta V_{e,N}' - i \frac{n'}{n} \delta V_{e,N} + k_{\perp} \delta V_{e,M} \right), 
\label{necont}
\end{equation}
where $n' = \partial n/\partial N$.
By substituting equations (\ref{dVeM})--(\ref{dVeNp}) into equation (\ref{necont}) we find that 
\begin{equation}
\delta n_e \approx \frac{n k_{\perp}}{B (\omega - k_{\perp} V_{e,M})} \left( \frac{n'}{n} - \frac{B'}{B} \right) \delta \phi. 
\label{necont2}
\end{equation}
We note that for frozen in electrons $-i \delta V_{e,N}' + k \delta V_{e,M} \approx 0$ when $B'/B$ is small. 
For lower hybrid-like waves $k_{\perp} \gg k_{\parallel}$, so for simplicity we have neglected the contribution to 
$\delta n_e$ from $k_{\parallel} \delta V_{e,L}$. 
Based on observations $\omega - k_{\perp} V_{e,M} < 0$, $n'/n > 0$ and $B'/B < 0$, so $\delta n_e$ is predicted to be anti-correlated with $\delta \phi$. Since $|\delta {\bf V}_{e}| \gg |\delta {\bf V}_{i}|$ the 
current density is $\delta {\bf J} = - e n_e \delta {\bf V}_e$, which is given by
\begin{equation}
\delta {\bf J} = \left( \frac{e^2 n_e k_{\parallel} \delta \phi}{m_e (\omega - k_{\perp} V_{e,M})}, \frac{e n_e \delta \phi'}{B} , -\frac{i e n_e k_{\perp} \delta \phi}{B} \right).
\label{dJLH}
\end{equation}

The magnetic field fluctuations can be calculated using Ampere's law: 
\begin{equation}
\nabla \times \delta {\bf B} = \mu_0 \delta {\bf J}. 
\label{ampere2}
\end{equation}
This yields three equations:
\begin{equation}
i k_{\perp} \delta B_N - \delta B_M' = \frac{\omega_{pe}^2 k_{\parallel} \delta \phi}{c^2 (\omega - k_{\perp} V_{e,M})}.
\label{BLLH}
\end{equation}
\begin{equation}
\delta B_{L}' - i k_{\parallel} \delta B_N = \frac{\mu_0 e n_e \delta \phi'}{B} ,
\label{BMLH}
\end{equation}
\begin{equation}
i k_{\parallel} \delta B_M - i k_{\perp} \delta B_{\parallel} = -\frac{i \mu_0 e n_e k_{\perp} \delta \phi}{B},
\label{BNLH}
\end{equation}
Since we have assumed $k_{\perp} \gg k_{\parallel}$ and $\delta B_L \gtrsim \delta B_M$ equation (\ref{BNLH}) reduces to
\begin{equation}
\delta \phi = \frac{B \delta B_{L}}{\mu_0 e n_e},
\label{phiBapp}
\end{equation}
which is used to determine the wave potential from the fluctuating magnetic field. Similarly, equation (\ref{BMLH}),
reduces to equation (\ref{phiBapp}) for $\delta B_{L}' \gg k_{\parallel} \delta B_N$.
Thus, the single-spacecraft method does not require the lower hybrid wave to satisfy the plane-wave approximation,
and still holds when the wave is localized in the direction perpendicular to ${\bf k}$ and ${\bf B}$.

From equations (\ref{necont2}) and (\ref{phiBapp}) we estimate 
\begin{equation}
(\delta n_e/n_e)/(\delta B_L/B) \approx \frac{f_{ce}}{f_{pe}} \frac{c k_{\perp} d_e}{(\omega - k_{\perp} V_{e,M})} \left( \frac{n'}{n} - \frac{B'}{B} \right).
\label{dnndBBB}
\end{equation}
This suggests qualitatively that $(\delta n_e/n_e)/(\delta B_L/B)$ decreases toward the magnetopause from the magnetospheric side, which is consistent with observations. 

Equation (\ref{phiBapp}) can become invalid if the following occur: 

(1) Ion velocity fluctuations become comparable to the electron fluctuations. 

(2) Density perturbations become sufficiently large to invalidate $\delta {\bf J} = - e n \delta {\bf V}_e$. 

(3) Thermal electron effects become large enough for electrons to deviate significantly from $\delta {\bf E} \times {\bf B}$ drift. 

For lower hybrid waves at the magnetopause we find that they propogate in the $\pm {\bf M}$ direction in 
the spacecraft frame. We 
therefore determine $\delta \phi$ and $v_{ph}$ from $\delta E_M$, so the relevant current to consider is 
$\delta J_N$. Regarding point (1) we conclude that $\delta {\bf V}_i \ll \delta {\bf V}_e$ because $f \gg f_{ci}$. 
This is supported by observations of the 37.5~ms ion moments, which show that the $|\delta {\bf V}_i|$ is only a few 10's of km~s$^{-1}$.
We can therefore approximate the normal current as $\delta J_N = -e n \delta V_{e,N} - e \delta n_e \delta V_{e,N}$. The deviation in 
$\delta J_N$ due to density fluctuations, point (2), is proportional to $\delta n/n$. In both events we find that 
$|\delta n/n|$ is typically $0.1$ (with peak values of $|\delta n/n| \approx 0.2$), so we might expect an uncertainty in equation (\ref{phiBapp}) of $\sim 10 \%$. 
Regarding point (3), for thermal electrons the electron velocity in the ${\bf N}$ direction can be approximated by
\begin{equation}
\delta V_{e,N} \approx - \frac{e}{m_e} \frac{\Omega_{ce}}{\Omega_{ce}^2 + v_e^2 k_{\perp}^2/2} \delta E_M,
\label{VeNthermal}
\end{equation}
where $v_e = \sqrt{2 k_B T_e/m_e}$ is the electron thermal speed. 
Thus, the effect of finite $T_e$ is to reduce $\delta V_{e,N}$ and thus likely reduce $v_{ph}$ estimated 
from single-spacecraft methods. For example, for $k = 4 \times 10^{-4}$~m$^{-1}$, $B = 25$~nT, and 
$T_e = 100$~eV (from section \ref{14Dec2015}) we find that $\delta V_{e,N}$ decreases by $\approx 15 \, \%$ 
compared to the $T_e \rightarrow 0$ limit. 
Overall, the observations suggest that these effects are relatively minor and do not invalidate the single spacecraft methods used to estimate $v_{ph}$. This is evident in section \ref{28nov2016}, where the lower hybrid waves can be calculated from four-spacecraft measurements without assumptions. 

From equation (\ref{BLLH}) we can estimate the relative amplitudes of parallel and perpendicular magnetic
field fluctuations. If we assume that the waves are approximately planar, we can estimate the amplitude of
magnetic fields perpendicular to ${\bf B}$ to be
\begin{equation}
\delta B_{\perp} \approx \frac{\Omega_{ce}}{(\omega - k_{\perp} V_{e,M})} \frac{k_{\parallel}}{k_{\perp}} \delta B_{\parallel},
\label{Bperp}
\end{equation}
where $\Omega_{ce}$ is the angular electron cyclotron frequency. 
This equation allows $k_{\parallel}$ to be estimated, although it depends on the wave frequency, 
which may differ from the observed frequency in the spacecraft reference frame.
The electron and magnetic field energy densities are then given by
\begin{equation}
W_e = \frac{1}{2} n_e m_e \delta V_e^2 = \frac{1}{2} \frac{n_e m_e}{B_0^2} \left( 1 + \frac{\Omega_{ce}^2}{(\omega - k_{\perp} V_{e,M})^2}
\frac{k_{\parallel}^2}{k_{\perp}^2} \right) k_{\perp}^2 \delta \phi^2,
\label{We}
\end{equation}
\begin{equation}
W_B = \frac{1}{2} \frac{\delta B^2}{\mu_0} = \frac{1}{2} \left( 1 + \frac{\Omega_{ce}^2}{(\omega - k_{\perp} V_{e,M})^2} \frac{k_{\parallel}^2}{k_{\perp}^2} \right) \frac{\delta \phi^2 \mu_0 e^2 n_e^2}{B_0^2}.
\label{WB}
\end{equation}
By taking the ratio of $W_e$ and $W_B$ we estimate the dispersion relation in the spacecraft reference frame
using the following
\begin{equation}
\frac{W_e(\omega)}{W_B(\omega)} = d_e^2 k_{\perp}^2(\omega) \rightarrow k_{\perp}(\omega) = \frac{1}{d_e} \sqrt{\frac{W_e(\omega)}{W_B(\omega)}},
\label{WBWe}
\end{equation}
where $d_e = c/\omega_{pe}$ is the electron inertial length. Equations (\ref{We}) and (\ref{WB}) reduce 
to equations (\ref{We1}) and (\ref{WB1}) for $V_{e,M} = 0$ or $\omega = k_{\perp} V_{e,M}/2$.

Finally, we consider the case when electron velocity fluctuations associated with lower hybrid waves cannot be measured
directly. In this case we assume ${\bf V}_{e,\perp} = \delta {\bf E} \times {\bf B}/B^2$. We then calculate
the electron kinetic energy perpendicular to ${\bf B}$ and the magnetic field energy density parallel to ${\bf B}$
\begin{equation}
W_{e,\perp} = \frac{1}{2} n_e m_e V_{e,\perp}^2 = \frac{1}{2} n_e m_e \frac{k_{\perp}^2 \delta \phi^2}{B^2},
\label{Weperp}
\end{equation}
\begin{equation}
W_{B,\parallel} = \frac{1}{2} \frac{\delta B_{\parallel}^2}{\mu_0} = \frac{1}{2} \frac{\mu_0 e^2 n_e^2 \delta \phi^2}{B^2}.
\label{WBpar}
\end{equation}
From equations (\ref{Weperp}) and (\ref{WBpar}) we obtain
\begin{equation}
\frac{W_{e,\perp}(\omega)}{W_{B,\parallel}(\omega)} = d_e^2 k_{\perp}^2(\omega) \rightarrow
k_{\perp}(\omega) = \frac{1}{d_e} \sqrt{\frac{W_{e,\perp}(\omega)}{W_{B,\parallel}(\omega)}}.
\label{WBWe2}
\end{equation}
Thus, the dispersion relation of lower hybrid waves can be approximated from the fluctuating electric and magnetic fields, without high-resolution electron moments. In both cases the wavelength is computed using
$\lambda = 2 \pi/k_{\perp}$.

\begin{figure*}[htbp!]
\begin{center}
\includegraphics[width=90mm, height=60mm]{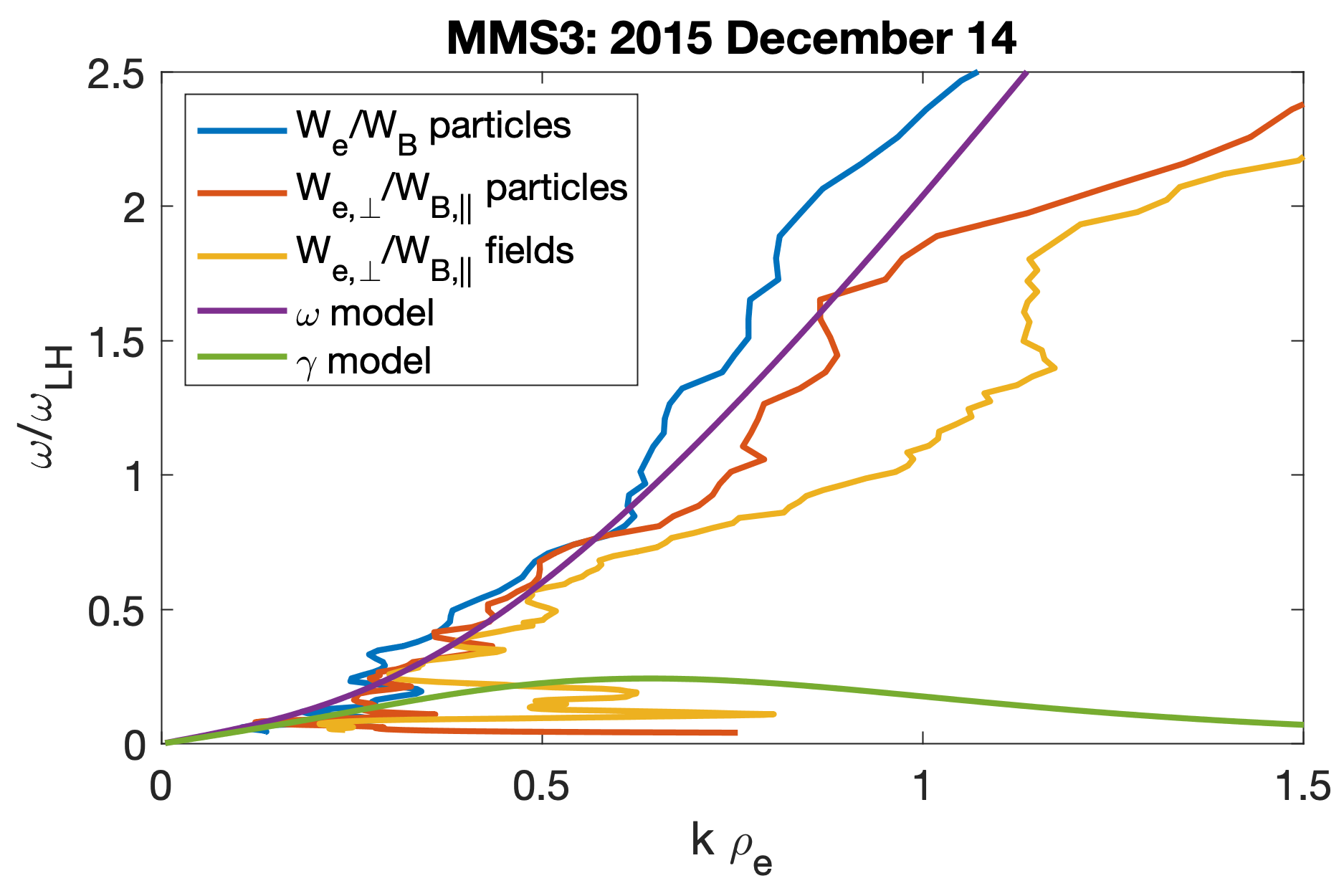}
\caption{Lower hybrid wave dispersion relations calculated using equations (\ref{WBWe}) and (\ref{WBWe2}) for the lower hybrid waves observed on 2015 December 14 from Figure \ref{Disprel14Dec}.
The blue, red, and yellow curves are the dispersion relations computed equation (\ref{WBWe}) using electron
moments, equation (\ref{WBWe2}) using electron moments, and equation (\ref{WBWe2}) using only fields data.
The purple and green lines are the modeled LHDI dispersion relation and growth rate.}
\label{Obsdisprel}
\end{center}
\end{figure*}

Figure \ref{Obsdisprel} compares the dispersion relations computed using equations (\ref{WBWe}) and
(\ref{WBWe2}). The blue line show the dispersion relations using equation (\ref{WBWe}) and electron moments (reproduced from Figure \ref{Disprel14Dec}),
the red line shows the dispersion relation computed from equation
(\ref{WBWe2}) using electron moments for $W_{e,\perp}$, and the yellow line shows the dispersion from 
(\ref{WBWe2}) using ${\bf E}$ to estimate $W_{e,\perp}$. All three methods predict similar $k$ for $\omega/\omega_{LH} \sim 0.5$, where the electric field power peaks (Figure \ref{Disprel14Dec}). At higher frequencies larger $k$ are calculated when only fields are used,
and the smallest $k$ are predicted when equation (\ref{WBWe}) is used. For $\omega/\omega_{LH} \lesssim 0.25$,
there is a large increase in $k$ when only fields are used. However, in this frequency range spatial changes
due to the motion of the magnetopause with respect to spacecraft may result in field or particle powers that
are not associated with waves. Overall, similar qualitative results are found for the three methods, and agree
well with the model dispersion relation, and similar $k$ are found for the lower hybrid frequencies where the
electric field power peaks.

\end{document}